\documentclass[reprint,amsmath,amssymb,aps,nofootinbib,longbibliography]{revtex4-2}
\usepackage{graphicx}
\usepackage{amsmath}
\usepackage{dcolumn}
\usepackage{bm}

\usepackage{mathtools}
\usepackage{blkarray}
\usepackage{footnote}
\usepackage{hyperref}
\hypersetup{
    colorlinks,
    citecolor=blue,
    filecolor=blue,
    linkcolor=blue,
    urlcolor=blue,
    pdfdisplaydoctitle=true}
\usepackage[toc,page]{appendix}
\usepackage{adjustbox}
\usepackage{float}
\usepackage{color}

\usepackage{scalerel}
\usepackage{tikz, braket}
\usetikzlibrary{svg.path}

\definecolor{orcidlogocol}{HTML}{A6CE39}
\tikzset{
  orcidlogo/.pic={
    \fill[orcidlogocol] svg{M256,128c0,70.7-57.3,128-128,128C57.3,256,0,198.7,0,128C0,57.3,57.3,0,128,0C198.7,0,256,57.3,256,128z};
    \fill[white] svg{M86.3,186.2H70.9V79.1h15.4v48.4V186.2z}
                 svg{M108.9,79.1h41.6c39.6,0,57,28.3,57,53.6c0,27.5-21.5,53.6-56.8,53.6h-41.8V79.1z M124.3,172.4h24.5c34.9,0,42.9-26.5,42.9-39.7c0-21.5-13.7-39.7-43.7-39.7h-23.7V172.4z}
                 svg{M88.7,56.8c0,5.5-4.5,10.1-10.1,10.1c-5.6,0-10.1-4.6-10.1-10.1c0-5.6,4.5-10.1,10.1-10.1C84.2,46.7,88.7,51.3,88.7,56.8z};
  }
}
\newcommand\orcidicon[1]{\href{https://orcid.org/#1}{\mbox{\scalerel*{
\begin{tikzpicture}[yscale=-1,transform shape]
\pic{orcidlogo};
\end{tikzpicture}
}{|}}}}

\begin{document}

\title{ Superconductivity Studied by Solving \textit{Ab Initio} Low-Energy Effective Hamiltonians for Carrier Doped CaCuO$_2$, Bi$_2$Sr$_2$CuO$_6$, Bi$_2$Sr$_2$CaCu$_2$O$_8$, and  HgBa$_2$CuO$_4$}
\author{
Michael Thobias Schmid$^{1}$ \orcidicon{0000-0003-2724-0621}, 
Jean-Baptiste Mor\'ee$^{1,2}$ \orcidicon{0000-0002-0710-9880}, 
Ryui Kaneko$^{1,3}$ \orcidicon{0000-0001-7994-6381},
Youhei Yamaji$^{4}$ \orcidicon{0000-0002-4055-8792}, 
and Masatoshi Imada$^{1,3,5}$ \orcidicon{0000-0002-5511-2056}
}

\affiliation{%
$^{1}$ Waseda Research Institute for Science and Engineering, Waseda University, 3-4-1, Okubo, Shinjuku, Tokyo 169-8555, Japan\\
$^{2}$ RIKEN Center for Emergent Matter Science, 2-1 Hirosawa, Wako, Saitama 351-0198, Japan\\
$^{3}$ Department of Engineering and Applied Sciences
Sophia University, 7-1, Kioi-cho, Chiyoda, Tokyo 102-8554, Japan\\
$^{4}$ Research Center for Materials Nanoarchitectonics (MANA) and Center for Green Research on Energy and Environmental Materials (GREEN), National Institute for Materials Science (NIMS), Namiki, Tsukuba-shi, Ibaraki, 305-0044, Japan\\
$^{5}$ Toyota Physical and Chemical Research Institute, 41-1, Yokomichi, Nagakute, Aichi 480-1192, Japan
}      

\begin{abstract}
Understanding the materials dependence together with the universal controlling parameter of superconductivity (SC) in copper oxide superconductors is one of the major challenges in condensed matter physics.
Here, we numerically analyze SC by using {\it ab initio} low-energy effective Hamiltonians consisting of the antibonding (AB) combination of Cu $3d_{x^2-y^2}$ and O $2p_{\sigma}$ orbitals without adjustable parameters.
We have performed the state-of-the-art variational Monte Carlo calculations for the four carrier doped cuprates with diverse experimental optimal SC critical temperature $T_{c}^{\rm opt}$: CaCuO$_2$ ($T_{c}^{\rm opt} \sim 110$ K), Bi$_2$Sr$_2$CuO$_6$ ($T_{c}^{\rm opt} \sim 10$-$40$ K), Bi$_2$Sr$_2$CaCu$_2$O$_8$ ($T_{c}^{\rm opt} \sim 85$-$100$ K), and  HgBa$_2$CuO$_4$ ($T_{c}^{\rm opt} \sim 90$ K). 
Materials and hole doping concentration ($\delta$) dependencies of the SC order parameter $F_{\rm SC}$ and the competition with spin and charge orders 
show essential and quantitative agreement with the available experiments on the four materials
in the following points:
(1) In a wide range  $0.05 \leq \delta \leq 0.25$, the ground state is commonly the uniform SC state, which is severely competing with the charge, spin stripe and antiferromagnetic (AFM) states.
(2) $F_{\rm SC}$ at the optimum doping shows amplitude consistent with the superfluid density measured in the muon spin resonance ($\mu$SR) and its dome structure found in $\delta$ dependence shows consistency with that of the SC gap in the tunneling and photoemission measurements.
Based on the confirmed materials dependence, we further find insights into the universal SC mechanism:  
(I) $F_{\rm SC}$ increases with the ratio $U/|t_1|$ {within the available realistic materials}, indicating that $U/|t_1|$ is the principal component controlling the strength of the SC {in the real materials dependence}.
Here, $U$ and $t_1$ are the on-site Coulomb repulsion and the nearest neighbor hopping, respectively, in the {\it ab initio} Hamiltonians.
(II) A universal scaling $T_{c}^{\rm opt}\sim 0.16 \lvert t_1 \rvert F_{\rm SC}$ holds.
(III) SC is enhanced and optimized if $U$ is increased beyond the real available materials, and it is further enhanced when the off-site interaction is reduced, {while the presence of the off-site interaction is important to make the SC ground state against other competing states.}
The present findings provide useful clues for the design of new SC materials with even higher $T_{c}^{\rm opt}$. 
\end{abstract}
\maketitle

\section{Introduction}
\label{sec:introduction}
The mechanism and origin of the large superconducting (SC) gap, high superfluid density, and high critical temperatures $T_c$ observed in high-$T_{c}$ superconductors, such as copper oxides, remain a central challenge in condensed matter physics.
In these copper oxides, the $d$-wave SC state is severely competing with other orders, such as spin and charge stripes or antiferromagnetic (AFM) states, and the observed $T_{c}$ widely ranges from above $130\, \text{K}$ to below $10\, \text{K}$. 
Understanding and reproducing these diverse phenomena without relying on adjustable parameters is hence desirable, especially when clarifying their origin. 
When {\it ab initio} calculations are able to reproduce systematic materials dependence quantitatively by solely relying on their crystal structures, it  provides us with valuable insight into the universal mechanism behind and into the principal components for the enhancement of SC beyond existing materials.

Many studies have suggested severe competitions of the SC with charge/spin stripe and AFM states theoretically based on simplified Hubbard-like or $t$-$J$ Hamiltonians as models of the cuprate superconductors~\cite{PhysRevLett.113.046402,PhysRevB.85.081110,zhao2017,doi:10.1126/science.aam7127,darmawan2018,ido2018,PhysRevX.13.011007,10.21468/SciPostPhys.12.6.180}. 
A positive correlation between $U$ and $T_{c}$ or SC tendency was also pointed out by taking $U$ as an adjustable parameter in the Hubbard type Hamiltonians~\cite{yokoyama2013}. 
However, {\it ab initio} studies without adjustable parameters are few and it is not clear whether the diversity of the materials dependence can be accounted for. There still exists a limited number of {\it ab initio} studies: The phase diagram including the SC phase was reproduced by solving the {\it ab initio} Hamiltonian for a particular case of Hg compound~\cite{ohgoe2020}. 
{\it Ab initio} Hamiltonians were derived by Nilsson {\it et al.} for several cuprate compounds, which reported an empirical observation without solving the Hamiltonians that the experimental optimal $T_{c}$ is generally higher for larger $U/|t|$ in their parameters~\cite{PhysRevB.99.075135,moree2022}.  
The relation of the charge transfer energy to $T_c$ was also pointed out~\cite{Weber_2012,PhysRevResearch.3.033157}. 
Aside from the cuprates, there exist some {\it ab initio} studies on strong-coupling superconducting materials such as the iron-based superconductors~\cite{Misawa2014-ym}, fullerene~\cite{Nomura15Ful}, and nickelates~\cite{Kitatani2020-ra} to discuss the superconducting properties. 
However, to our knowledge, there exist no systematic studies on the SC properties by solving solely {\it ab initio} Hamiltonians without adjustable parameters with the help of accurate many-body solvers to reveal the origin of the diverse materials dependence, especially for the challenging cuprates. 
Unless reproducing the materials dependent properties quantitatively, the universal mechanism would also not be able to be identified confidently  either. 

In this paper we show properties of typical cuprate superconductors calculated by solving the {\it ab initio} Hamiltonians of four families of materials, namely carrier doped CaCuO$_2$, HgBa$_2$CuO$_4$ (abbreviated as Hg1201 hereafter), Bi$_2$Sr$_2$CuO$_6$ (Bi2201 hereafter), and Bi$_2$Sr$_2$CaCu$_2$O$_8$ (Bi2212 hereafter)~\cite{moree2022}, by applying a state-of-the-art quantum many-body solver based on the variational Monte Carlo (VMC) algorithm~\cite{misawa2019,tahara2008}, including the combination with neural network~\cite{nomura2017,nomura2021} elaborated from earlier variational algorithms~\cite{YokoyamaShiba1987,GrossJoyntRice1987,Gross1988,Capriotti2001}.  
It is experimentally known that the optimum critical temperature $T_{c} ^{\rm opt}$ is realized at around $\delta=0.1$-0.15~\cite{Azuma1992,yamamoto2000} for doped CaCuO$_2$ ($T_{c}^{\rm opt}\sim$ 110K) and Hg1201 
($T_{c}^{\rm opt}\sim$ 90K), and at around $\delta=0.15$-0.25~\cite{koike1989,fang1992,Hobou2009} for doped Bi2212 ($T_{c}^{\rm opt}\sim$ 85-100K) and Bi2201 ($T_{c}^{\rm opt}\sim$ 10-40K). 
We elucidate that similarity and diversity among the four families, especially the amplitude of the SC order parameters and  $T_{c}^{\rm opt}$ in the experiments are accounted for by using the present {\it ab initio} results, which provides us with insights into the materials dependence and the universal mechanism:

{We emphasize that our {\it ab initio} analyses contain essential differences  from most of the Hubbard models studies. One important difference  is the presence of the realistic off-site interactions. We will clarify that this crucially stabilizes the charge uniform SC state without clear stripe long-range order.} 

The dominance of the SC for all the four families is successfully demonstrated. In addition, the $\delta$ dependence of $F_{\rm SC}$ has a dome structure with the peak at $\delta\sim 0.05$-$0.1$ consistently with the experimental indications. On the other hand, the dome peak of $T_c$ appears at larger $\delta>0.1$ in the experiment. This shift from the dome peak of $F_{\rm SC}$ is understood from the decreasing renormalization factor with decreasing doping, which does not affect $F_{\rm SC}$ but $T_c$.

Although $T_{c}^{\rm opt}$ has a variety among these four families, we show universally that (I) the higher SC order parameter $F_{\rm SC}$ at the optimal doping basically results from a larger ratio $U/|t_1|$, where $U$ is the on-site repulsive Coulomb interaction and $t_1$ is the nearest neighbor hopping in  our {\it ab initio} parameters of single-band effective Hamiltonian for the AB orbital of strongly hybridized Cu $3d_{x^2-y^2}$ and O $2p_{\sigma}$ orbitals. Furthermore, we show that (II) $T_{c}^{\rm opt}$ is determined by the scaling $T_{c}^{\rm opt}\sim 0.16 \lvert t_1 \rvert F_{\rm SC}$.

The $\delta$ dependence of the local energy suggests a universal superconducting mechanism: Though the bare interaction is strongly repulsive, the Mottness converts it to the strong effective attraction required for the Cooper pairing. 

Despite monotonic increase of $F_{\rm SC}$ with $U/|t_1|$ within the existing materials, we further show that (III) larger $U/|t_1|$ beyond the {\it ab initio} values makes the peak of $F_{\rm SC}$ followed by the reduction. 
We find that $F_{\rm SC}$ can be roughly 30\% more enhanced than the {\it ab initio} case when $U$ is 20\% increased beyond the {\it ab initio} value by preserving the transfer and other off-site interaction parameters.
We also show that $F_{\rm SC}$ is further enhanced to as much as the double of the existing material by the additional reduction of the off-site Coulomb interaction.
These searches beyond the {\it ab initio} parameters for the existing materials offer a guide for future experimental materials design.

This paper is organized as follows. 
Section \ref{sec:methods} presents the methods and computational details: First, the effective Hamiltonians studied in this paper are summarized in Sec. \ref{sec:effeciveham}.
Then, we give the numerical methods in Sec.~\ref{sec:numerocalmethods} and define the physical quantities in Sec.~\ref{sec:quantities}.
We present in Sec.~\ref{sec:ab initio results} the results for each of the four families of compounds.
Based on the obtained {\it ab initio} results, in Sec.~\ref{sec:results beyond ab initio}, we further explore the direction to enhance and optimize the SC order parameter by controlling the effective interaction parameters beyond the {\it ab initio} values, to gain insights into the future materials design. In Sec.~\ref{sec:discussion}, we discuss our analyses. Summary and conclusion are given in Sec.~\ref{sec:summary}.

\section{Methods}
\label{sec:methods}

\subsection{{\it Ab initio} effective Hamiltonian }
\label{sec:effeciveham}

Within this paper, we solve the \textit{ab initio} single-band effective Hamiltonians for CaCuO$_2$, Hg1201, Bi2201, and Bi2212 as derived in Ref.~\cite{moree2022}. {It should be noted that the single band is constructed from the AB orbital of strongly hybridized Cu 3$d_{x^2-y^2}$ and O 2$p_{\sigma}$ orbitals and not from the atomic single orbital of Cu 3$d_{x^2-y^2}$. This is justified by very large hybridization gap of the AB and bonding (B) or nonbonding (NB) orbitals. See Appendix D of Ref.~\cite{moree2022} and the last paragraph of Sec.~\ref{sec:discussion}  in this paper.}  Here the transfer and interaction parameters are derived at values close to the experimental optimum hole concentration (at 10\% doping for CaCuO$_2$ and Hg1201 and at 20\% for the two Bi compounds). 
This choice is appropriate in this paper, because properties at optimum hole concentration are the central subject.
The effective Hamiltonians have the form
\begin{align}
	\mathcal H &= \mathcal H_{\rm kin} + \mathcal H_{U} + \mathcal H_{V} \label{eq:Ham}
\end{align}
with
\begin{eqnarray}
\mathcal H_{\rm kin} &=& \sum_{i, j, \sigma} t_{ij} c^{\dagger}_{i\sigma} c^{\phantom{\dagger}}_{j\sigma}, \nonumber \\
\mathcal H_{U} &=& \sum_{i} U n_{i \uparrow} n_{i \downarrow}, \nonumber \\
\mathcal H_{V} &=& \frac{1}{2}\sum_{i \neq j} V_{ij} n_{i} n_{j}. \nonumber 
\end{eqnarray}
Here $i,j$ are the unit cell indices of the maximally localized Wannier function~\cite{marzari1997maximally, souza2001maximally}  and $c^{\dagger}_{i\sigma}$ ($c^{\phantom{\dagger}}_{i\sigma}$) is the corresponding creation (annihilation) operator of an electron of spin $\sigma$ at the site $i$. 
The number operator is given by $n_{i} = \sum_\sigma n_{i\sigma}$ and $n_{i\sigma}=c^{\dagger}_{i\sigma} c^{\phantom{\dagger}}_{i\sigma}$. 
Hopping amplitudes $t_{ij}$ in the kinetic energy $\mathcal H_{\rm kin}$ depend on the relative coordinate vector  $\bm r_i - \bm r_j$ by assuming the translational invariance of the crystal structure. 
The direct effective Coulomb interaction given by $\mathcal H_U$ is scaled by the on-site interaction $U$, and off-site interaction $\mathcal H_V$ is the sum over the combination of the site $i$ and $j$, which is proportional to $V_{ij}$.
Leading values for all of the four materials are listed in Table~\ref{tab:ham}. For longer ranged part of $t_{ij}$ and $V_{ij}$, see Sec.~S1A in Supplemental Material (SM)~\cite{SM_Michael}. Note that the Hamiltonian parameters for Hg1201 in Ref.~\cite{moree2022}, which we employ, are improved from Ref.~\cite{hirayama2019}. 
It results in slightly different physical quantities on the quantitative level in the present solution in comparison to Ref.~\cite{ohgoe2020}. 

{From four different materials, we learn that the realistic range of available {\it ab initio} Hamiltonian parameters is estimated to be $6\lesssim U/|t_1|\lesssim 10, 0.2\lesssim|t_2/ t_1|\lesssim 0.3, 1.2\lesssim|V_1/ t_1|\lesssim 2.0, 0.5\lesssim|V_2/ t_1|\lesssim 1.2$ etc. (see also Table~\ref{tab:pdsum3}). In this paper, we investigate whether the diversity of the SC properties can be understood within this range of parameters. }
\begin{table}[tbh!]
    \caption{\textit{Ab initio} single-band effective Hamiltonian for CaCuO$_2$, Hg1201, Bi2212, and Bi2201 taken from Ref.~\cite{moree2022}. $U$ is the on-site interaction. The $n$th-neighbor hopping amplitude and Coulomb interaction are denoted as $t_n$ and $V_n$, respectively. Interlayer hoppings and interactions are neglected here. All values are given in eV.}
    \label{tab:ham} 
    \begin{ruledtabular}
        \begin{tabular}{ l c c c c c }
            & $t_1$ &  $t_2$ & $t_3$ & $t_4$  & $t_5$ \\
            \colrule
            CaCuO$_2$ &-0.521 & 0.132 & -0.047 & 0.008 & 0.000  \\
            Hg1201    &-0.544 & 0.111 & -0.043 & 0.010 & 0.000 \\
            Bi2212    &-0.452 & 0.135 & -0.053 & -0.001 & 0.007  \\
            Bi2201    &-0.527 & 0.140 & -0.042 & 0.009 & -0.007  \\
            \colrule
            & $U$ & $V_1$  & $V_2$  & $V_3$  & $V_4$  \\
            \colrule              
            CaCuO$_2$  & 4.221 & 0.969 & 0.539 & 0.380 & 0.316 \\
            Hg1201  & 3.999 & 1.002 & 0.596 & 0.448 & 0.389    \\
            Bi2212  & 4.226 & 0.915 & 0.518 & 0.366 & 0.312   \\
            Bi2201  & 4.393 & 1.030 & 0.602 & 0.450 & 0.395   \\
        \end{tabular}
    \end{ruledtabular}
\end{table}

We note that the effective Hamiltonian parameters in Eq.~(\ref{eq:Ham}) are restricted to a single CuO$_2$ layer.
However, in the case of multilayer cuprates CaCuO$_2$ and Bi2212, the distance between CuO$_2$ layers is comparable to the cell parameter along $x$ direction, so that interlayer coupling parameters (given in Table~\ref{tab:bi2212Vinter} in Appendix for Bi2212) also exist in the effective Hamiltonian \cite{moree2022}, and its amplitude $V_n^l \lesssim 0.6$ eV is comparable to that of the intralayer off-site interaction $V_n \lesssim 0.9$ eV.
This interlayer coupling is ignored in Eq.~(\ref{eq:Ham}), but potentially plays a role in the SC properties.
This role is actually examined in Sec.~\ref{sec:Bi2212} in the case of Bi2212, which ensures that the SC order parameter $F_{\rm SC}^{\infty}$ (defined later in Sec.~\ref{sec:quantities}) and physical quantities are essentially not affected by the interlayer coupling within the present case of CaCuO$_2$ and Bi2212 as we discuss in Sec~\ref{sec:Bi2212}.
Thus, we restrict to Eq.~(\ref{eq:Ham}) even in the case of CaCuO$_2$ and Bi2212.
We employ this ``single-layer approximation" for all the four materials throughout this paper unless otherwise noted.

\subsection{Numerical Methods}
\label{sec:numerocalmethods}
We solve the Hamiltonian in Eq.~(\ref{eq:Ham}) by applying the many-variable variational Monte Carlo (mVMC) method~\cite{tahara2008,misawa2019} with a trial wave function of the form 
$\ket{\psi} = \mathcal P^\mathrm{G} \mathcal P^\mathrm{J} \mathcal P^\mathrm{dh} \ket{\phi^\mathrm{pair}}$. 
Here we consider the Gutzwiller factor 
$\mathcal P^\mathrm{G} = \exp(- g \sum_i  n_{i\uparrow} n_{i \downarrow})$~\cite{gutzwiller1963}, 
the Jastrow correlation factor 	$P^\mathrm{J} = \exp( \sum_{i<j} \alpha_{ij} n_in_j)$~\cite{jastrow1955, capello2005}, 
the doublon-holon correlation factor 
$\mathcal P^\mathrm{dh} = \exp[-\sum_{m=0}^{4} \sum_{\ell = 1,2} \alpha^{(\ell)}_{m} \sum_{i} \xi^{(\ell)}_{i(m)}]$~\cite{yokoyama1990}, 
and a generalized pairing wave function of the form 
$\ket{\phi}^{\mathrm{pair}} = (\sum_{i\sigma, j \sigma^\prime} f_{i\sigma,j \sigma^\prime} c^\dagger_{i\sigma} c^\dagger_{j\sigma^\prime})^{N_\mathrm{e}/2}\ket{0}$. 
The variational parameters are $g$, $\alpha_{ij}$, $\alpha^{(\ell)}_{m}$, and $ f_{i\sigma,j \sigma^\prime}$.

We also supplement the mVMC method with the restricted Boltzmann machine (RBM)~\cite{nomura2017} and the first-order Lanczos step to improve the wave function and also to take the zero limit of the variance extrapolation to improve the estimate by following the variance extrapolation method~\cite{imada2000,kashima2001,sorella2001} and by using the simple mVMC, mVMC +Lanczos, and mVMC+RBM results (see Appendices~\ref{app:RBM_Lanczos}, \ref{app:VarExt}, and Ref.~\cite{ido2022} for the detailed procedure. 

Competing states with spin and/or charge order or strong fluctuations can be studied by imposing the mean-field order at the initial trial wave function~\cite{tahara2008}. 
The correlated metallic state without any symmetry breaking can also be studied by using the ground-state wave function of the noninteracting system as an initial state. 
These initial states are then relaxed to lower the energy by optimizing the variational parameters. 
If the competing states exist, the optimization leads to multiple locally stable solutions. 
The true ground state is determined by comparing the total energy after taking the variance extrapolation described in Appendix~\ref{app:VarExt} and if possible after careful size extrapolation to see the thermodynamic limit.

The computational details are the following.
For all numerical solutions of finite-size lattices subsequently presented in this paper, we assumed the antiperiodic-periodic boundary condition on a $N= L \times L$ square lattice of length $L$, where $N$ is the number of sites on the single-layer system by ignoring the interlayer coupling except for Bi2212.  For Bi2212, we examine the two-layer system to examine the effect of interlayer coupling, because a unit cell of Bi2212 contains two layers and the interlayer coupling could play roles in the SC. 
Within a layer, hoppings and interactions were taken into account up to $t_9$ and $V_9$, i.e., up to the 2D distance $\bm R = (3,3)$ in the unit of the Cu-Cu distance, while contributions smaller than $0.001$ eV were ignored.  
Unless explicitly mentioned this is applied throughout the whole paper. 
We concentrate solely on hole doped cases, where the hole doping is given via $\delta = 1 -  N_e/N$  and $N_e$ is the number of electrons in the system. 

\subsection{Physical Quantities}
\label{sec:quantities}

The physical quantities discussed in this paper are defined as follows:
The total energy per site $ E/N = \braket{\mathcal H}/N$ is calculated after the variance extrapolation as it is summarized in Appendix~\ref{app:VarExt}.
To see whether the state has spin and charge order, we compute the spin structure factor
\begin{equation}
S_s(\bm q) = \frac{1}{N} \sum_{i,j} \braket{ \bm S_i \cdot \bm S_j } \mathrm{e}^{\mathrm{i} \bm q (\bm r_i - \bm r_j)}.
\end{equation} 
and
the charge structure factor 
\begin{equation}
S_c(\bm q) = \frac{1}{N} \sum_{i,j} \braket{ n_i n_j } \mathrm{e}^{\mathrm{i} \bm q (\bm r_i - \bm r_j)},
\end{equation}
where $\bm S_i=(S_i^x,S_i^y,S_i^z)=\sum_{\sigma, \sigma'} c_{i\sigma}^{\dagger}\bm \sigma_{\sigma\sigma'} c_{i\sigma'} /2$ is the spin-1/2 operator and  $\bm \sigma_{\sigma\sigma'}$ is the Pauli matrix $(=(\sigma^x_{\sigma\sigma'},\sigma^y_{\sigma\sigma'},\sigma^z_{\sigma\sigma'})$).
The long-range order is determined whether $S_s(\bm q)/N$ or $S_c(\bm q)/N$ remains after taking  the limit $N\rightarrow \infty$. 
The SC long-range order is measured by the $d$-wave SC correlation function 
\begin{align}
    P_d(\bm r) = \frac{1}{2N} \sum_{\bm r_i} \braket{ \Delta^{\dagger}_{d}(\bm r_i)  \Delta^{\phantom{\dagger}}_{d}(\bm r_i + \bm r)  + \text{h.c.}}. 
\end{align}
Here $\Delta^{\dagger}_{d}(\bm r_i)$ describes the order parameter of the form
\begin{align}
    \Delta^{\phantom{\dagger}}_{ d}(\bm r_i)  = \frac{1}{\sqrt{2}} \sum_{\bm r} f_{d}(\bm r) (c_{\bm r_i \uparrow} c_{\bm r_i + \bm r \downarrow} {-} c_{\bm r_i \downarrow} c_{\bm r_i + \bm r \uparrow}),
\end{align}
where the $d_{x^2 -y^2}$-wave symmetry is included via the form factor
\begin{align}
    f_d(\bm r)  = \delta_{r_y,0} ( \delta_{r_x,1} + \delta_{r_x,-1})  - \delta_{r_x,0} ( \delta_{r_y,1} + \delta_{r_y,-1}). 
\end{align}
Then, we deduce $P_d(\bm r)$ over the long-range part, as
\begin{equation}
    \bar{P}_{d}(L)  = \frac{1}{M} \sum_{\sqrt{2} L/4 < \lvert \bm r \rvert} P_d(\bm r), 
\end{equation}
where $\bm{r} = (r_x, r_y)$ includes all sites within the square $( -L/2, L/2]^2$ and $M$ is the number of lattice points satisfying $\sqrt{2} L/4 < \lvert \bm r \rvert < \sqrt{2} L/2$. 
We define the SC order parameter in the thermodynamic limit, $F_{\rm SC}^{\infty}$ by 
\begin{eqnarray}
F_{\rm SC}^{\infty}& =&\lim_{L\rightarrow \infty} F_{\rm SC}(L) = \sqrt{\bar{P}_{d}^\infty},
 \label{eq:FSC} \\
F_{\rm SC}(L)&\equiv &\sqrt{\bar{P}_{d}(L)}
\nonumber \\
\bar{P}_d^{\infty}&=& \lim_{L\rightarrow \infty} \bar{P}_{d}(L). \nonumber    
\end{eqnarray}

\section{{\it Ab initio} results}
\label{sec:ab initio results}

In this section we present our calculated {\it ab initio} results on the four families of compounds. We first analyze the results for doped CaCuO$_2$ in detail and then compare it with the result of doped Hg1201. 
Doped Bi2201 and Bi2212 suffer from the experimental uncertainty of the apical oxygen position due to the supermodulation of the crystal structure. 
Since it generates a variance of the Hamiltonian parameters if one assumes the translational symmetry of the Hamiltonian parameters, we show the properties by indicating this range. 
All of the four materials show dominance of the $d$-wave SC in the doping concentration dependence and the calculated results reproduce the experimental materials dependence of the strength of SC, which makes it possible to extract the universal properties and systematic trends as well.

\subsection{Doped CaCuO$_2$}
\label{CaCuO2}

Here, we present the results of our calculation for the doped CaCuO$_2$ in the following order: \\
(1) $\delta$ dependence of SC properties; in particular, $\bar{P}^\infty_{d}$ and the $d$-wave SC order parameter $F^\infty_{\rm SC}$. (See the definitions in Sec.~\ref{sec:quantities}.) \\
(2) Competition between the SC state and other competing states including stripe states.
\begin{figure*}[htb]
    \includegraphics[width=\linewidth]{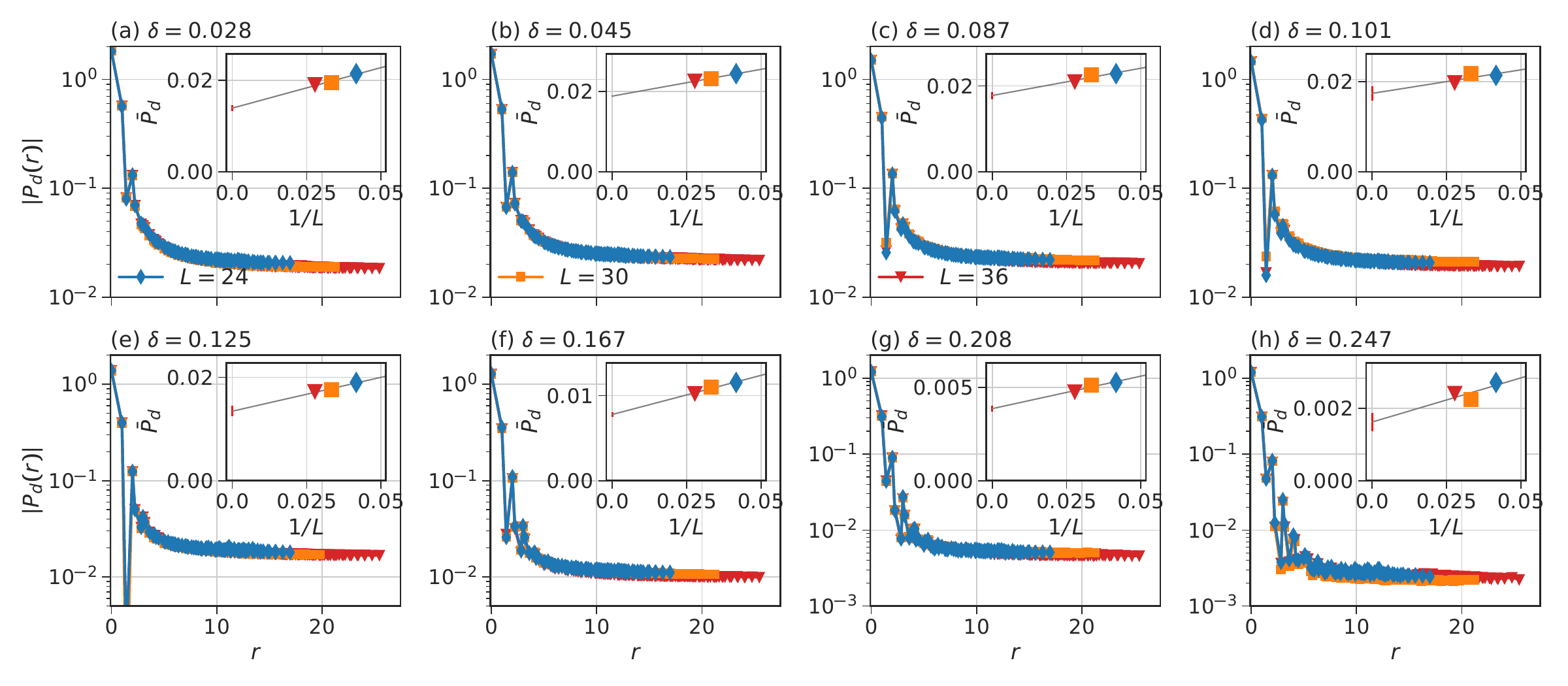} 
    \caption{SC correlation function $P_{d}(r)$ for CaCuO$_2$ at different hole dopings: (a) $\delta = 0.028$, (b) $\delta = 0.045$, (c) $\delta = 0.087$, (d) $\delta = 0.101$, (e) $\delta = 0.125$ (here $L = 28$ instead of $L=30$), (f) $\delta = 0.167$, (g) $\delta = 0.208$, (h) $\delta = 0.247$. 
    In each case, we show $P_{d}(r)$ for the square lattice size $L=24, 30, 36$. For the same distance $r$ we employ the largest value of correlation. We perform the same procedure in later figures. 
    Inset of each panel: Size extrapolation of $\bar P_{d}(L)$ to the thermodynamic limit $\bar P_{d}^{\infty}$, whose numerical value is listed in Table~\ref{tab:pdsum}. 
    The gray line shows the linear approximation. Statistical errors originating from the Monte Carlo sampling are smaller than the symbol size.} 
    \label{fig:PddCaVarSizeExt}
\end{figure*}
%
\subsubsection{Properties of superconducting phase} 
First, we discuss the $\bm{r}$ dependence of the pairing correlation $P_d(r)$ for $L\times L$ lattice and its long-ranged part $\bar{P}_{d}(L)$:
Fig.~\ref{fig:PddCaVarSizeExt} shows $P_{d}(r)$ and $\bar{P}_{d}(L)$ for several choices of square-lattices with the linear size $L$ from $24$ to $36$ and hole doping $\delta$ from $0.028$ ($2.8\, \%$) to $0.247$ ($24.7\, \%$).  
For each value of $\delta$, we observe that $\bar{P}_{d}$ does not significantly depend on $L$, suggesting the existence of a strong SC long-range order in the thermodynamic limit in this ground-state candidate.
This is indeed confirmed by a size extrapolation, i.e., plot of $\bar{P}_{d}(L)$ as a function of $1/L$ to estimate $\bar{P}_{d}^\infty$ in the limit $L \rightarrow \infty$ via linear regression, as shown in the insets of  Fig.~\ref{fig:PddCaVarSizeExt}(a)-(h). 
The linear $1/L$ scaling was employed in Ref.~\cite{ohgoe2020} and is expected to work because of Dirac-type linear dispersion for the quasiparticle excitation of the $d$-wave superconductor at the nodal points. {Here, we have shown the data calculated from the transfer and interaction parameters in the Hamiltonian fixed at 10\% hole doping for simplicity as we addressed above. However we can test its robustness by taking $\delta$ dependent transfer and interaction parameters. The result is shown in Appendix~\ref{app:delta_dependent_Hamiltonian} and the difference is small.}

\begin{figure}[tb]
\includegraphics[width=\linewidth]{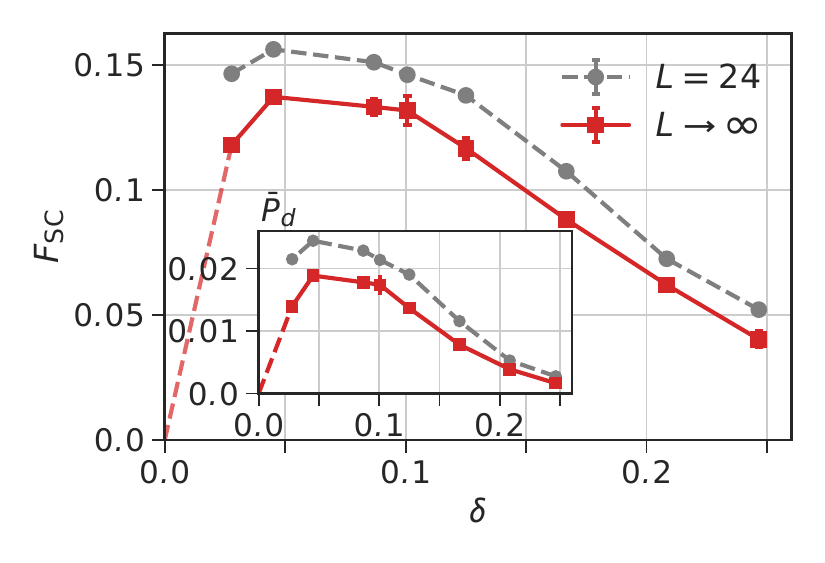} 
\caption{The SC order parameter $F_{\rm SC}$ as a function of $\delta$ for doped CaCuO$_2$. The gray filled circles show the values of $F_{\rm SC}(L)$ at $L=24$ square lattice calculated from $\bar{P}_d(L)$ shown in Fig.~\ref{fig:PddCaVarSizeExt} by using $F_{\rm SC}(L) =\sqrt{\bar{P}_d(L)}$ . 
The red squares are the size extrapolated values $F_{\rm SC}^{\infty}$ calculated from $\bar{P}_d^{\infty}$. 
Inset shows the corresponding $\bar{P}_d(L)$ at $L=24$ and  $\bar{P}_d^{\infty}$.}
\label{fig:CaCuOPdd}
\end{figure}
After taking the size extrapolation to the thermodynamic limit, we show the $\delta$ dependence of the order parameter $F_{\rm SC}^{\infty}$ calculated from Eq.(\ref{eq:FSC}) and $\bar{P}_{d}^{\infty}  = \lim_{L\to\infty}\bar{P}_{d}(L)$ in Fig.~\ref{fig:CaCuOPdd} and the numerical values in Table \ref{tab:pdsum}. 
This shows a rapid increase of $F_{\rm SC}^{\infty}$ from 0 at $\delta=0$ up to $\delta\sim 0.05$ as a function of $\delta$ followed by a plateau around $0.05\leq\delta \leq 0.1$ and monotonic decrease with further increasing $\delta$ above around 0.1 in the thermodynamic limit.

The dome structure ubiquitously observed for $T_{c} $ in the cuprates is qualitatively similar to the $\delta$ dependence in $F_{\rm SC}^{\infty}$, but the peak for $F_{\rm SC}^{\infty}$ is located at somewhat lower $\delta\sim 0.05$ than the case of experimental $T_{c} $, where the optimum $\delta$ is observed to be $\delta \sim 0.12$~\cite{hiroi1993}.{We will discuss this discrepancy in Sec.~\ref{sec:discussion}.} 
However, the monotonic decrease of $F_{\rm SC}^{\infty}$ with increasing $\delta$ for $\delta \ge 0.1$ is consistent with the universal trend of $\delta$ dependence of the SC gap identified from the angle-resolved photoemission spectra (ARPES) and the scanning tunneling microscope (STM) of the cuprates in general~\cite{Tanaka2006,Alldredge2008}, though the SC gap in the experimental estimate contains an ambiguity associated with the contribution from the pseudogap.
\begin{table}[tb]
	\caption{Size extrapolated SC correlation function $\bar P_d^\infty$ and order parameter $F_{\mathrm{SC}}$ for doped CaCuO$_2$ for several choices of doping $\delta$. 
        The fitting error is of the order of $\sim 10^{-4}$ for $\bar P_d^\infty$ and  $\sim 10^{-3}$  for $F_{\mathrm{SC}}^{\infty}$.}
	\label{tab:pdsum} 
	\begin{ruledtabular}
		\begin{tabular}{ l c c c c c }
			 $\delta$ & $\bar P_d^\infty$  &  $F_{\mathrm{SC}}^{\infty}$ \\
			\toprule
			 $0.028$ & $0.0139(6)$ & $0.118(2)$ \\
                      $0.045$ & $0.0188(2)$ & $0.137(1)$ \\
			         $0.087$ & $0.0177(8)$ & $0.133(3)$ \\
                      $0.101$ & $0.0174(16)$ & $0.132(6)$ \\
                      $0.125$ & $0.0136(9)$ & $0.116(4)$ \\
			         $0.167$ & $0.0078(2)$ & $0.088(1)$ \\
			         $0.208$ & $0.0039(2)$ & $0.062(1)$ \\
                      $0.247$ & $0.0016(2)$ & $0.040(3)$ \\
			\colrule
		\end{tabular}
	\end{ruledtabular}
\end{table}

In the mean-field picture, the SC gap is the product of the order parameter $F_{\rm SC}^{\infty}$ and the effective attractive interaction. If we consider the experimentally observed maximum gap $\sim 50$~meV~\cite{Tanaka2006,Alldredge2008} and $F_{\rm SC}^{\infty}\sim 0.13$, the characteristic scale of attractive interaction is as large as $\sim 0.4$ eV. 
This imposes a constraint on theories for the SC mechanism.

The sharp increase of $F_{\rm SC}^{\infty}$ between $\delta=0$ and 0.05 and the subsequent dome structure are similar to the earlier study by a VMC method and a cluster dynamical mean field study~\cite{yokoyama2013,Sakai2018} for the Hubbard model, where rapidly increasing $F_{\rm SC}$ from 0 at $\delta=0$ already reaches $F_{\rm SC} \sim 0.1$  at $\delta=0.03$ in the present notation.
In case of the Hubbard model, however, it was argued that the ground state is actually not SC but stripe-ordered states~\cite{doi:10.1126/science.aam7127,darmawan2018,ido2018}.

\subsubsection{Competition of SC, stripe, and AFM states}
\label{sec:competition}
\begin{figure}[tb]
\includegraphics[width=\linewidth]{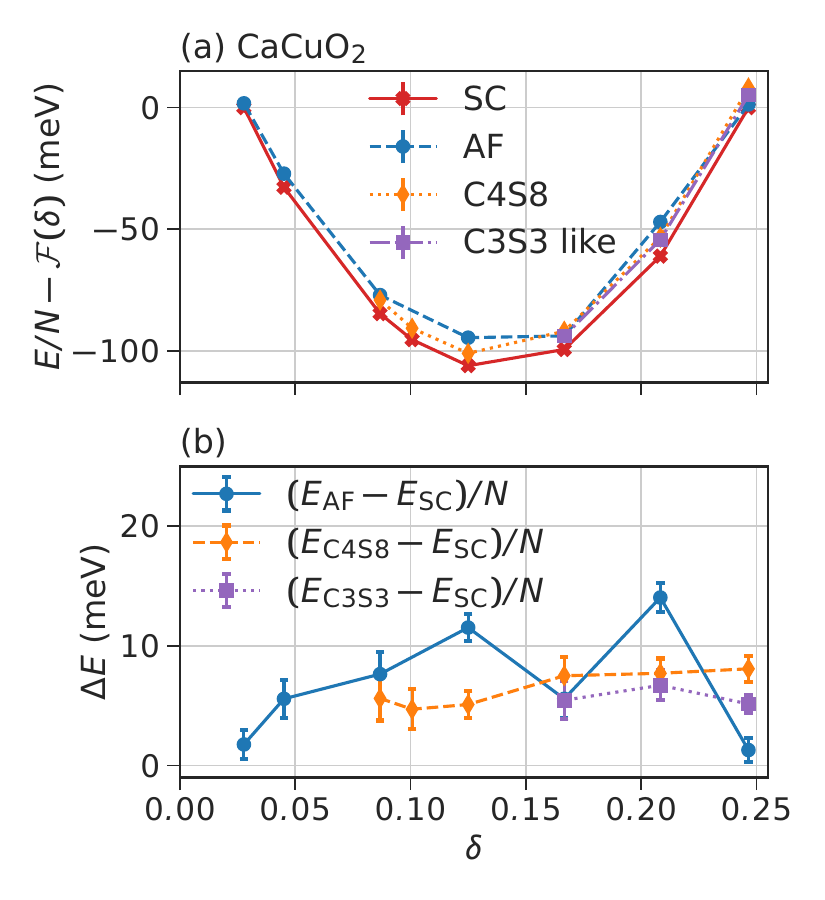} 
\caption{(a) Variance extrapolated energies of doped CaCuO$_2$ for various ground state candidates, SC, charge and spin stripe states C3S3 and C4S8, and AFM state as a function of hole doping $\delta$ on a $L=24$ square lattice. 
All energies are subtracted by the function $\mathcal F(\delta) =  -12.76470 \cdot \delta + 6.44626$ for better visibility.
(b) Energy difference $\Delta E$ for the variance extrapolated data from (a).} 
\label{fig:EnergyOverDeltaCa}
\end{figure}

Now, we analyze the competition between the SC and other states.
The energies of the SC state and other states at $L=24$ are given in Fig.~\ref{fig:EnergyOverDeltaCa}. 
We see that the SC state has the lowest energy in the region from $\delta=0.05$ to $\delta=0.25$, indicating that the SC phase is dominant in the ground state of doped CaCuO$_2$. 
The SC ground state, however, is severely competing with the C4S8, and C3S3-like stripe states, and AFM-type state within the energy difference of $5$-$10$ meV. 
Here, C$m$S$n$ denotes the charge and spin ordered stripe state with the periodicity of $m$ lattice spacing for the charge modulation and the period $n$ for the spin order.
Spin and charge real-space patterns and structure factors are explicitly illustrated in Secs. S2-S4 of SM~\cite{SM_Michael} for the C4S8, C3S3-like, and AFM-like excited states. {In the region studied here, we find only C4S8 and C3S3 as candidates of the competing stripe order, which has similarity to an earlier study of the simple Hubbard model with the next-nearest-neighbor transfer in the range $0.2<|t_2/t_1|<0.3$ and $0.05<\delta<0.25$~\cite{ido2018} and an {\it ab initio}  study~\cite{ohgoe2020}. Note that C4S8 at $\delta=1/8$ and C3S3 at $\delta=1/6$ have a particular commensurability energy gain. In Fig.~\ref{fig:EnergyOverDeltaCa} (b), a dip exists in the AFM states at $\delta= 0.167$. At the moment, the origin of the dip is not clear.}

Figures~\ref{fig:CaCuOSsqScqOverLAF0000} and \ref{fig:CaCuOSsqScqOverLStripes} in Appendix~\ref{app:ThermoLimitCaCuO} show size dependence of spin and charge structure factors.
Although the demanding computation cost does not allow larger system calculation, the available size dependence supports that the AFM states up to $\delta=0.16$ and the C4S8 stripe state at finite doping around $\delta\sim 0.12$ do have the long-range order as one can see in  Figs.~\ref{fig:CaCuOSsqScqOverLAF0000}~(a) and \ref{fig:CaCuOSsqScqOverLStripes}, respectively. 
However, although the initial trial states are ordered mean-field states, long-ranged order seems absent or is very weak after the optimization in VMC calculations for the C3S3 stripe states and seems replaced by well-developed short-ranged correlations at $\delta \ne 0$. 
This is the reason why we add ``-like" for the case of C3S3. 
The AFM and stripe states are in any case excited states of the SC ground states at $\delta\ge 0.05$.

{On the comparison between SC and stripe or AFM states, similar severe competitions were reported in Hubbard models. However, the present {\it ab initio} results have a crucial difference, where the charge uniform SC phase is the dominant ground state, while the ground states of Hubbard models irrespective of the presence or absence of $t_2$ mostly have the stripe long-range order. The reason for this difference originates from the presence of realistic off-site interaction in the {\it ab initio} case as was already pointed out in Ref.~\cite{ohgoe2020}. We discuss this point in the comparison to the Hubbard models in Appendix~\ref{app:Comparison_to_Hubbard}   }
\subsection{Doped Hg1201 Compared with CaCuO$_2$}
\label{Hg1201}
Now, we present our results in the case of Hg1201, and compare it to CaCuO$_2$. 
In this comparison, we find (i) the positive correlation between $F_{\rm SC}^\infty$ and $U/|t_1|$, (ii) the relation $T_{c} ^{\rm opt} \sim 0.16 |t_1|F_{\rm SC}^\infty$ from the comparison with the experimental $T_c$.

\begin{figure}[tbh!]
    \includegraphics[width=\linewidth]{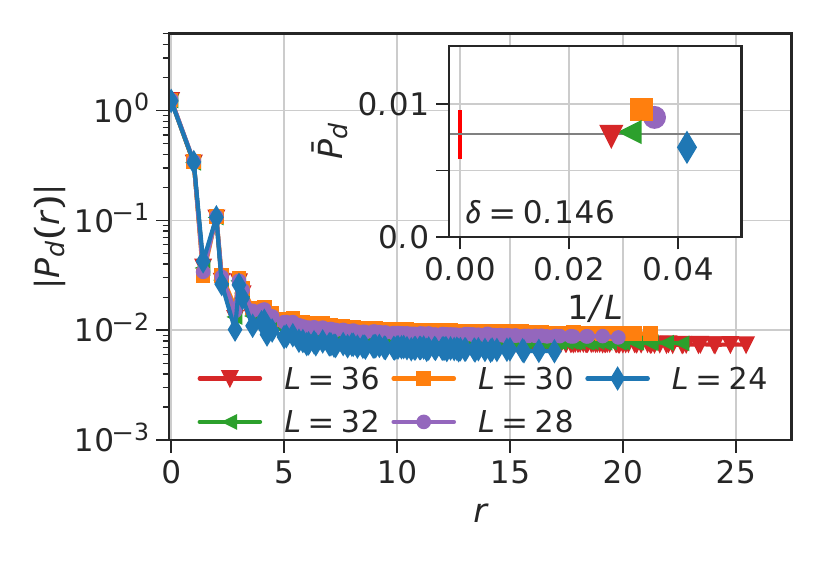}  
    \caption{SC correlation function $P_{d}(r)$ for Hg1201 at $\delta=0.146$\, which is close to the experimental optimal doping, at the square lattice size $L = 24, 28, 30, 32$ and $36$. 
    Error bars are smaller than the symbol size.
    Inset: Size extrapolation of $P_{d}$ to the thermodynamic limit $L\rightarrow \infty$. 
    Because of relatively scattered data we employ the average of the two biggest lattice sizes ($L = 32, 36$) for the size extrapolation $L\rightarrow \infty$. In fact, the value at the largest sizes is consistent with the systematic $\delta$ dependence observed near $\delta=0.146$ after the size extrapolation (not shown).
    The error bar at $1/L = 0$ is estimated as the biggest difference to $F_{\rm SC}^\infty$ from the given data. }
    \label{fig:PddHgVarSizeExt}
\end{figure}

The pairing correlation for Hg1201 Hamiltonian and the size extrapolation are shown in Fig.~\ref{fig:PddHgVarSizeExt} for $\delta=0.146$, indicating the existence of the SC long-range order. 
The size of the order parameter is $F_{\rm SC}^\infty\sim 0.09$ as compared with $\sim 0.116$ for doped CaCuO$_2$ at $\delta\sim 0.12$, respectively, which are both close to each optimal concentration. 
The difference in $F_{\rm SC}^\infty$ between Ca and Hg compounds can be compared with the difference in $U/|t_1|=8.10$ for CaCuO$_2$ and 7.35 for Hg1201. 
As we discuss later, $F_{\rm SC}^\infty$ rapidly increases with $U/|t_1|$ if we monitor the effect of $U/|t_1|$ beyond the {\it ab initio} value around $U/|t_1|=7$-$8$. 
Therefore, $F_{\rm SC}^\infty$ amplifies the increase in $U/|t_1|$ while effects of other parameters are minor: Namely, $F_{\rm SC}^\infty$ should have a functional form $F_{\rm SC}^\infty(U/t_1,V_i/t_1,t_i/t_1)$ with $1\le i \le 9$ in general, but $V_i/t_1$ and $t_i/t_1$ dependencies are {weaker as compared to dependence on $U/|t_1|$ in the realistic parameter range. See Appendix~\ref{app:t2_dependence} for the example of $|t_2/t_1|$ dependence. See also Fig.~\ref{fig:CaCuOPdScalingZoom} for the $V_1/t_1$ dependence. In both cases, the change in $F_{\rm SC}$ at the optimum doping is at most 10\% in the realistic parameter range. In fact, in the comparison of Bi2212, Bi2201, CaCuO$_2$, and Hg1201, $|t_2/t_1|$ is $\sim 0.30, 0.27, 0.25$ and 0.20, respectively, which does not have systematic correlation with $T_c^{\rm opt}$.  Appendix~\ref{app:t2_dependence} shows tiny anticorrelation of $|t_2/t_1|$ and $F_{\rm SC}$, but is practically negligible at the optimal doping.
After careful examination of other parameters as well, the difference of $F_{\rm SC}^\infty$ in these four compounds studied is concluded to be ascribed to the difference in $U/|t_1|$. }
 
The materials dependent $F_{\rm SC}^\infty$ may also be compared with the difference in $T_{c} ^{\rm opt}\sim 110$ and 90 K for CaCuO$_2$ and Hg1201, respectively, because $T_{c} $ may be proportional to the order parameter $F_{\rm SC}^\infty$. Since $T_{c} $ has the dimension of energy and should also be scaled by the overall characteristic energy scale $t_1$, $T_{c} $ may be proportional to $|t_1|F_{\rm SC}^\infty$. 
In fact, the ratio of $T_{c} ^{\rm opt}/|t_1|F_{\rm SC}^{\infty}$ as a non-dimensional quantity is $\sim 0.16(1)$ at the optimal doping $\delta\sim 0.12$ for CaCuO$_2$ and $\sim 0.16(2)$ at the optimal point $\delta\sim0.15$ for Hg1201 as we show in Table~\ref{tab:pdsum2}, supporting the hypothesis that $T_{c} ^{\rm opt}$ is universally given from the relation 
\begin{equation}
    T_{c} ^{\rm opt} \sim 0.16 |t_1|F_{\rm SC}^\infty
\label{eq:Tcscaling}
\end{equation}
at the optimal doping.

 In the Uemura plot~\cite{Uemura1989}, it was observed from the muon spin resonance ($\mu$SR) measurement that $T_{c} ^{\rm opt}$ is proportional to the ratio between the superfluid density $n_s$, here interpreted as $F_{\rm SC}^{\infty}/\sqrt{2}$, and the effective mass $m^*$. 
 Since the mass enhancement from the bare band mass $m_0$, namely $m^*/m_0$ at the optimal hole density may be similar in the cuprates, $T_{c} ^{\rm opt}$ is indeed expected to be roughly proportional to $|t_1|F_{\rm SC}^{\infty}$ according to the Uemura plot, because the inverse band mass is essentially determined by the dominant transfer $t_1$.  
 In addition, $n_s$ estimated from the relaxation rate $\sigma\sim 2$ $\mu s^{-1}$ from the $\mu$SR measurement for the cuprates with $T_{c} \sim 80-100$ K corresponds to $n_sm_0/m^* \sim 4\times 10^{21}$ cm$^{-3}$.
 If we cut out the volume including one Cu atom with the $c$ axis length $\sim 6$ \AA \ as in \cite{Uemura1989} irrespective of the unit cell volume to compare with $F_{\rm SC}^{\infty}$ defined as the value per Cu atom, this corresponds to $F_{\rm SC}^{\infty}\sim 0.10$ by considering the definition $F_{\rm SC}^{\infty}=\sqrt{2}n_s$ and $m^*/m_0\sim 5$ assumed in Ref.~\cite{Uemura1989}. 
 Then it is also quantitatively consistent with the present result of $F_{\rm SC}^{\infty}\sim 0.10$ at the optimum doping. 
 These indicate that our results indeed capture the realistic situation more or less quantitatively. 
 We will show in Sec.~\ref{sec:discussion} that $F_{\rm SC}^{\infty}$ also agrees with the estimate from the angle resolved photoemission spectra (ARPES) in case of Bi2212 and Bi2201.
\begin{table*}[tbh]
    \caption{Comparison of long-ranged SC correlation $\bar{P}_d^{\infty}$ and the order parameter $F_{\mathrm{SC}}^{\infty}$ with $U/|t_1|$ as well as comparison between $|t_1|F_{\mathrm{SC}}^{\infty}$ and $T_{c} ^{\rm opt}$ for doped CaCuO$_2$ at $\delta=0.125$, and Hg1201 at $\delta=0.146$ (these values of $\delta$ are chosen in accordance with those closest to experimental optimal values, $\delta\sim 0.12$ and 0.15, respectively, and to allow the numerical size extrapolation easier).
    Note that the ratio  $T_{c} ^{\rm opt}/(|t_1|   F_{\mathrm{SC}}^{\infty})$ is given as a nondimensional quantity by using 1 eV $=1.16\times 10^4$ K. 
    The parentheses in the last digit indicate the error bar.}
    \label{tab:pdsum2} 
    \begin{ruledtabular}
        \begin{tabular}{ l c c c c c c  }
             & $U/|t_1|$ 
             & $\bar{P}_d^{\infty}$  &  $F_{\mathrm{SC}}^{\infty}$ & $|t_1|F_{\mathrm{SC}}^{\infty}$(eV) & $T_{c} ^{\rm opt}$ (K) & $T_{c} ^{\rm opt}/(|t_1|   F_{\mathrm{SC}}^{\infty})$ \\
            CaCuO$_2$ & $8.10$ 
            & $0.0136(9)$ & $0.116(4)$ & $0.060(2)$ & $110$ & $0.16(1)$\\
            Hg1201 & $7.35$ 
            & $0.008(2)$ & $0.09(1)$ & $0.048(5)$ & $90$ & $0.16(2)$\\
        \end{tabular}
    \end{ruledtabular}
\end{table*}

\subsection{Doped Bi2201 and Bi2212}
\label{Bi2201 and Bi2212}
In this subsection, we discuss {\it ab initio} results for Bi2212 and Bi2201 and compare them each other.  Unfortunately, in these two compounds, an uncertainty exists in the experimental crystal parameters that causes the uncertainty in the effective Hamiltonian parameters as well. Especially, the distance $d^{z}_{\rm Oap}$ between an apical oxygen and the nearest Cu atom is not fully precisely determined and the available experimental data have considerable variations~\cite{Torrance1988,Torardi1988,Ito1998,Schloegl1993,Beskrovnyi1990,Cicco1993}. This uncertainty is also related to the structural distortion and long-period modulation of the CuO$_2$ plane arising from the effect from the BiO layer~\cite{yamamoto1990,slezak2008} as we discuss in Sec.~\ref{sec:discussion}. 
 Recent {\it ab initio} studies have clarified that this uncertainty leads to a possible variety of effective Hamiltonian parameters, especially owing to the variation of the apical oxygen position~\cite{moree2022}. 
 
 In principle, the structural optimization in {\it ab initio} calculations is desired to predict the stable atomic position. However, such an optimization in strongly correlated electron systems is at the moment not necessarily accurate enough and we leave this task for future studies. Instead, in this paper, we admit a range of Hamiltonian parameters and discuss the consequence. 
 
 As analyzed in Ref.~\cite{moree2022}, the apical oxygen position sensitively affects the effective Hamiltonian parameters, primarily the value of $U$. For Bi2212, the value $U\sim 4.2$ eV in Table~\ref{tab:ham} is intermediate and the uncertainty range is between 4.0 and 4.7 eV for $U$ by considering that $d^{z}_{\rm Oap}$ may range from 2.25 \AA \ to 2.45 \AA. On the other hand, the value $U\sim 4.4$ eV for Bi2201 in Table~\ref{tab:ham} is the upper bound and the uncertainty ranges from $U\sim 4.4$ to 3.5 eV by considering that $d^{z}_{\rm Oap}$ may range from $d^{z}_{\rm Oap}=$ 2.6 to 2.45 \AA. 
 We first present in Secs.~\ref{sec:Bi2212} and \ref{Bi2201} the results obtained from the parameters shown in Table~\ref{tab:ham}, namely $U=4.2$ eV for Bi2212 and $U=4.4$ eV for Bi2201 and then discuss in Sec.~\ref{Uncertainty} the possible range of SC properties originating from this uncertainty later.

\subsubsection{Bi2212}\label{sec:Bi2212}
\begin{figure}[tb]
	\includegraphics[width=\linewidth]{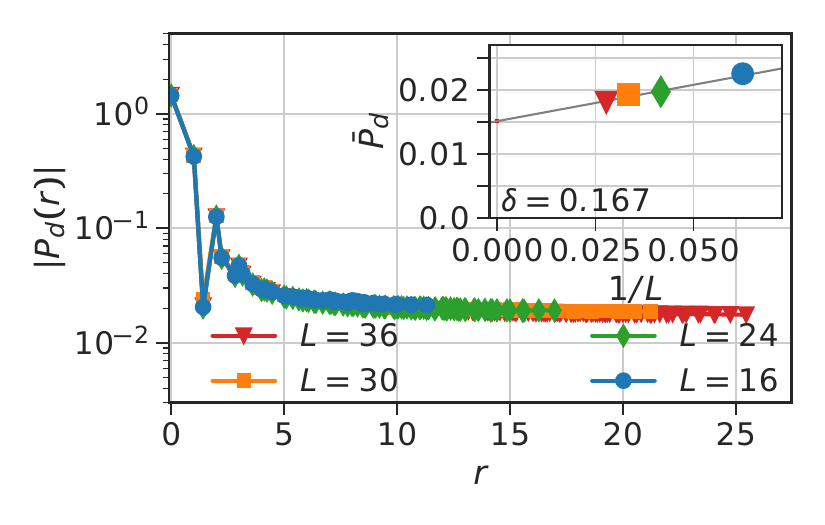} 
    \caption{SC correlation function $P_{d}(r)$ for Bi2212 at $\delta=0.167$\ for the square lattice size $L = 16, 24, 30, 36$. 
    Inset: Size extrapolation of $\bar P_{d}$ to the thermodynamic limit. The gray line shows the linear extrapolation to $1/L\rightarrow 0$ by using the values for $L = 16, 24, 30, 36$. Error bars are smaller than the symbol size.}
	\label{fig:Pdbi2212}
\end{figure}

We begin with the results for Bi2212.
Figure~\ref{fig:Pdbi2212} shows $P_{d}(r)$ and $\bar{P}_{d}(L)$ at $\delta  = 0.167$ for $L$ from $16$ to $36$ by switching off the interlayer transfers and interactions. Namely we first show the results obtained by solving the single-layer Hamiltonian despite the actual two-layer unit cell of Bi2212.
{The case of $\delta$ dependent Hamiltonian is also shown in Appendix~\ref{app:delta_dependent_Hamiltonian}, where the difference is small.}
Similarly to CaCuO$_2$, we identify a SC ground state with large $\bar{P}_{d}$ and it does not change significantly by increasing $L$, where the size extrapolation gives $\bar{P}_{d}^{\infty} \sim 0.0151$ as shown in the inset of Fig.~\ref{fig:Pdbi2212}, which corresponds to a SC order parameter $F_{\mathrm{SC}}^{\infty} \sim 0.12$. 
This relatively strong value of $F_{\mathrm{SC}}^{\infty}$ is understandable, because the $U/|t_1|$ ratio has a value of $U/|t_1| \sim 9.4$, which is the strongest among all four considered compounds.
Again the enhanced $F_{\mathrm{SC}}^{\infty}$ originates from the larger $U/|t_1|$ in accordance with the observation in the comparison of Hg1201 and CaCuO$_2$ in Sec.~\ref{Hg1201}.
This large $F_{\mathrm{SC}}^{\infty}$ is also consistent with the high $T_{c} $ (up to $\sim 100$ K)~\cite{Hobou2009}.
We will discuss more intricate aspect in Sec.~\ref{sec:discussion}. 

The competition with other phases is seen in $\delta$ dependence of the total energy shown in Fig.\ref{fig:EnergyOverDeltaBi2212} in Appendix~\ref{app:Bi2212-01}. 
Similarly to CaCuO$_2$, the SC state is the ground state in most of the doping concentration, while it is severely competing with spin and charge ordered states.

\begin{figure}[tbh]
    \includegraphics[width=\linewidth]{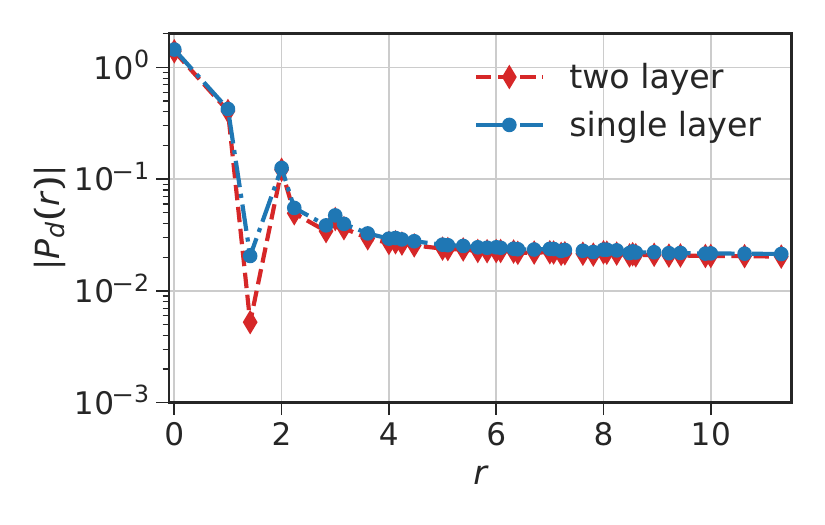} 
    \caption{SC correlation function $P_{d}(r)$ for Bi2212 at $\delta=0.167$\, in the cases of the single- and the explicit two-layer calculations. In the single-layer case a $16 \times 16$ square lattice was considered, which corresponds to a $16 \times 16 \times 2$ lattice in the two-layer case. Error bars are smaller than the symbol size.}
    \label{fig:Pdbi2212Two}
\end{figure}

Now, we extend the calculation by switching on the interlayer terms and solve the two-layer Hamiltonian obtained in Ref.~\cite{moree2022} to examine the effects of interlayer coupling.  
For the calculations  we take two identical layers (in terms of intralayer parameters for $t_{ij}$, $V_{i}$, and $U$ where $i$ and $j$ are intralayer combination), coupled by the interlayer terms listed in Table~\ref{tab:bi2212Vinter} (Appendix~\ref{app:interlayer}).
The interlayer contributions are restricted to the leading interlayer hopping term of size $t_0^l= -0.098$ eV and nearest- and next-nearest-neighbor interlayer interaction of size  $V_{0}^{l} = 0.643$ eV and $V_{1}^{l} = 0.463$ eV.
See Ref.~\cite{moree2022} for more details.
A comparison between the single- and two-layer cases of $P_{d}(r)$ for $L = 16$ and $\delta = 0.167$ is shown in Fig.~\ref{fig:Pdbi2212Two}. 
We see that the long-range average of $P_d(\bm r)$ in the two-layer case is close to that of $P_d(\bm r)$ in the single layer case, which demonstrates that $P_d(\bm r)$ is not significantly affected by the interlayer Coulomb interaction and hopping parameters.
Indeed, for the single layer we found a long-range average of the SC correlation function of  $\bar{P}_{d}^{\text{single}} = 0.0225$, while for the two-layer compound the average is $\bar{P}_{d}^{\text{two}} = 0.0213$. 
The corresponding values of the SC order parameter are $F_{\mathrm{SC}}^{\text{single},\infty}=0.150$ and $F_{\mathrm{SC}}^{\text{two},\infty}=0.146$, which differ by only $\sim 2.8\%$.
The essentially same behavior between the single- and two-layer cases may not depend on the system size in accordance with the result of a two-layer Hubbard model at the optimum doping~\cite{iwano2022}.
This similarity may be attributed to (i) the relatively small leading interlayer hopping parameter of $t_0^l = -0.098$ eV and also (ii) the robustness of the SC solution against the interlayer Coulomb interaction parameter, because the pairing occurs essentially only within a layer.

\subsubsection{Bi2201}
\label{Bi2201}
In the case of Bi2201, we again solve the single-layer Hamiltonian. 
The competition with other phases are seen in $\delta$ dependence of the total energy shown in Fig.\ref{fig:EnergyOverDeltaBi2201} in Appendix~\ref{app:Bi2212-01}.
\begin{figure}[tbh!]
    \includegraphics[width=\linewidth]{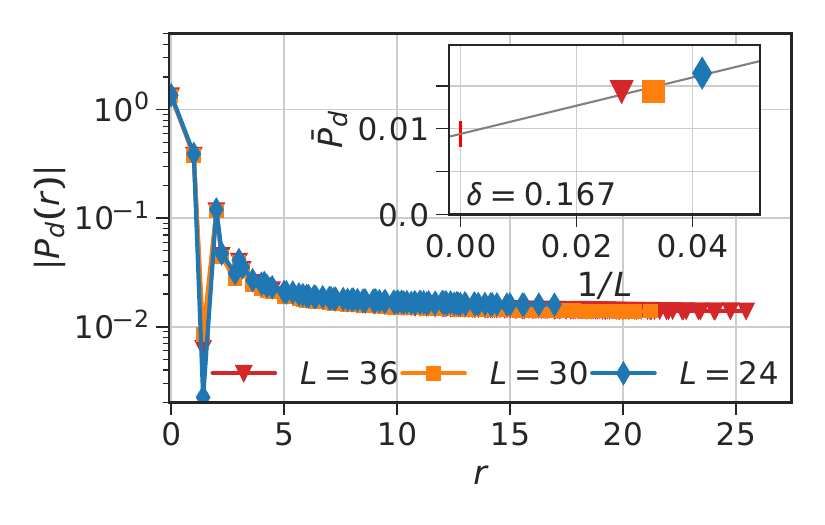} 
    \caption{SC correlation function $P_{d}(r)$ for Bi2201 at $\delta=0.167$\, at the square lattice sizes $L= 24, 30, 36$. 
    Inset: Size extrapolation of $\bar P_{d}$ to the thermodynamic limit. 
    The gray line shows the linear extrapolation to $1/L\rightarrow 0$ by using the values for $L = 24, 30, 36$. Error bars are smaller than the symbol size.}
    \label{fig:PddBi2201VarSizeExt}
\end{figure}
Figure~\ref{fig:PddBi2201VarSizeExt} shows $P_{d}(r)$ and its extrapolation to the thermodynamic limit, which suggests the stable long-ranged SC order. 
Again, the obtained value $F_{\rm SC}^{\infty}\sim 0.10$ is consistent with the rule that larger $U/|t_1|$ leads to larger $F_{\rm SC}^{\infty}$ because $U/|t_1|$ is the second largest among the four materials in the estimate shown in Table~\ref{tab:ham}.
The smaller $U/|t_1|$ relative to Bi2212 leads to weaker SC. 
However, on a more quantitative aspect, we need to be careful about the uncertainty of the Hamiltonian parameter. We will discuss this issue below.

\subsubsection{Effect of structural uncertainty on possible variation of SC properties}
\label{Uncertainty}
Since we have the uncertainty of the Hamiltonian parameters particularly for the interaction as we discussed above, we here monitor the effects of modifying the effective interactions for Bi2212 and Bi2201, which well represent the effect of variant apical oxygen position as shown in Appendix C of Ref.~\cite{moree2022}. 
Namely, Table~\ref{tab:ham} with preserved transfer parameters fixed at each {\it ab initio} value, together with interaction scaling represents most of the effect of the apical oxygen shift and we scale the Hamiltonian (\ref{eq:Ham}) such that
\begin{align}
    \mathcal{H}(\alpha,\xi) &= H_{\mathrm{kin}} + \alpha   H_U  + \xi   H_V,
\label{int_scale}
\end{align}
where $\alpha=\xi=1$ represents the {\it ab initio} case given in Table~\ref{tab:ham}, $\alpha$ scales the on-site Coulomb interaction term $H_U$, and $\xi$ scales the remaining off-site interactions $V_i$. 
Since the apical oxygen shift alters the interaction parameters in the way $\alpha\sim \xi$~\cite{moree2022}, we examine the dependence on $\alpha=\xi$ below.

\begin{figure}[tbh]
    \includegraphics[width=\linewidth]{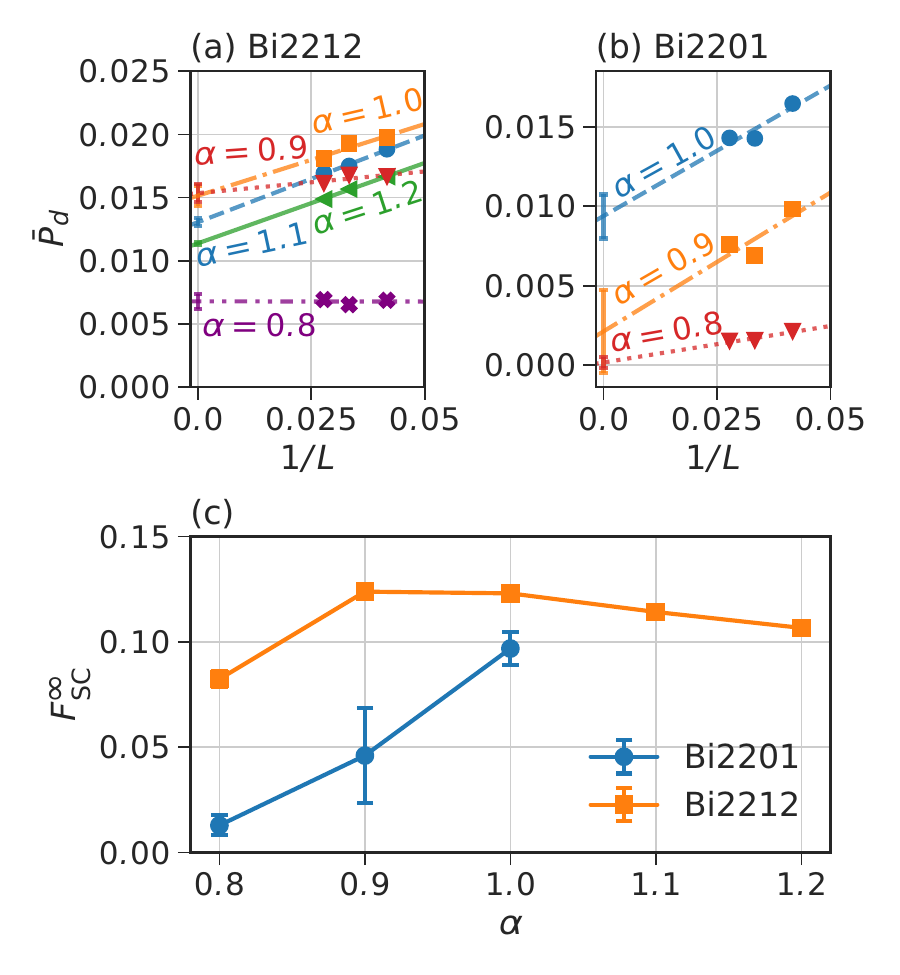} 
    \caption{Variation of SC order arising from uncertainty of the apical oxygen position via $\alpha = \xi$ scaling for Bi2212 and Bi2212. The realistic range is $0.95\le \alpha=\xi \le 1.1$ for Bi2212 and $0.8\le \alpha=\xi \le 1.0$ for Bi2201. 
    Size extrapolation of $\bar P_{d}$ for: (a) Bi2212 at $\delta=0.167$\ and $\alpha \in \{0.8, 0.9, 1.0, 1.1, 1.2 \}$, (b) Bi2201 at $\delta=0.167$ and $\alpha \in \{0.8, 0.9, 1.0 \}$. 
    (c) $F^\infty_{\mathrm{SC}}$ as a function of $\alpha=\xi$ for Bi2212 (orange) and Bi2201 (blue). }
    \label{fig:BiCompundsApicalUncertainty}
\end{figure}

Figures~\ref{fig:BiCompundsApicalUncertainty}~(a) and~(b) show the size scaling and $\alpha=\xi$ dependence of $P_d^{\infty}$ for Bi2212 and Bi2201 at $\delta=0.167$, which is close to the experimental optimum doping.
Further details are given in Sec.S5 of SM~\cite{SM_Michael}.
Figure~\ref{fig:BiCompundsApicalUncertainty}~(c) shows the corresponding $\alpha=\xi$ dependence of the SC order parameter $F_{\rm SC}^{\infty}$.
For Bi2212, by taking the realistic uncertainty range $4.0$ eV $\le U \le 4.7 $ eV (corresponding to $0.95\le \alpha = \xi \le 1.1$) obtained from $2.25\, \text{\AA}$  \ $\le d_{\rm Oap}^z \le 2.45\, \text{\AA}$, we find the range of $0.011 \lesssim P_d^{\infty} \lesssim 0.015$, namely $0.10 \lesssim F_{\rm SC}^{\infty}\lesssim 0.12$. 
On the other hand, the uncertainty for Bi2201 obtained from $2.45\, \text{\AA}$ $\le d_{\rm Oap}^z \le 2.6\, \text{\AA}$  leads to $3.5$ eV $\le U \le 4.4$ eV (corresponding to $0.8\le \alpha = \xi \le 1.0$), which results in $0.00017 \lesssim P_d^{\infty} \lesssim 0.0094$ ($0.013 \lesssim F_{\rm SC}^{\infty} \lesssim 0.10$). 
The lower bound for Bi2201 causes a fatal damage to the SC order. 

Table~\ref{tab:pdsum3} summarizes the size-extrapolated $\bar P_d^\infty$ and corresponding $F_{\mathrm{SC}}^{\infty}$ for the two Bi compounds when we use the Hamiltonian parameters listed in Table~\ref{tab:ham} and when we admit the uncertainty range of the interaction parameters for Bi2212 and Bi2201. 
The scaling Eq.(\ref{eq:Tcscaling}) proposed for CaCuO$_2$ and Hg1201 is also valid in the Bi compounds and the experimental $T_c^{\rm opt}$ is within the inferred range. 
We conclude that its materials dependence is well captured for the four studied materials (see also Fig.~\ref{fig:TcComparisionExperiment} in Sec.~\ref{sec:summary}).
\begin{table*}[tbh]
    \caption{Comparison of long-ranged SC correlation $\bar{P}_{d}^{\infty}$ and the order parameter $F_{\mathrm{SC}}^{\infty}$ with $U/|t_1|$ for Bi2212 and Bi2201 at $\delta=0.167$, which is chosen in accordance with the experimental optimal values. 
    The range of values represents the uncertainty range arising from the uncertainty of the apical oxygen position. This estimate helps the inference of the correct apical oxygen position (see text).}
    \label{tab:pdsum3} 
    \begin{ruledtabular}
        \begin{tabular}{ l c c c  c c c }
             & $U$ & $U/|t_1|$ & $\bar{P}_d^{\infty}$  &  $F_{\mathrm{SC}}^{\infty}$ & $0.16|t_1|F_{\mathrm{SC}}^{\infty}$ (K) & $T_{c} ^{\rm opt}$ (K) \\
            Bi2212 & 4.0-4.7 & 8.8-10.4 & $0.011$ - $0.015$  & $0.10$ - $0.12$ & 83-101 &85-100 \\
            Bi2201 & 3.5-4.4 & 6.7-8.4 & $0.00017$ - $0.0094$ & $0.013$ - $0.096$ & 12-93 & 10-40 \\
        \end{tabular}
    \end{ruledtabular}
\end{table*}
 
We realize that the fragility and diversity of $T_{c} ^{\rm opt}$ experimentally observed in the range $10<T_{c} ^{\rm opt}<40$ K for Bi2201 is accounted for by the range of actual apical oxygen position. 
This range may be caused by the type of dopant atoms, impurities, and the spatial inhomogeneity caused by the supermodulation, which may depend on samples and the amplitude of the modulation. 
In fact, it was observed that the $d_{\rm Oap}^z$ periodically varies as much as 6\% in accordance with the supermodulation for Bi2212 and a comparable modulation may exist for Bi2201 as well, which can be the origin of the experimental uncertainty~\cite{yamamoto1990,slezak2008}. 
The basic origin of this diversity and relatively low $T_{c} $ among the four families of compounds is attributed to relatively small $U/|t_1|$ in the lower uncertainty range, at which the SC order becomes sensitively damaged by a slight decrease of $U/|t_1|$. 
We discuss more general aspects of the interaction dependence in Sec.~\ref{sec:results beyond ab initio}.
Even when we admit the uncertainty range, the general trend about the weaker SC of Bi2201 than those of Bi2212 is well explained by this {\it ab initio} result. It can also safely be addressed that Bi2212 has one of the strongest SC and $T_{c} ^{\rm opt}$ among the four families comparably to CaCuO$_2$.

{The effects of apical oxygen position on Hamiltonian parameters are discussed in Sec.~\ref{sec:discussion} and in Appendix~\ref{app:ApicalO_effect}.}

\section{Results beyond {\it ab initio}: interaction dependence of superconducting order}
\label{sec:results beyond ab initio}
 We now study SC properties beyond the {\it ab initio} results.  {\it Ab initio} results in the previous section successfully reproduce the experimental trend and have revealed that $U/|t_1|$ is the principally important Hamiltonian parameter to control the SC order parameter. 
Therefore, it is intriguing to examine the optimum Hamiltonian parameters to maximize the order parameter and hence the optimum $T_{c} $ beyond the existing materials for the purpose of materials design to seek for higher $T_{c} $ superconductors.   
We present $U/\lvert t_1\rvert$ dependence of $F_{\rm SC}$ as well as dependence on off-site interaction when tuning the interaction parameters artificially away from the {\it ab initio} value while keeping transfer parameters fixed at {\it ab initio} values. We here take an example of CaCuO$_2$ Hamiltonian at $\delta=0.167$ and monitor the effect of $\alpha$ and $\xi$ dependencies defined in Eq.~\eqref{int_scale}.
 
\begin{figure}[tb]
    \includegraphics[width=\linewidth]{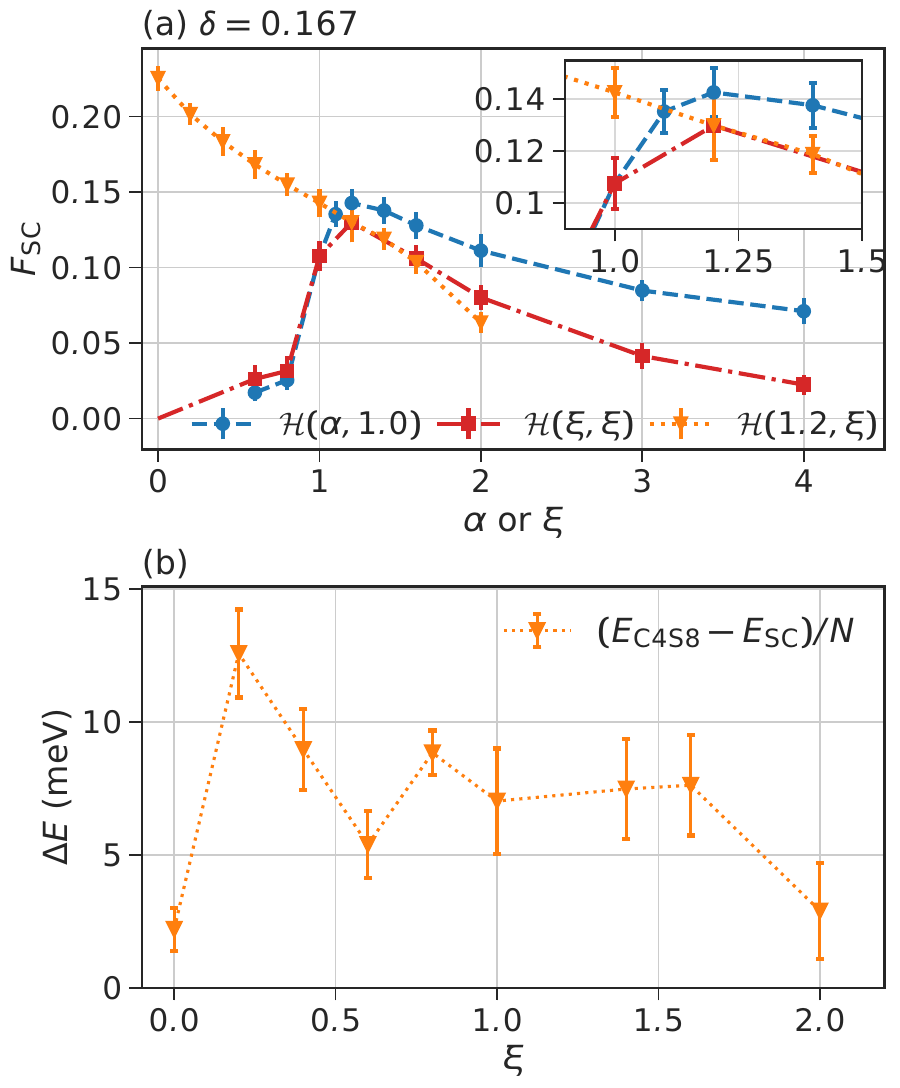} 
    \caption{ (a) $F_{\rm SC}$ over scaling $\alpha$ or $\xi$ (effective scaling of $U/\lvert t_1 \rvert$ and $V_i/U$) of the SC state on the $L = 24$ square lattice and $\delta = 0.167$ hole doping. 
    The inset is an enlarged plot around the peak $0.9 < \alpha,\xi < 1.5$.
    For further details see the main text.
    (b) Variance extrapolated energy difference $\Delta E$ between the SC and C4S8 as a function of  $\xi$ in the case $\mathcal H(\alpha = 1.2, \xi)$.  } 
    \label{fig:CaCuOPdScalingZoom}
\end{figure}

We examine three types of scaling to go beyond the {\it ab initio} Hamiltonian: (i) Scale only the on-site Coulomb interaction by $\alpha$ with fixed $\xi = 1.0$ ($\mathcal H(\alpha, 1.0)$), (ii) Scale the full interaction part equally by using $\alpha = \xi$ (namely, $\mathcal H(\alpha, \alpha)$), and (iii) Fix $\alpha = 1.2$ and scale the off-site Coulomb interaction uniformly via $\xi$ by employing $\mathcal H(1.2, \xi)$, by considering the fact that (ii) shows the maximum SC order at $\alpha=\xi=1.2$. 
For the cases (i) and (ii) we chose scaling values $\alpha$ ranging from $0.6$ up to $4.0$ while for the third case the range from $0.0$ to $2.0$ is chosen.  
Since the size dependence is not appreciable, we study $L=24$ lattice.
The SC order parameter $F_{\rm SC}$ are shown  in Fig.~\ref{fig:CaCuOPdScalingZoom} (a) at $\delta = 0.167$ hole doping. 

The results show that the SC order parameter can be enhanced with the amount of around 30\% from the {\it ab initio} value when the interaction parameter is tuned to $\alpha\sim 1.2$ for (i) and around 20\% at $\alpha=\xi\sim 1.2$ for (ii), which may allow $T_{c} ^{\rm opt}$ as much as $\sim 130\text{--}140\, \text{K}$, when compared to the {\it ab initio} results for CaCuO$_2$.

In the tuning (iii), we find that the order parameter further increases up to $F_{\rm SC}^{\infty}\sim 0.22$ by decreasing $\xi$, which is twice as large as the {\it ab initio} case.
However, we keep in mind that on-site and off-site interactions cannot independently be controlled in the usual experimental conditions. The present result offers a guide to enhance the SC in designing artificial structure and metamaterials including surface and interface, where quicker screening of the off-site interaction is desirable by keeping the on-site interaction at the optimal value (in this case $\alpha=1.2$).

Another limitation to be considered is the competition with the stripe and AFM order.
As far as we restrict the on-site interaction within the {\it ab initio} range, the SC energy is always lower than that of the stripe state as one sees in Fig.~\ref{fig:CaCuOPdScalingZoom}~(b).
Here, we show the competition with the C4S8 because it is established that the most severe competitor is the C4S8 state.
However, the decreasing of energy difference, such as at ($\alpha=1.2$, $\xi=0$), where the difference is $\leq 2\, \text{meV}$, may result in the thermal destruction of the SC order.
For $\xi\ge 0.2$ the SC is still a stable ground state, while at small $\xi$, $F_{\rm SC}$ becomes nearly twice of the {\it ab initio} value for the doped CaCuO$_2$.

The order parameter decreases when $U/|t_1|$ is too large {beyond the realistic range of the cuprates  studied in this paper} as we see in Fig.~\ref{fig:CaCuOPdScalingZoom}~(a). 
This reduction was already pointed out on the level of the Hubbard model~\cite{yokoyama2013}. 
{The reduction at large $U/|t_1|$ is studied in Appendix~\ref{app:strong_coupling}, which shows nontrivial power-law dependence of $F_{\rm SC}$ on $U/|t_1|$.}
The reduction itself may be easily understood on a qualitative level from the suppression of charge fluctuation with increasing $U/|t_1|$, which also suppresses the quantum entanglement caused by the suppression of both spin singlet fluctuation and dynamical exciton generation as was reported in the literature~\cite{yokoyama2013,imada_suzuki2019}.
It was pointed out that the enhanced quantum entanglement can be achieved by the fractionalization of electrons~\cite{imada_suzuki2019,imada_review2021}, which may be maximized at the optimum $U/|t_1|$. 

\section{Discussion}
\label{sec:discussion}

By assuming the same ratio $T_{c} /(|t_1|   F_{\rm SC}^{\infty}) \sim 0.16$ with CaCuO$_2$ and Hg1201, we can infer the range of $T_{c} $ arising from the uncertainty of the apical oxygen position and resultant uncertainty of $U/|t_1|$ listed in Table~\ref{tab:pdsum3}.  
The range of inferred $T_{c} $ listed in Table~\ref{tab:pdsum3} for Bi2212 is consistent with the experimentally observed range of $T_{c} ^{\rm opt}$ within the error bars. 
This suggests that the sample dependence of $T_{c} ^{\rm opt}$ may be accounted for by the sample dependence of the apical oxygen position.

We note that the order parameter $F_{\rm SC}^{\infty}\sim 0.10$ obtained here for Bi2212 and Bi2201 also shows consistency with the result obtained by using the machine learning of the angle-resolved photoemission spectroscopy data for Bi2212 and Bi2201, respectively, at the optimum doping~\cite{PhysRevResearch.3.043099}, which gave $\langle c_{k\uparrow}c_{-k\downarrow}\rangle\sim 0.065$ at the antinodal point for Bi2212 and the momentum averaged value $\sim 0.063$ for Bi2201 at the optimum doping. 
These are translated commonly to $F_{\rm SC}^{\infty}\sim 0.09$ in the present definition of $F_{\rm SC}^{\infty}$ because of the relation $F_{\rm SC}^{\infty}=\sqrt{2}\langle c_{k\uparrow}c_{-k\downarrow}\rangle$.

In the case of Bi2201, the estimated $0.16 |t_1| F_{\rm SC}^{\infty}$ is also listed in Table~\ref{tab:pdsum3}. The comparison with the sample dependence of experimental $T_{c}$ suggests that the true apical oxygen position is distributed near the lower bound $d^z_{\rm Oap}\sim 2.45$\AA, if it is spatially uniform. Alternatively if the supermodulation exists, the lower bound of  $d^z_{\rm Oap}$ in the modulation may be close to 2.45 \AA, because it governs the SC order as the bottleneck.
It is desired to test this inference by precise and simultaneous measurements of the relation between $T_{c} $ and $d_{\rm Oap}^{z}$ for Bi2212 and Bi2201.

The order parameter $F_{\rm SC}^{\infty}$ increases with decreasing hole doping for $\delta>0.05$ as we find in Fig.~\ref{fig:CaCuOPdd}, which follows the same trend as the SC gap as we discussed, but is slightly different from the dome structure known for $T_{c} $ in the cuprates, where the peak of the dome is located at higher $\delta$. Complete and quantitative understanding of this different trend is not the scope of this paper and is left for future studies. However, the origin of this difference can be inferred to be attributed to the increasing damping and incoherence of electrons in the underdoped region toward the metal-insulator transition as was analyzed before~\cite{Tanaka2006,Alldredge2008}, which is represented as the enhanced self-energy of the normal electrons toward the Mott insulator. 
{In the experimental conditions, the atomic doping/substitution introduces atomically spatial inhomogeneity, which is ignored in the present study and may cause the localization of carriers, which could also be the origin of slower increase of $T_c$ upon doping.}

In Appendix~\ref{app:n(k)}, we show a qualitative difference of the momentum distribution between the optimal and underdoped hole concentrations, which suggests a signature of the increased damping at lower carrier concentration within the present {\it ab initio} study. 
{We also discuss in Appendix~\ref{app:n(k)} the subtlety and complexity in the underdoped region due to the pseudogap formation, involved in several quantities, which is not the scope of this paper.}

The $\delta$ dependence of the energy decomposed to the kinetic, on-site and off-site interaction energies are analyzed in Appendix~\ref{app:d_dep_energy}. 
It should be noted that the on-site interaction energy $E_U=\langle \mathcal{H}_U \rangle$ has a convex curvature as a function of $\delta$, which contributes to the effective attraction of electrons despite the original strong repulsion $U$. 
{In fact, the $U/|t_1|$ dependence of the local attraction is qualitatively consistent with $F_{\rm SC}$ in Fig.~\ref{fig:CaCuOPdScalingZoom}. See Appendix~\ref{app:d_dep_energy} for more details.}
The local effective attraction may cause the Cooper pairing as well as the stripe formation. It was pointed out that the convex curve of the local energy generates bistable excitations, one in the underdoped side and the other in the overdoped side, inducing the electron fractionalization and the enhanced quantum entanglement through the quantum tunneling of the two excitations~\cite{imada_suzuki2019,imada_review2021}. {This line of further research is an important future subject.}

The importance of the apical oxygen position has been pointed out from various viewpoints~\cite{pavarini2001,mori2008,weber2010,sakakibara2010}. 
In this paper, we have elucidated the crucial role of controlling $U$ in general in the single-band description, which quantitatively explains the variation of $T_c^{\rm opt}$ and its uncertainty in the Bi compounds. 
In addition, the modulation of the SC gap with the modulation of the apical oxygen position in Bi2212 has indicated that the longer $d_{\rm Oap}^{z}$ induces the smaller SC gap~\cite{slezak2008}. 
This is consistent with the trend of $F_{\rm SC}^{\infty}$, which decreases when $\alpha$  is increased in the realistic range as shown in Fig.~\ref{fig:BiCompundsApicalUncertainty}. 
This indicates that Bi2212 is located already slightly above the peak in the $\alpha$ dependence of $F_{\rm SC}^{\infty}$, which corresponds to $\alpha=1.2$ for doped CaCuO$_2$ shown in Fig.~\ref{fig:CaCuOPdScalingZoom}. 
The present observation is also in accordance with the effect observed by laser irradiation aiming at the displacement of the apical oxygen position~\cite{hu2014}. 
The control of $d_{\rm Oap}^{z}$ if possible in a spatially uniform fashion may help to optimize the SC in which the disturbance and pair breaking by the randomness caused by the inhomogeneous supermodulation in the case of the Bi compounds could be avoided.

{The effects of the displacement of the apical oxygen position on the Hamiltonian parameters were discussed in Ref.~\cite{moree2022}. In Appendix~\ref{app:ApicalO_effect}, we readdress this issue in relation to earlier work.}

{We have mainly focused on the quantities at optimal doping to clearly extract the diversity of the materials dependence, where the experimental subtlety due to the effects from extrinsic randomness as well as the complexity of physical quantities arising from the pseudogap formation in the underdoped region is irrelevant. The behavior of suppressed $T_c$, the SC carrier density $n_s$, and the coherent spectral weight arising from the pseudogap formation in the underdoped region are left for future studies. See Appendix~\ref{app:n(k)} for more details.}

{In general, the $d$-wave SC correlation has a dip at (1,1) distance. The origin is speculated as follows: The singlet pair between electrons at (0,0) and (1,0) sites dynamically interfere with the singlet pair between (1,1) and (1,0), because these two singlets are incompatible. The same is true for the singlets, which share (0,1) site. This double interfere makes the SC correlation of the pair between (0,0) and (1,1) smaller very generally. However, at the moment, we do not know the origin of particularly large dip around $\delta=1/8$, which is left for future study.}

{There exist studies proposing the possibility of the pair-density-wave (PDW) states~\cite{Himeda2002}. However, in the present {\it ab initio} Hamiltonians, the PDW correlation remains small as is shown in Fig. S7 of SM~\cite{SM_Michael},
where peak is absent at nonzero momentum.
If the stripe long-ranged order coexists in the SC ground state, the PDW order must be trivially accompanied. However, as one sees in Figs.~\ref{fig:CaCuOSsqScqOverLAF0000}(b) and S6, the stripe correlation exists but remains small and is scaled to zero in the thermodynamic limit.}

{In this paper, we have employed the single-band Hamiltonian for the AB orbital, because the hybridization gap between AB and B or NB orbitals is so large ($\sim 8-9$ eV) that the B and NB bands are more or less completely filled and inactive (see Fig. 10(b) of Ref.~\cite{hirayama2019} and Appendix D of Ref.~\cite{moree2022}). When one starts from the three bands constructed from Cu 3$d_{x^2-y^2}$ and O 2$p_{\sigma}$ atomic orbitals, the analysis would be more complicated. See {Appendix \ref{app:multi-band}} for this issue.}

\section{Summary and conclusions}
\label{sec:summary}
We have studied the superconductivity in the {\it ab initio} Hamiltonians for CaCuO$_2$, Hg1201, Bi2201, and Bi2212 derived by using the experimental crystal structure in Ref.~\cite{moree2022} without adjustable parameters . 
The dominance of SC order against severely competing stripe states and  AFM state in a wide range of hole concentration is shown in the solutions for the ground state of all four materials obtained from the variational Monte Carlo calculations, which agrees with the experimental results. 

The SC order parameter $F_{\rm SC}^{\infty}$ at the optimal doping shows consistency with the superfluid density measured in the $\mu$SR
and the machine learning analysis of the ARPES data for Bi2212 and Bi2201. $F_{\rm SC}^{\infty}$ decreases with increasing doping for the doping concentration $\delta>0.05$, showing a similarity to the SC gap reported in the STM and ARPES measurements. On the other hand, $F_{\rm SC}^{\infty}$ quickly decreases to zero toward $\delta=0$ for $\delta<0.05$ forming a dome structure which has a similarity to experimental $T_{c}$, but the dome peak appears at slightly lower $\delta$ for the calculated $F_{\rm SC}^{\infty}$. 
This may be attributed to the reduced renormalization factor suggested by the broadened momentum distribution.

\begin{figure}[tbh!]
    \includegraphics[width=\linewidth]{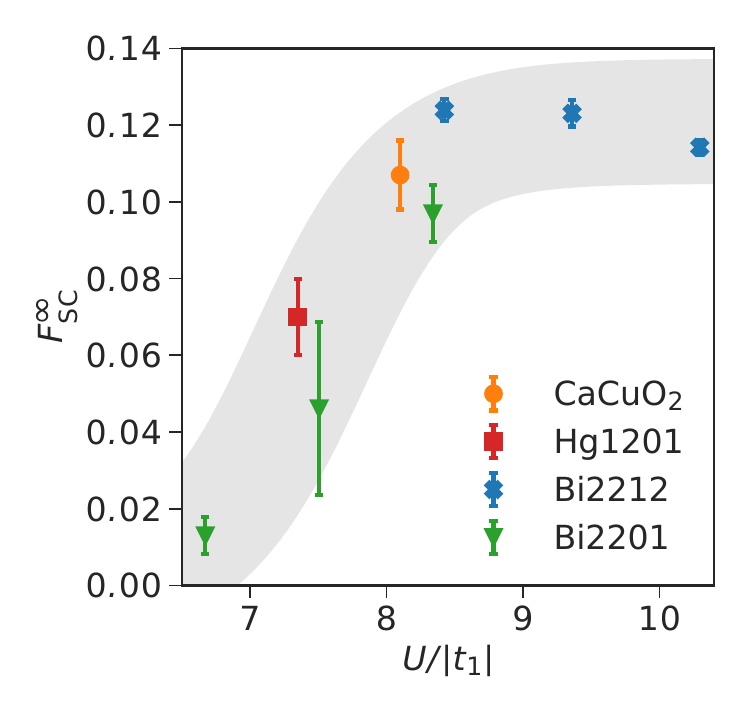}
    \caption{$F_{\rm{SC}}^{\infty}$ as a function of $U/|t_1|$ for the four cuprate compounds at $\delta = 0.167$ plotted from the list in Tables~\ref{tab:pdsum2} and~\ref{tab:pdsum3}.}
    \label{fig:FSC_vs_U/t}
\end{figure}

\begin{figure}[tbh!]
    \includegraphics[width=\linewidth]{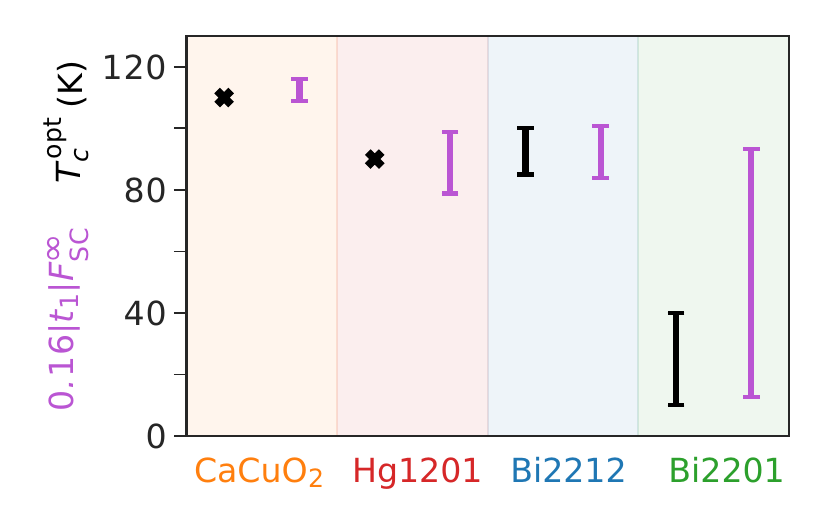}
    \caption{Experimental $T_{c}^{\rm{opt}}$ (black crosses or bars) in comparison to the $ T_{c}  = 0.16  \lvert t_1 \rvert  F_{\rm{SC}}^{\infty}$ scaling for each compound (purple bars). 
     $T_{c}$ is taken from Tables~\ref{tab:pdsum2} and~\ref{tab:pdsum3}.}
    \label{fig:TcComparisionExperiment}
\end{figure}

From the comparison of the four materials, we have revealed that $U/|t_1|$ is a crucial parameter to control the strength of the SC order; larger $U/|t_1|$ materials show larger SC order parameter $F_{\rm{SC}}^{\infty}$ in the realistic materials. 
This explains that $T_{c} $ and the SC gap at the optimum doping are larger for CaCuO$_2$ than Hg1201, where $T_{c} ^{\rm opt}$ is well scaled by $\lvert t_1 \rvert  F_{\rm SC}^{\infty}$ as $T_{c} ^{\rm opt}\sim 0.16   \lvert t_1 \rvert   F_{\rm{SC}}^{\infty}$. 
Though the experimental uncertainty in the crystal structure prohibits a quantitative comparison, $F_{\rm SC}^{\infty}$ is also larger for Bi2212 than Bi2201 at least qualitatively, in agreement with the experimental indications. 
When we apply the same scaling $T_{c}^{\rm opt} \sim 0.16   \lvert t_1 \rvert   F_{\rm SC}^{\infty}$ to the two Bi compounds with the calculated order parameter, it also well explains the experimental sample dependence of $T_{c}$. The strong dependence of $F_{\rm SC}^{\infty}$ on $U/|t_1|$ for real materials are summarized in Fig.~\ref{fig:FSC_vs_U/t}: In the range of $7.0\le U/|t_1|\le 8.0$, $F_{\rm SC}^{\infty}$ sharply increases and the calculated sensitive materials dependence of  $F_{\rm SC}^{\infty}$ is well captured within this range. {This simply means that, except for Bi2212,  most of the realistic materials we have studied are positioned in the weak-coupling  side, where the SC order parameter rapidly increases with increasing $U/|t_1|$. }
The good scaling of $T_{c}^{\rm opt}$ by $0.16 |t_1| F_{\rm{SC}}^{\infty}$ is also summarized in Fig.~\ref{fig:TcComparisionExperiment}, which indicates that the detailed difference of $U/|t_1|$ within the range of $7<U/|t_1|<9$ in the {\it ab initio} parameters reproduces the diverse materials dependence of $T_c^{\rm opt}$. 
Since the larger variance for the theoretical prediction on Bi2201 is ascribed to the experimental uncertainty of the apical oxygen, it is desired to precisely determine the apical oxygen position in the experiments. 

Based on the successful reproduction of the materials dependent properties, the underlying superconducting mechanism is identified by the effective local attraction emerging from the Mottness, which converts the original strong repulsion to the attraction.  

The SC order parameter has the maximum above the {\it ab initio} values of $U/|t_1|$ at 20\% larger value of $U/|t_1|$ with the enhancement of 20\%-30\%. If one can control on-site and off-site interaction independently, further optimization of the SC order parameter as much as the factor 2 larger value  beyond the available compounds synthesized so far without falling into other competing states can be achieved as the theoretical maximum value in the present mechanism. 
By increasing $|t_1|$ as well as the whole parameter values uniformly, $T_{c}^{\rm opt}$ should obviously increase accordingly. 
These offer a clue for the materials design in the future.  

\section*{Acknowledgements}
The authors thank Atsushi Fujimori for useful discuussions. 
This research was supported by MEXT as ``Program for Promoting Researches on the Supercomputer Fugaku"(Basic Science for Emergence and Functionality in Quantum Matter - Innovative Strongly Correlated Electron Science by Integration of Fugaku and Frontier Experiments -, JPMXP1020200104 and JPMXP1020230411). 
M.I. acknowledges the support of MEXT KAKENHI, Grant-in-Aid for Transformative Research Areas, Grant Nos. JP22H05111, JP22H05114, and JSPS KAKENHI Grant Numbers JP16H06345, JP19H00658.
We thank the Supercomputer Center, the Institute for Solid State Physics, The University of Tokyo for the use of the facilities.
We also thank the computational resources of supercomputer Fugaku provided by the RIKEN Center for Computational Science (Project ID:  hp200132, hp210163, hp220166, hp230207, hp230169). 
J.-B.M acknowledges financial support from the Special Postdoctoral Researcher Program at RIKEN.

\renewcommand{\thesection}{A\arabic{section}}
\appendix
\section{Lanczos method and restricted Boltzmann machine procedure}
\label{app:RBM_Lanczos}
\subsection*{Lanczos method}
To further improve the accuracy of the VMC calculations or the variance extrapolation of the competing ground state candidates, we apply the Lanczos method~\cite{ido2022}.
To do so the optimized mVMC wave function $\ket{\Psi}$ is extended by
\begin{equation}
    \ket{\psi_n} = \left( 1 + \sum_{n = 1}^M \alpha_n \mathcal{H}^n \right) \ket{\Psi}.
\end{equation}
The factor in front of $\ket{\Psi}$ can be regarded as an additional projection operator with variational parameters $\alpha_n$, which--when chosen appropriately--further reduce the energy. 
Although one could systematically improve the wave function by increasing $M$, the computational cost increases exponentially with $M$, too. 
Hence we employ the Lanczos method only up to the first step ($n=1$) within this work. 

\subsection*{Restricted Boltzmann machine procedure}
To further improve the mVMC wave function we apply a restricted Boltzmann machine as a variational wave function $\ket{\Psi}$, as first suggested in Ref.~\cite{carleo2017}. 
Here we follow the notations given in Refs.~\cite{nomura2017,ido2022}.
The variational wave function including RBM takes the following form
\begin{equation}
    \ket{\Phi} = \mathcal P^\mathrm{G} \mathcal P^\mathrm{J} \mathcal P^\mathrm{dh} \mathcal{N} \ket{\phi^\mathrm{pair}},
\end{equation}
where $\mathcal N$ is the additional RBM correlator. 
The RBM correlator
\begin{align}
    \mathcal{N}
    &= \prod_{k} 2 \cosh\left( b_k + \sum_i W_{ik} \sigma_i\right)  \mathrm{e}^{\sum_i a_i \sigma_i}
\end{align}
introduces the additional variational parameters $b_k$ (hidden layer), $a_i$ (visible layer), and $W_{ik}$ (network).  

In practice we apply the additional RBM projection after the wave function was already optimized via mVMC calculations, i.e., the variational parameters of the optimized wavefuction $\ket{\phi} = \mathcal P^\mathrm{G} \mathcal P^\mathrm{J} \mathcal P^\mathrm{dh} \ket{\phi^\mathrm{pair}}$  are kept fixed during the RBM procedure. 

The accuracy of the wave function depends on the number of hidden and visible parameters ($N_h$, $N_v$)~\cite{carleo2017}. 
Hence we can define a hidden variable density as  $\alpha_{\rm RBM} = N_h/N_v$ as measure for the accuracy. 
Note that the number of RBM variational parameters increases with $\alpha$, too.

Within this work the RBM procedure was applied with $\alpha_{\rm RBM} = 4$.

\section{Variance Extrapolation Procedure}
\label{app:VarExt}
The total energy per site $E/N=\braket{\mathcal H}/N$ is calculated after the variance extrapolation.
The true ground state as a function of the variance $(\delta E)^2 = (\braket{\mathcal H^2} - \braket{\mathcal H}^2)/\braket{\mathcal H}^2$ is obtained by taking the limit $\delta E\rightarrow 0$, because the true eigenstate satisfies $\delta E=0$~\cite{imada2000,kashima2001,sorella2001}. 
If several different states such as SC, spin- and charge-ordered as well as normal metallic states are competing, the ground state is determined from the lowest-energy state after the variance extrapolation.
In practice, the variance extrapolation is a linear regression of the energies per site $E/N$ (e.g.\ obtained from combinations of mVMC, Lanczos, or RBM) over the variance $\delta E$ for a specific state, to approximate the zero variance value of the energy. 

\begin{figure}[tbh!]
    \includegraphics[width=\linewidth]{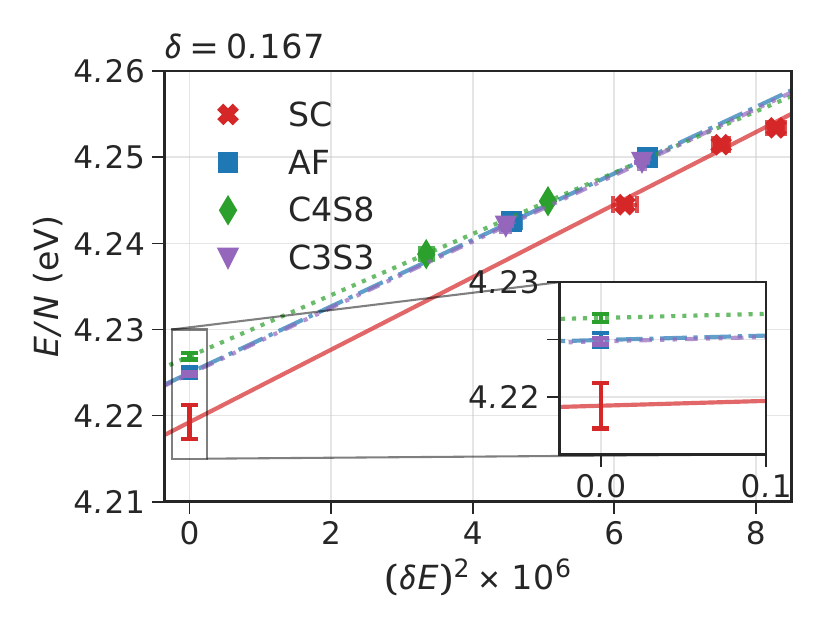}
    \caption{Variance extrapolation for the example of CaCuO$_2$ at $\delta=0.167$ on a $L=24$ lattice. 
    The inset shows an enlarged plot around $(\delta E)^2 = 0$.}
    \label{fig:CaCuOVarianceExtrapolation24}
\end{figure}
As an example the variance extrapolation for $\delta = 0.167$ in CaCuO$_2$ is shown in Fig.~\ref{fig:CaCuOVarianceExtrapolation24}.
From the inset we confirm the lowest energy of the SC state with severely competing AFM and stripe states very close to it (the difference is around $5\,\text{meV}$). 
However,  the overlap of the error bars is small and in comparison to AFM, C4S8 and C3S3 states, the SC state definitely has lower energy.

\section{Results of $\delta$ Dependent Hamiltonian}
\label{app:delta_dependent_Hamiltonian}
{The hole density dependence of $F_{\rm SC}$ at $L=24$ is compared between the case presented in the main text and that calculated by the $\delta$ dependent Hamiltonian given in Sec. S1B of SM~\cite{SM_Michael}. 
The results for CaCuO$_2$ in Fig.~\ref{fig:Pdd_Ca11_Bi2212_L24}(a) and for Bi2212 in Fig.~\ref{fig:Pdd_Ca11_Bi2212_L24}(b) show that the difference of the two Hamiltonians is small and essential feature can be analyzed by taking the Hamiltonians analyzed in the main text.
\begin{figure}[tbh!]
    \includegraphics[width=\linewidth]{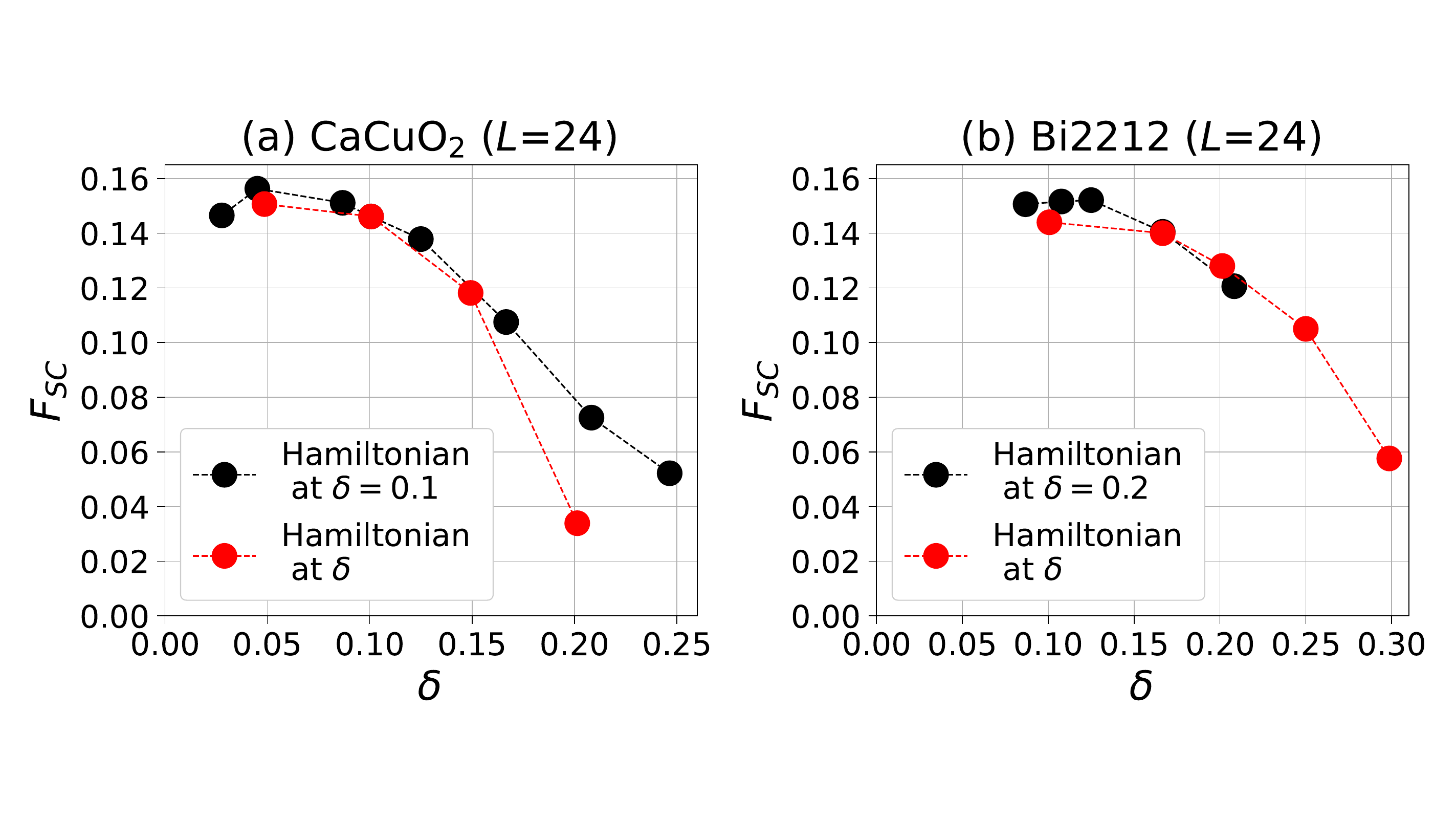}
    \caption{{Comparison of $F_{\rm SC}$ between that obtained by using $\delta$ dependent Hamiltonians (red) and the fixed Hamiltonian (black) as in the main text.}}
    \label{fig:Pdd_Ca11_Bi2212_L24}
\end{figure}
}

\section{Thermodynamic limit of the trial wave function after optimization for CaCuO$_2$}
\label{app:ThermoLimitCaCuO}

Here we discuss the stability of the AFM, C3S3, and C4S8 states in the thermodynamic limit for CaCuO$_2$.
To do so, each state is stabilized on different lattice sizes and the spin and charge structure factors ($S_s(\bm q)$ and $S_c(\bm q)$) are calculated. 

\begin{figure}[tbh!]
    \includegraphics[width=\linewidth]{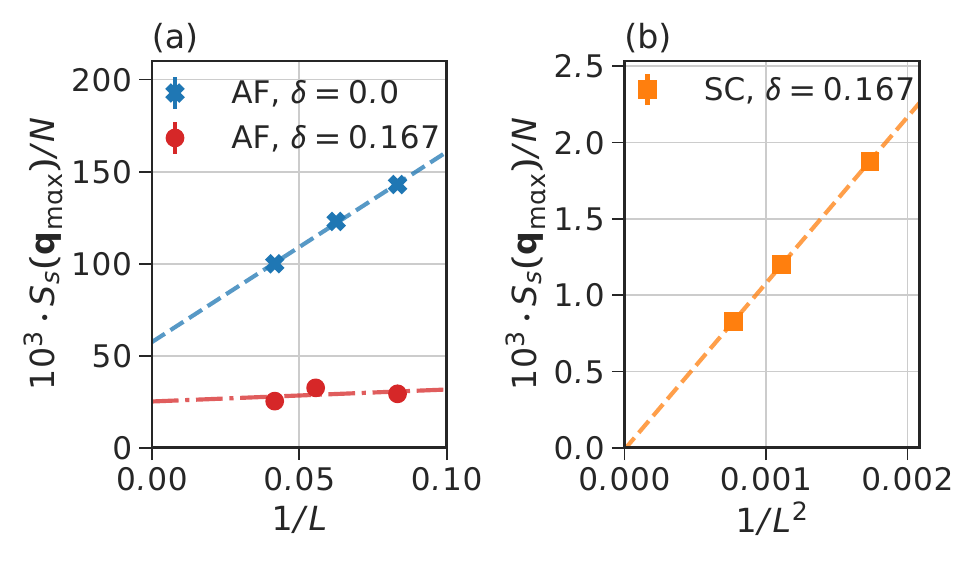}
    \caption{Size dependence of the spin structure factor $10^3 \times S_s(\bm q_{\mathrm{max}})/N$ at $(\pi,\pi)$ for (a) the AFM state after optimization at half filling ($L = 12,16,24$) and $\delta = 0.167$ ($L = 12,18,24$), and for (b) the SC state at $\delta=0.167$ for different lattice sizes ($L = 24,30,36$).}
    \label{fig:CaCuOSsqScqOverLAF0000}
\end{figure}

We plot $S_s(\bm q)$ as a function of $1/L$ or $1/L^2$ depending on the cases of the presence or absence, respectively, of the AFM order by following the convention and perform a linear extrapolation.
The result is shown in Fig.~\ref{fig:CaCuOSsqScqOverLAF0000}, where in  Fig.~\ref{fig:CaCuOSsqScqOverLAF0000}(a) the peak height of the spin structure factor follows a linear trend with a nonzero offset, indicating a stable long-ranged AFM order.
In  Fig.~\ref{fig:CaCuOSsqScqOverLAF0000}(b), the peak of the spin structure is scaled to zero in the thermodynamic limit, which indicates the absence of the long-ranged AFM order in the SC state at this doping. 
We did not find the coexistence of the SC and AFM order at other doping, either.

\begin{figure}[tbh!]
    \includegraphics[width=\linewidth]{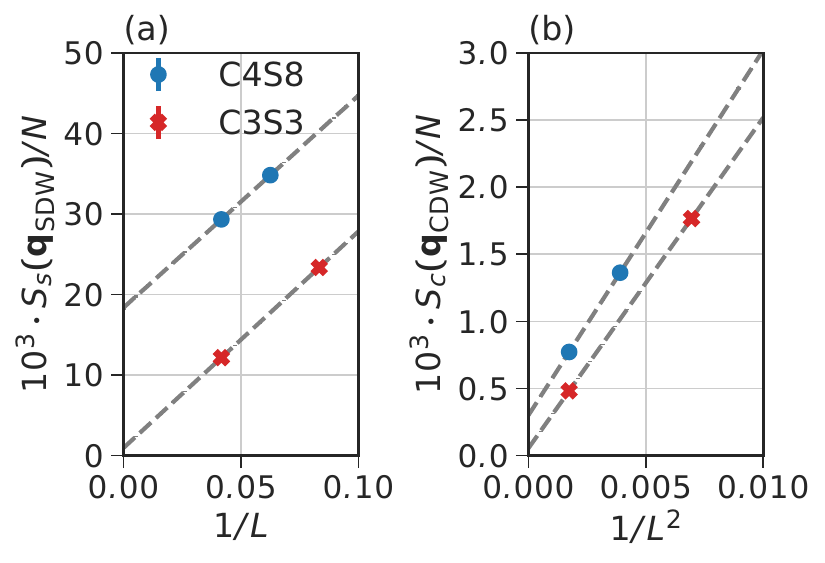}
    \caption{Size dependence of the charge and spin stripe state after the optimization of variational parameters for different lattice sizes and fillings. 
    For C3S3 (red crosses) $\delta = 0.207$ and $L= 12,24$ and for C4S8 (blue dot) $\delta = 0.125$ and $L = 16, 24$. 
    (a) $10^3 \times S_s(\bm q_{\mathrm{SDW}})/N$ vs $1/L$. 
    (b) $10^3 \times S_c(\bm q_{\mathrm{CDW}})/N$ vs. $1/L^2$. ${\bf q}_{\rm SDW}$ and ${\bf q}_{\rm CDW}$ are the wave numbers at  the peak of $S_s$ and $S_c$, respectively, for the spin density wave (SDW) and charge density wave (CDW) modulations, respectively. The dashed lines are only to guide the eyes.}
    \label{fig:CaCuOSsqScqOverLStripes}
\end{figure}

The size scaling of the charge-stripe states is shown in Fig.~\ref{fig:CaCuOSsqScqOverLStripes}. 
Although the plot of the size dependence is not sufficient due to very demanding computational cost for larger sizes, the trend suggests that only the C4S8 state shows a clear long-ranged order of spin and charge at around $\delta=0.125$ in the thermodynamic limit.
The C3S3 state seems to collapse to a paramagnetic state at the chosen doping of $\delta = 0.207$, at which C3S3 has relative stability. 
Hence we use the labeling ``C3S3-like" instead of C3S3 in the main text. 
We do not go into details of the size dependence for the stripe-like states, because they are in any case excited states.

\section{Comparison to Hubbard model studies}
\label{app:Comparison_to_Hubbard}

{Here we discuss the crucial difference of the {\it ab initio} results from the extensively studied Hubbard models without the off-site interaction.
In the Hubbard model studies, irrespective of the presence of the next-nearest transfer $t_2$ or absence of it, in the hole doped region, a broad consensus seems to be formed, where the ground state has stripe type long-ranged charge order~\cite{zhao2017,darmawan2018,ido2018,Ponsioen2019,Tocchio2019,Qin2020,Sorella2021,Marino2022,Xu2023}. The charge and spin stripe states were also suggested to coexist with weak SC order in some cases~\cite{darmawan2018,Xu2023}, but other studies did not find the SC order~\cite{ido2018,Tocchio2019,Qin2020,Marino2022}.}

{In the Hubbard model studies, the numerical methods have a variety including the present VMC, density matrix renormalization group, constrained path quantum Monte Carlo, tensor network and density matrix embedding theory, which have their own advantages and disadvantages and they are complementary.
When they can be compared with reliable solutions, the above Hubbard model calculations were benchmarked, which have shown comparable accuracy when compared between each state-of-the-art updated version. Even in quantum spin models, the level of accuracy of the above methods is roughly similar~\cite{nomura2021}. We also refer to recent thorough comparisons~\cite{Wu2023}.}

{On the other hand, when realistic off-site interaction is included, the ground state is reported to be charge homogeneous superconducting state~\cite{ohgoe2020}. The off-site interaction substantially suppresses the SC order but the charge and spin stripe states are more damaged by the frustration introduced by the off-site interaction. The role of off-site interaction for the stabilization of the SC state relative to the stripe state was directly demonstrated in Ref.~\cite{ohgoe2020} by the comparison of {\it ab initio} result and that of the Hamiltonian obtained by switching off only the off-site interactions. We confirmed the similar behavior for the present Hamiltonians.
The absence of developed stripe correlations in the SC ground state is seen in Fig.~\ref{fig:CaCuOSsqScqOverLAF0000} herein and in Fig.S6 of SM~\cite{SM_Michael}.}

\section{$t_2$ dependence}
\label{app:t2_dependence}
\begin{figure}[tbh]
    \includegraphics[width=\linewidth]{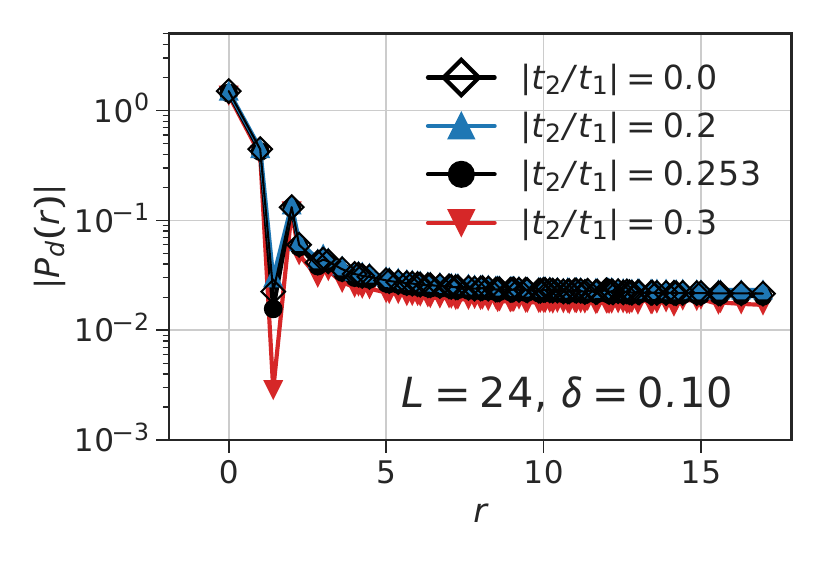} 
    \caption{{SC correlation function $P_d(r)$ for modified $t_2$ from the CaCuO$_2$ Hamiltonian for $L=24$ lattice.}
    } 
    \label{fig:Ca_Pd_compare_t2overt1_02_03}
\end{figure}
\begin{figure}[tbh]
    \includegraphics[width=\linewidth]{
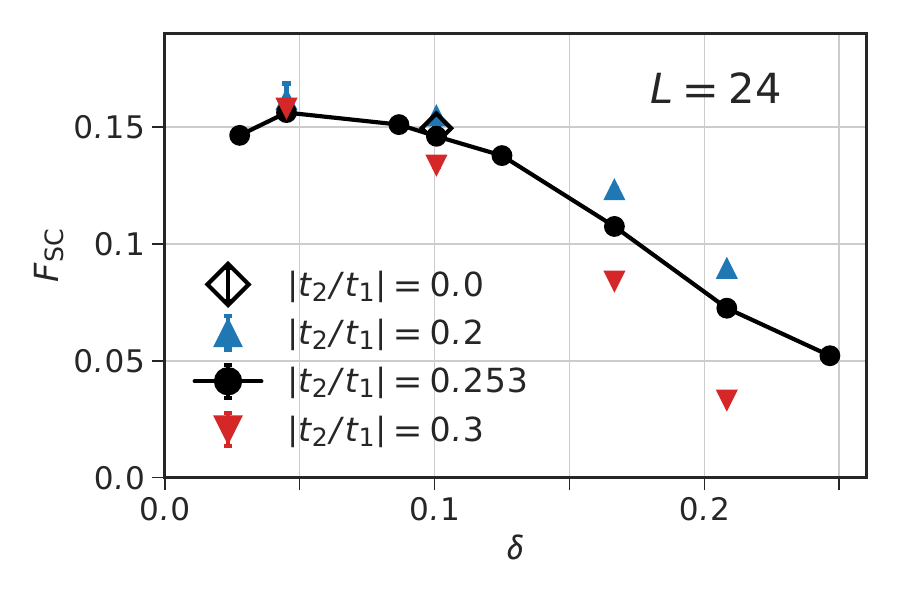}
\caption{{SC order parameter $F_{\rm SC}$ as a function of $\delta$ calculated for three choices of $t_2$ modified from from CaCuO$_2$ Hamiltonian.
   } } 
    \label{fig:PdOverDopingCaCuO_2}
\end{figure}
{Here we show $t_2$ dependence of $F_{\rm SC}$ by starting from the {\it ab initio} Hamiltonian for hole doped CaCuO$_2$ with other Hamiltonian parameters fixed, where the effect of $t_2$ is monitored beyond the {\it ab initio} value primarily within the realistic range of $|t_2/t_1|$ ($0.2\leq |t_2/t_1|\leq 0.3$) in Figs.~\ref{fig:Ca_Pd_compare_t2overt1_02_03} and \ref{fig:PdOverDopingCaCuO_2} for $L=24$ lattice. $F_{\rm SC}$ slightly decreases with increasing $t_2$, which is qualitatively consistent with a different approach~\cite{sakakibara2010}. However, the variation of $F_{\rm SC}$ is at most 10\% near the optimum doping in the realistic range. Furthermore, $F_{\rm SC}$ has essentially no $t_2$ dependence in the range $0.0\leq |t_2/t_1|\leq 0.2$. {On the other hand, the period of the stripe order is known to sensitively depend on $t_2$~\cite{ido2018,darmawan2018,Xu2023,Tocchio2019,Qin2020,Marino2022} in the ground states of Hubbard models and it may alter the superconductivity if it coexists, while the present charge uniform superconducting ground state is quite different.}}

\section{Analysis of $\delta$ dependence of energy}
\label{app:d_dep_energy}
\begin{figure}[tbh]
    \includegraphics[width=\linewidth]{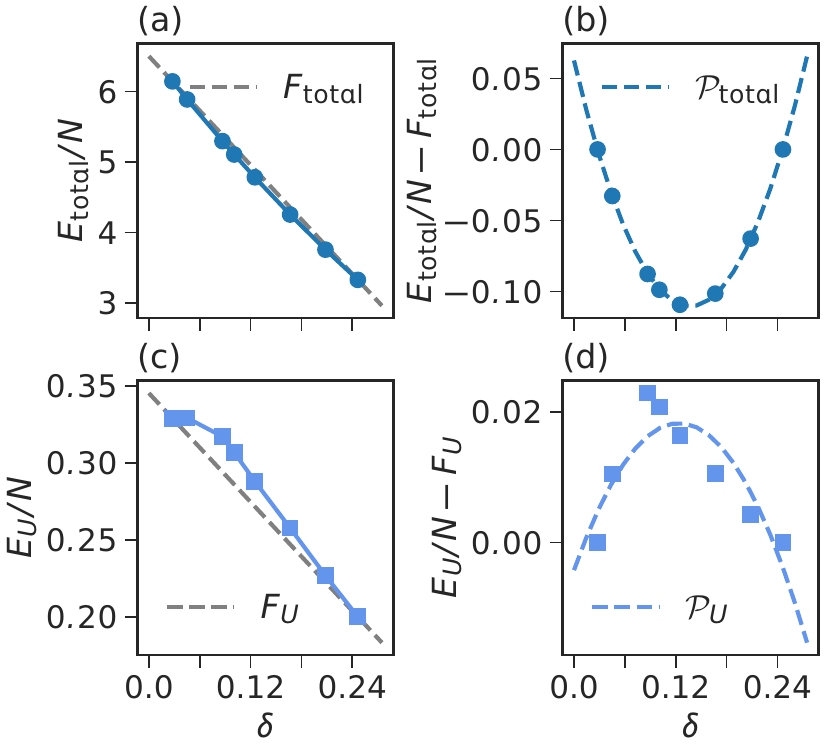} 
    \caption{
    Energy per site as a function of $\delta$ in the SC state on a $L=24$ square lattice. Total energy $E_{\mathrm{total}}/N$ (a), and $E_{\mathrm{total}}/N-F(\delta)$ (b) as well as
    on-site Coulomb energy $E_U/N$ (c), and $E_U/N-F(\delta)$ (d) are plotted. Here, a $\delta$-linear function $F(\delta)$ defined below is subtracted in (b) and (d) for better visibility. 
    Note that $F(\delta)$ is different depending on the type of the energy as seen in Table~\ref{tab:CaCuOL24EnergiesFitParameter}.
    The gray line in the left column indicates $F(\delta) =   b_0 + b_1 \delta $, which is subtracted in (b) and (d). 
    Right column: Energies after subtraction of $F(\delta)$ are fitted via a quadratic function $\mathcal P(\delta) =  c_0 + c_1 \delta + c_2 \delta^2$.
    Fitting parameters are given in Table~\ref{tab:CaCuOL24EnergiesFitParameter}.} 
    \label{fig:CaCuOL24Energies}
\end{figure}

In Fig.~\ref{fig:CaCuOL24Energies} the total energy per site $E_{\rm tot}$ and the on-site Coulomb part $E_U$ (see panel~(a),~(c),  respectively) are shown for doped CaCuO$_2$.
Each energy contribution is subtracted by a linear function $F(\delta) =   b_0 + b_1 \delta$ for better visibility, where $b_0$ and $b_1$ are listed in Table~\ref{tab:CaCuOL24EnergiesFitParameter} (see gray lines in~(a),~(c)) and are shown in~(b),~(d). 
The subtracted energies are fitted by a quadratic function $\mathcal P(\delta) =  c_0 + c_1 \delta + c_2 \delta^2$ to examine the curvature. 
Explicit values of the parameters are given in Table~\ref{tab:CaCuOL24EnergiesFitParameter}.

The result shows that the total energy is concave as a function of $\delta$ of course, which is required from the thermodynamic stability, while only $E_U$ exhibits convex behavior with $c_2<0$. Because the effective particle interaction is given by the $\delta^2$ term, we find that the local quantity $E_U$ is the origin of the attraction while the nonlocal energies contribute to $c_2>0$ in the total energy ensuring the thermodynamic stability. 
This supports the idea that the local strong correlation (repulsion) called Mottness turns to originate the effective attraction of the electrons, which is the underlying mechanism of both of the Cooper pairing and charge segregation such as the stripes. 
The attraction is understood from the following Mottness: The local energy is retained high in the Mott insulator because of $U$. However, if the carrier is doped, this is released by acquiring the itinerancy in a nonlinear fashion as a function of $\delta$ which yields $c_2<0$ and the attraction.  {In accordance with the $\alpha$ dependence of $F_{\rm SC}$, $c_2$ shows a similar behavior as  $-c_2=1.47, 2.24, 2.13, 1.85$ and 1.53 at $\alpha =1.0, 1.1, 1.2, 1.3$ and 1.4, respectively with a peak around $\alpha\sim 1.1$-1.2. }

\begin{table}[tbh]
    \caption{Fitting parameters of linear correction function and quadratic fit function discussed in Fig.~\ref{fig:CaCuOL24Energies}.
            See the main text for the explanation.}
    \label{tab:CaCuOL24EnergiesFitParameter} 
    \begin{ruledtabular}
        \begin{tabular}{ l c c c c c }
              & $b_0$     & $b_1$     & $c_0$    & $c_1$   & $c_2$ \\
            \toprule
            $E_{\mathrm{total}}/N$ & $6.4995$ & $-11.9723$ & $0.0626$ & $-2.5274$  & $9.2307$\\
            $E_U/N$ & $0.3455$ & $-0.5900$ & $-0.0043$ & $0.3651$  & $-1.4749$\\
        \end{tabular}
    \end{ruledtabular}
\end{table}

\section{Interlayer contributions for Bi2212 }
\label{app:interlayer}
\begin{table}[h!]
\caption{Effective interlayer Hamiltonian parameters (in eV) for Bi2212 at $\delta=0.2$ taken from Ref.~\cite{moree2022}.} 
\label{tab:bi2212Vinter}
\begin{ruledtabular}
    \begin{tabular}{ccccccc}
        Bi2212 & & & & & & \\
        \toprule
        $V^{l}_{0}$ & $V^{l}_{1}$ & $V^{l}_{2}$ & $V^{l}_{3}$ & $V^{l}_{4}$ & $V^{l}_{5}$ & $V^{l}_{6}$ \\
        $0.643$ & $0.463$ & $0.368$ & $0.291$ & $0.262$ &  $0.220$ & $0.120$ \\
        $t^{l}_{0}$ & $t^{l}_{1}$ & $t^{l}_{2}$ & $t^{l}_{3}$ & $t^{l}_{4}$ & $t^{l}_{5}$ & $t^{l}_{6}$\\
        $-0.098$ & $-0.001$ & $0.019$ & $-0.010$ & $0.007$ & $0.000$ & $-0.003$ \\
    \end{tabular}
\end{ruledtabular}
\end{table}
The effective interlayer Hamiltonian parameters for Bi2212 are shown in Table~\ref{tab:bi2212Vinter}, where $V^l_n$ is the interlayer Coulomb interaction, and  $t^l_n$ the interlayer hoppings. 
As defined in Ref.~\cite{moree2022}, the notation $n=0$ represents the interaction or hopping between interlayer nearest-neighbor Cu atoms (located one above the other), and $n \geq 1$ represents the interaction or hopping between a Cu atom and its $(n+1)$th interlayer nearest-neighbor (located above or below its intralayer $n$th nearest-neighbor).

\begin{figure}[tb]
\includegraphics[width=\linewidth]{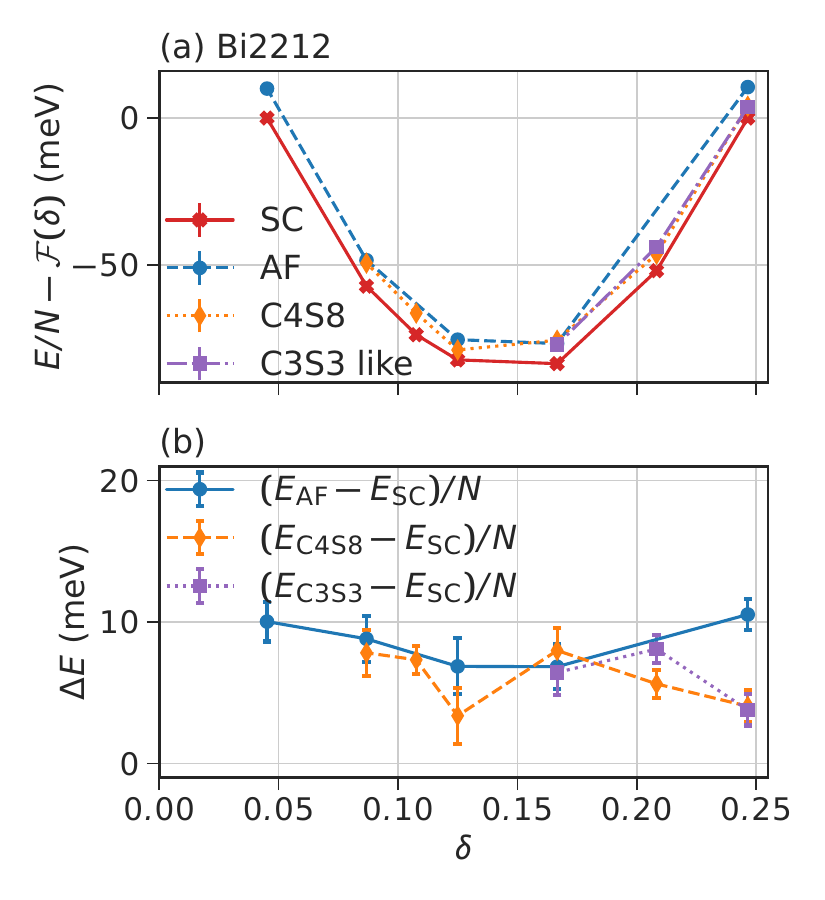} 
\caption{Variance extrapolated energies of Bi2212 for various ground-state candidates (SC, charge/spin stripe C3S3 and C4S8, and AFM states) as a function of hole doping $\delta$ on a $L=24$ square lattice. (a) Total energies per site subtracted by $\mathcal{F}(\delta)$.
All energies are subtracted by the function $\mathcal F(\delta) =  -12.34985 \cdot \delta +  6.36953$. 
(b) Energy difference $\Delta E$ for the variance extrapolated data from (a).} 
\label{fig:EnergyOverDeltaBi2212}
\end{figure}
\begin{figure}
\includegraphics[width=\linewidth]{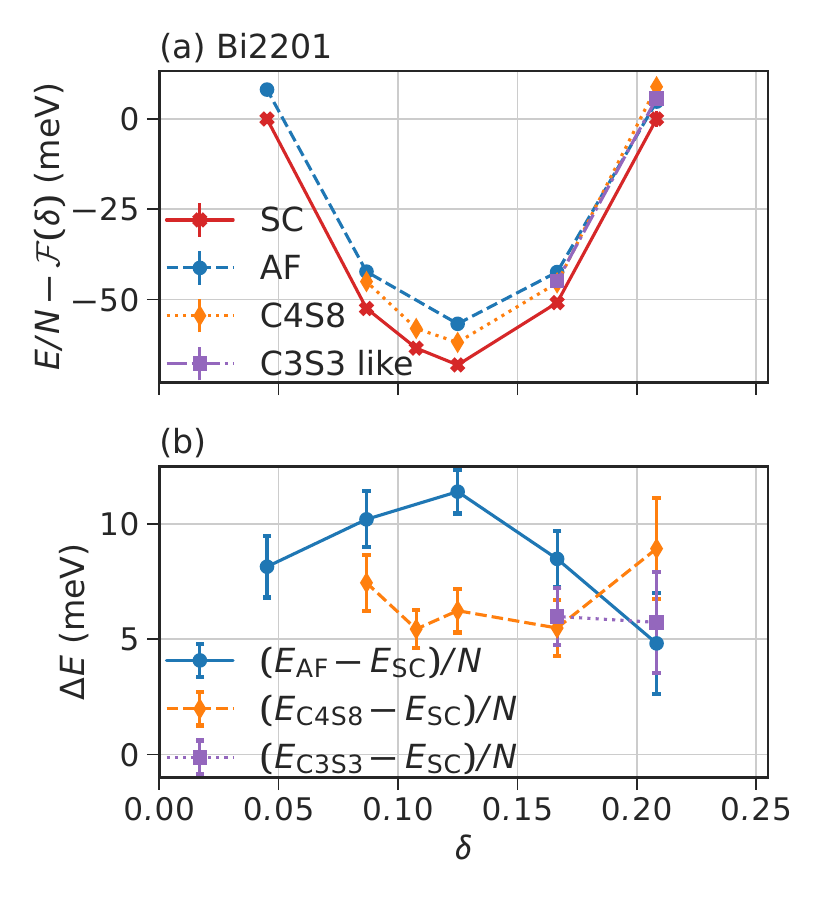} 
\caption{Variance extrapolated energies of Bi2201 for various ground-state candidates (SC, charge-spin stripe (C3S3 and C4S8), and AFM state) as a function of hole doping $\delta$ on a $L=24$ square lattice. (a) Total energies per site subtracted by $\mathcal{F}(\delta)$.
All energies are subtracted by the function $\mathcal F(\delta) =  -15.38797 \cdot \delta + 7.79753$. 
(b) Energy difference $\Delta E$ for the variance extrapolated data from (a).} 
\label{fig:EnergyOverDeltaBi2201}
\end{figure}
\section{Doping dependence of energy for Bi2201 and Bi2212}
\label{app:Bi2212-01}

The energy per site as a function of $\delta$ and the competition of SC, stripe and AFM states is shown in Fig.~\ref{fig:EnergyOverDeltaBi2212} for Bi2212 and Fig.~\ref{fig:EnergyOverDeltaBi2201} for Bi2201.

\section{Scaling of the SC order in strong-coupling region}
\label{app:strong_coupling}
\begin{figure}
    \includegraphics[width=\linewidth]{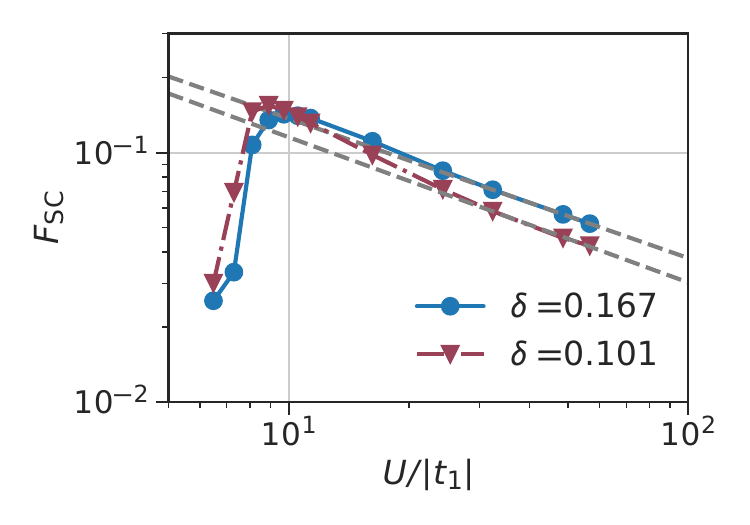} 
    \caption{{SC order parameter in strong-coupling region for $\delta=0.101$ and 0.167. The lattice size is $L=24$. The dashed lines are fitting $F_{\rm SC}\propto (|t_1|/U)^p$ with $p=0.58$ for $\delta=0.167$ and $p=0.60$ for $\delta=0.101$.}} 
    \label{fig:strong_coupling_scaling}
\end{figure}
The SC order parameter $F_{\rm SC}$ in the strong-coupling region is plotted in Fig.~\ref{fig:strong_coupling_scaling} by monitoring $U$ with other paramters of the Hamiltonian fixed at the {\it ab initio} values of CaCuO$_2$ for $L=24$ lattice.  $F_{\rm SC}$ is scaled as $F_{\rm SC}\sim (|t_1|/U)^p$ with $p \sim 0.6$ irrespective of the hole density. {As far as we know, there exists no theoretical argument in the literature including the role of the superexchange interaction $J\sim4|t_1|^2/U$ to understand this scaling.} The origin of this nontrivial power-law dependence imposes a severe constraint on the superconducting mechanism in the strong coupling region and is left for future studies.

\begin{figure*}
    \includegraphics[width=\linewidth]{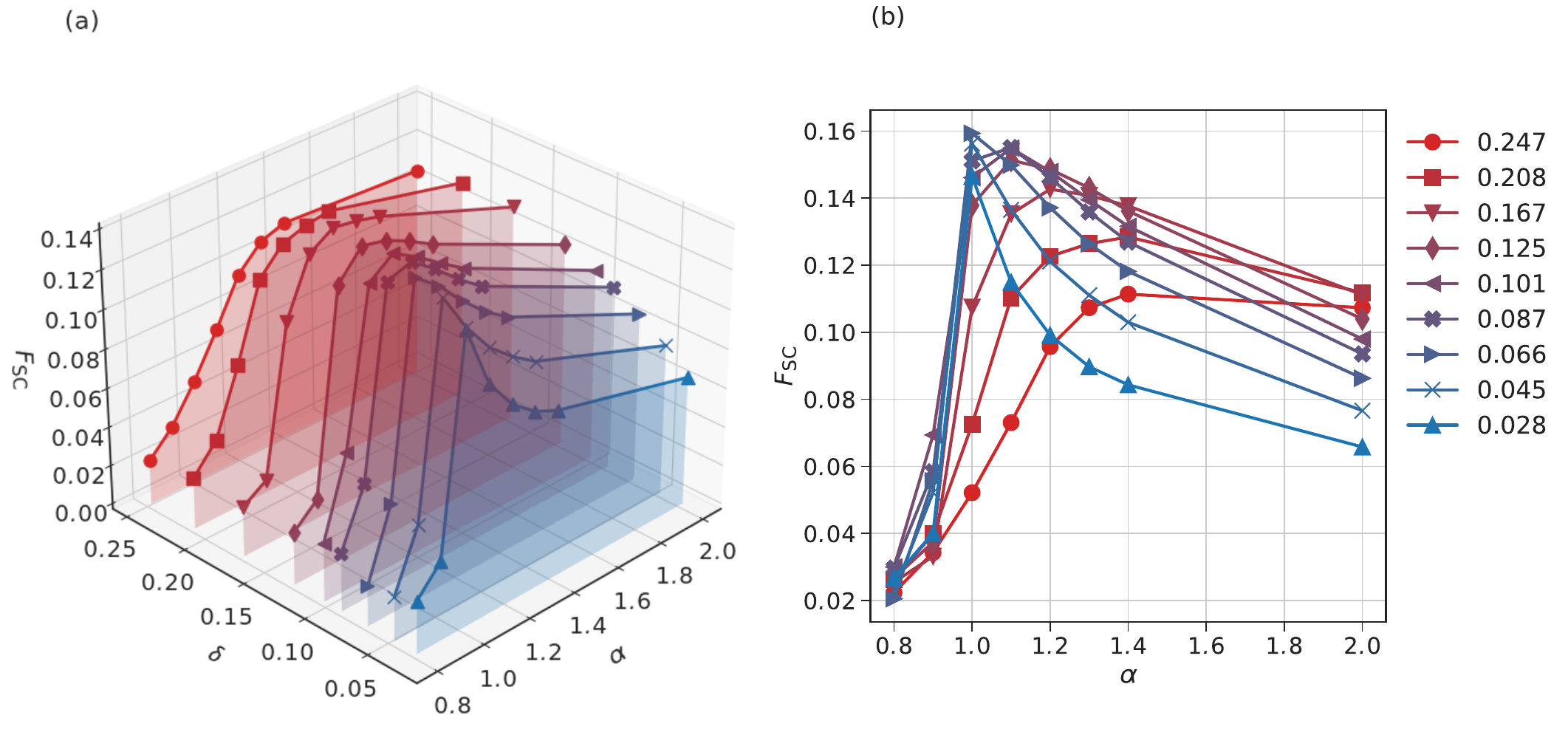} 
    \caption{{SC order parameter in the space of $\delta$ and $\alpha=(1/8.1)U/|t_1|$ modified from the {\it ab initio} CaCuO$_2$. Calculations were performed on $L=24$ lattice.}} 
    \label{fig:CaCuOFscOverDeltaAndAlpha}
\end{figure*}

{Overall $\delta$ and $\alpha$ dependences of $F_{\rm SC}$ are shown in Fig.~\ref{fig:CaCuOFscOverDeltaAndAlpha}. This shows that though the asymptotic behavior at large $\alpha$ is insensitive to $\delta$, the optimal $\delta$ and $\alpha$ depend on each other. We note here $\alpha$ corresponds to $(1/8.1)U/|t_1|$}

\section{Complexity in underdoped region and analysis on momentum distribution}
\label{app:n(k)}

\begin{figure*}
    \includegraphics[width=\linewidth]{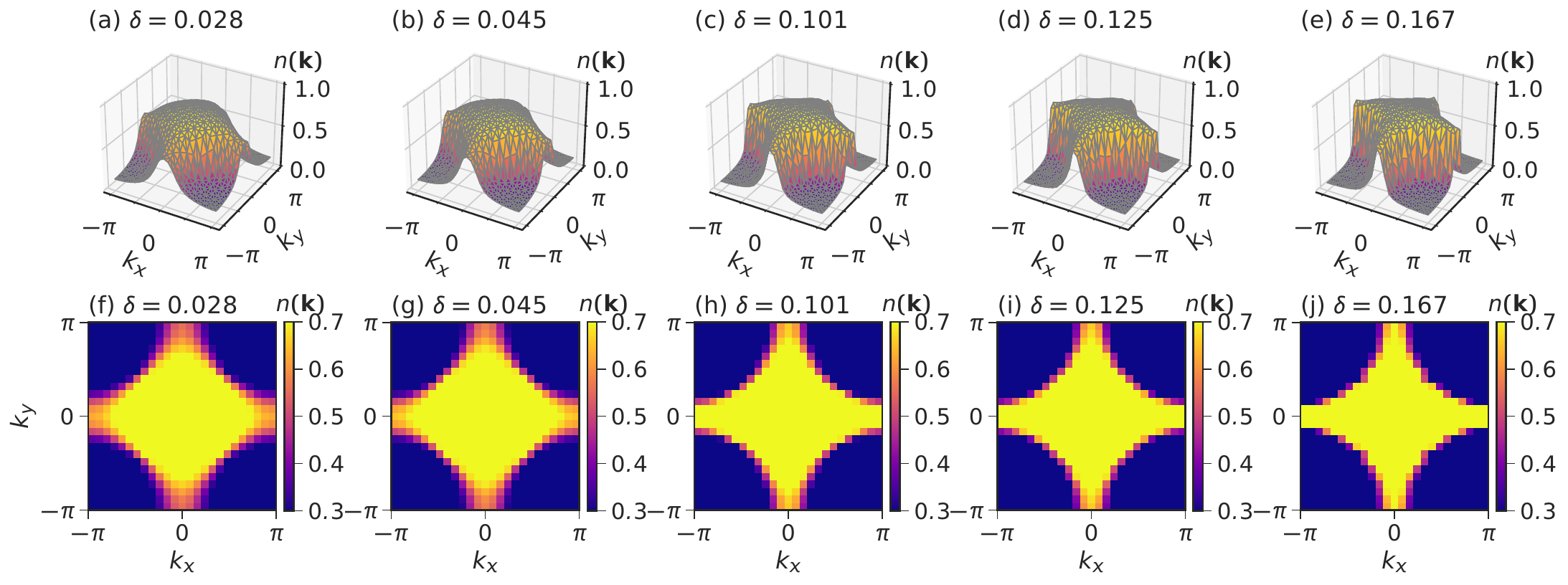} 
    \caption{Momentum distribution $n(\bm k)$ in the Brillouin zone for doped CaCuO$_2$ on a $L=24$ square lattice in the SC state. The doping concentration is $\delta=0.028$ for (a),(f), 0.045 for (b),(g), 0.101 for (c),(h), 0.125 for (d),(i) and 0.167 for (e),(j). 
    The lower panels ((f),(g) (h), (i) and (j)) are contour plots. } 
    \label{fig:CaCuONk0028To0167}
\end{figure*} 

We show calculated momentum distribution defined by 
\begin{equation}
n(\bm k)=\frac{1}{2N}\sum_{i,j,\sigma}\mathrm{e}^{\mathrm i \bm{k}(\bm r_i-\bm r_j)}\langle c_{i \sigma}^{\dagger}c_{j \sigma}\rangle,
\end{equation}
because the jump or singularity of $n(\bm k)$ at the Fermi level characteristic of metals measures the incoherence of the carrier. 
Here, we show $n(\bm k)$ for the case of doped CaCuO$_2$ in Fig.~\ref{fig:CaCuONk0028To0167} for the cases of $\delta = 0.028$ ((a),(f)), $0.045$~((b),(g)), $0.101$ ((c),(h)), $0.125$ ((d),(i)), and $\delta=0.167$ ((e),(j)) in the SC state on the $L=24$ square lattice.  
Although $F_{\rm SC}^{\infty}$ is similar between $\delta=0.028$ (or $0.045$) and $\delta=0.101$ (or $0.125$), $n(\bm k)$ shows substantially more smooth and rounded shape for $\delta=0.028$ and $0.04$ than $\delta=0.101$ and $0.125$, demonstrating that the effect of the larger normal damping at smaller $\delta$ is responsible for this rounded behavior.
The substantial increase in the damping with decreasing doping may be responsible for the suppression of $T_{c}$ in the underdoped region irrespective of the high $F_{\rm SC}^{\infty}$. More quantitative studies will be presented elsewhere. 
In SM~\cite{SM_Michael} Fig.~S9, we show the case of $\delta=0.028$ and $\delta=0.167$ on the $L=36$ square lattice to ensure that the size dependence is weak.

{The strong damping is most prominent in the underdoped region around the antinodal points. This region is under the strong influence of the pseudogap, which makes the relation of the physical quantities nontrivial. The different behavior of $T_c$ and $F_{\rm SC}$ is such an example. Deviation of the SC carrier density $n_s$ and the weight of quasiparticle coherence peak at the antinodal point from $F_{\rm SC}$ and the SC gap $\Delta_{\rm SC}$ against the naive expectation may be another example. Although  $F_{\rm SC}$ and the SC gap $\Delta_{\rm SC}$ grow on top of the pseudogap as we revealed in the case of $F_{\rm SC}$, $n_s$ seems to be severely suppressed by the pseudogap around the antinodal point. This trend is indeed seen in the comparison of the muon penetration depth and the SC gap measurement~\cite{Alldredge2008,Panagopoulos1999,Kondo2011,LeTacon2006}, which causes difficulty in the comparison of our calculated result and experimental indications in the underdoped region. Such a complexity is expected to be small at the optimum doping region, while the prominent materials dependence is seen most prominently at the optimum doping. This is the reason why we focus on the materials dependence at the optimum doping. }

\section{Effect of apical oxygen position on Hamiltonian parameters}
\label{app:ApicalO_effect}
{We extend analysis on the effect of apical oxygen position examined in Ref.~\cite{moree2022}. 
Studies on effects of the apical oxygen position represented by the distance $d_{\rm Oap}^z$ to Cu on effective Hamiltonian parameters are few~\cite{pavarini2001,mori2008,sakakibara2010}. 
In Table~\ref{tab:ApicalOham} we show the Hamiltonian parameters of Bi2201 when the apical oxygen is artificially shifted. 
\begin{table}[tbh!]
    \caption{\textit{Ab initio} single-band effective Hamiltonian for Bi2201 when the apical oxygen is shifted. The energy unit is eV.}
    \label{tab:ApicalOham} 
    \begin{ruledtabular}
         \begin{tabular}{ l c c c c c }
            $d_{\rm Oap}^z$ (\AA) & $t_1$ &  $t_2$  & $U$ & $V_1$  & $V_2$ \\
            \colrule
            2.58 & -0.527 & 0.140 & 4.393 & 1.030 & 0.602  \\
            2.53 & -0.513 & 0.159 & 3.994 & 0.837 & 0.447 \\
            \colrule
            & $U/|t_1|$ & $|t_2/t_1|$  & $V_1/|t_1|$  &   &   \\
            \colrule              
            2.58  & 8.336 & 0.266 & 1.954 & & \\
            2.53  & 7.789 & 0.310 & 1.632 & & \\
        \end{tabular}
    \end{ruledtabular}
\end{table}
}

{As is already addressed in Ref.~\cite{moree2022}, $U$ decreases with decreasing $d_{\rm Oap}^z$ because of increased screening from electrons at the apical oxygen, which is consistent with the claim in Ref.~\cite{mori2008}.
In Ref.~\cite{pavarini2001}, $|t_2/t_1|$ increases with 
increasing $d_{\rm Oap}^{z}$, for instance, in the comparison of Hg1201 and Bi2201. However, it shows opposite trend in Table~\ref{tab:ApicalOham} and in the results in Ref.~\cite{moree2022}.
References~\cite{pavarini2001,mori2008} focused on the effect of orbitals whose energies are above the Fermi level such as Cu 4$s$ and apical O 2$p_z$ orbitals.  The increasing $d_{\rm Oap}^z$ makes those levels lower because farther distance to the negative CuO$_2$ layer charge for the apical O 2$p_z$ orbital and farther distance to the negatively charged apical oxygen for the Cu$4s$ orbital makes the Madelung potential lower. This lowering induces a larger hybridization with the AB orbital located around the Fermi level constructed from Cu 3$d_{x^2-y^2}$ and in-plane O 2$p_{\sigma}$ orbital (our target band), which we call in-plane CuO$_2$ AB orbital. This increased hybridization especially enhances $t_2$. However, this is not the whole story. The effective next-nearest-neighbor hopping $t_2$ in the single-band Hamiltonian is also altered by the effect from the bands below the Fermi level such as $d_z$ orbital. Increasing $d_{\rm Oap}^z$ makes the lowering of the $d_z$ level and hence causes the decrease in the hybridization with in-plane CuO$_2$ AB orbital which cancels the effect of the increased hybridization of the orbital above the Fermi level as was pointed out in Ref.~\cite{sakakibara2010}.  More precisely, the apical O 2$p_z$ and Cu 3$d_z$  orbitals are strongly hybridized and they form bonding and antibonding bands and we need to consider all of these contributions, which are taken into account quantitatively in our calculations in the derivation of the effective Hamiltonian. As a consequence, the present estimate for $t_2/t_1$ has large difference from that in the estimates of Ref.~\cite{pavarini2001} in some cases. For instance, in the case of Hg1201, $|t_2/t_1|\sim~0.36$ for Hg1201 in Ref.~\cite{pavarini2001}, while $\sim 0.20$ in the present study. Although Ref.~\cite{pavarini2001} is based on complex  approximations to estimate $t_2/t_1$ only by taking into account the contribution from the band above the Fermi level, recent standard way employs the maximally localized Wannier orbitals and its Hamiltonian matrix elements for the estimate of hopping, by considering all the bands contribution near the Fermi level. This is much simpler, more straightforward and transparent for the estimate of the lattice fermion Hamiltonian parameters, which are used in Ref.~\cite{moree2022} as the basis of our VMC calculations. In fact, recent estimates of $|t_2/t_1|$ for Hg1201 are 0.20, and 0.23 in Refs.~\cite{Teranishi2018,Jang2019} (in the revised manuscript), respectively, which are consistent with the present 0.20. }

{As we already mentioned in Appendix~\ref{app:t2_dependence}, larger $|t_2/t_1|$ slightly but quantitatively suppresses $F_{\rm SC}$ in the realistic parameter range, which is opposite to the prediction in Ref.~\cite{pavarini2001} but is consistent with Ref.~\cite{sakakibara2010}. Furthermore, and most importantly, too small dependence of optimal $F_{\rm SC}$ on $|t_2/t_1|$ clarified in this paper makes the role of  $|t_2/t_1|$ on $F_{\rm SC}$ highly questionable. We find that the effect of the apical oxygen position on the superconductivity is primarily to control $U$.}

\section{Comparison to approach using multiband Hamiltonian}
\label{app:multi-band}

There exists recent work based on the atomic orbitals containing Cu 3$d_{x^2-y^2}$ and O 2$p_{\sigma}$ orbitals~\cite{Kowalski2021,Cui2022,Cui2023}.
Of course, multiorbital Hamiltonians should give essentially a similar answer if the derivation and the solving procedure are appropriate. 
On the other hand, the Hamiltonian becomes more complex with larger number of paramters, as it should be.

However, the AB band and NB or B band are well separated with the hybridization gap (band center separation is $\sim 9$ eV and the direct gap is 5-6eV for the cases we studied in this paper). See Appendix D of Ref.~\cite{moree2022} for detailed analyses. In this circumstance, we can safely start from the picture of single-band Hamiltonian derived from the AB band of strongly hybridized Cu 3$d_{x^2-y^2}$ and O2$p_{\sigma}$ orbitals only, because the B orbitals are more or less completely filled and inactive. See Appendix D of Ref.~\cite{moree2022}.  See also Fig.10(b) of Ref.~\cite{hirayama2019}, where the completely filled B bands are verified for the Hg-based cuprate through all the relevant hole densities and this is universal in the curate superconductors. In the single AB band description, the B degrees of freedom are downfolded and give the renormalization to the AB orbital description. Since the AB-B hybridization gap is large, the perturbative downfolding procedure to renormalize and eliminate B and NB degrees of freedom works well as a good approximation~\cite{moree2022}. This is based on the multiscale {\it ab initio} scheme for correlated electrons (MACE) with refined $GW$ approximation supplemented by the level renormalization feed back~\cite{hirayama2019}. Except for the AB band, all the bands are either well below or above the Fermi level so that they can be perturbatively taken by the partial trace summation to give renormalizations to the AB degrees of freedom.

Of course, one can start from the three-band Hamiltonian where the charge transfer gap and covalency are relevant parameters. However, it should end up with this AB/B description after the basis transformation if one focuses on the low-energy physics in the realistic situation of the cuprates. The effect of the parameters of the charge transfer energy and the $d$-$p$ covalency were taken into account in our downfolding procedure from the three-band to a single-band AB Hamiltonian in Ref.~\cite{moree2022}. For instance, larger charge transfer gap results in poorer screening and larger correlation ($U$) as is confirmed in Ref.~\cite{moree2022} [see the comparison of Table IV with Tables I and II in Ref.~\cite{moree2022}]. Therefore, the three-band parameters are encoded in $U/|t_1|$ and other parematers in the AB single-band description indirectly and systematically in a complex manner.

There exist several recent analyses based on multiband Hamiltonian containing Cu 3$d_{x^2-y^2}$ and O2$p_{\sigma}$ atomic orbitals for the cuprates or Hubbard-type models~\cite{Kowalski2021,Cui2022,Cui2023}. Three-band Hamiltonian constructed from the Cu 3$d_{x^2-y^2}$ and O2$p_{\sigma}$ atomic orbitals derived and listed in Table IV of Ref.~\cite{moree2022} shows rough consistency with the proposal by Ref.~\cite{Kowalski2021}, in which the authors claim stronger superconducting order for smaller charge transfer gap $\Delta E_{xp}$ and larger $d$-$p$ transfer $t_{xp}$ in the notation of Ref.~\cite{moree2022}. Naively, one would expect that smaller $\Delta E_{xp}$ makes stronger screening on the AB band and decreases effective $U$ in the single-band picture, while larger $t_{xp}$ directly leads to larger $t_1$. 
Both result in smaller $U/|t_1|$, which appears to contradict the statement claimed in the present paper that larger $U/|t_1|$ leads to larger superconducting order parameter in the realistic parameter region.  However, one needs to be careful about the parameter region employed in Ref.~\cite{Kowalski2021}. When one sees $U_d$ (the direct on-site repulsion  between atomic $d$ orbital) dependence of the superconducting order parameter in Fig.2 of Ref.~\cite{Kowalski2021}, one clearly finds that the superconducting order parameter decreases from $U_d=10$ to 18 or 14 in the energy scale of $t_{pp}$, transfer between neighboring O 2$p$ orbital. When one compares the parameters with those in Table IV of  Ref.~\cite{moree2022}, and compares with Tables I and II of Ref.~\cite{moree2022}, one notices that $U_d=10$ in Ref.~\cite{Kowalski2021} already corresponds to the region around the optimum of $U/|t_1|$ in the present single-band description and further increase of $U_d$ drives the system into the strong coupling region with larger $U$, where the superconducting order decreases with increasing $U/|t_1|$ as one can see in our result in Fig.9(a). This perfectly explains the $U_d$ dependence between $U_d=10$ and 18 in Ref.\cite{Kowalski2021} as well as the $t_{pd}= t_{xp}$ and $p$ level, $\epsilon_p$ dependences, because larger $t_{pd}= t_{xp}$  and smaller $\epsilon_p=\Delta E_{xp}$ both make smaller $U/|t_1|$ as was mentioned above and leads to larger superconducting order parameter in the strong-coupling region as is clarified in Fig. 9(a). It clarifies that the region studied in Ref.~\cite{Kowalski2021}  is outside of the real materials dependence of the cuprates studied here. 
We further need more detailed studies of the correspondence between multiorbital and the present single-band description in real materials systematically, which is left for future studies.  

\newpage

\bibliographystyle{apsrev4-2}

\clearpage

\noindent
{\Large \bf Supplementary Material} \\
\renewcommand{\theequation}{S\arabic{equation}}
\setcounter{equation}{0}
\renewcommand{\thesection}{S\arabic{section}}
\setcounter{section}{0}



\noindent
{\bf S1.  Full \textit{ab initio} Hamiltonians} \\
\label{sec:SMHamFull}

In Table~\ref{tab:hamFull} we present the full single band \textit{ab initio} Hamiltonian for all four compounds as derived in Ref.~\onlinecite{moree2022}. 

\begin{table*}[tbh!]
    \caption{\textit{Ab initio} effective Hamiltonian for CaCuO$_2$, Hg1201, Bi2212, and Bi2201. Hoppings and screened Coulomb interactions derived for the single-band effective Hamiltonians are taken from \cite{moree2022}. The nth neighbor hopping amplitude and Coulomb interaction are denoted as $t_n$ and $V_n$. Interlayer hoppings and interactions are neglected here. All values are given in eV.}
    \label{tab:hamFull} 
    \begin{ruledtabular}
        \begin{tabular}{ l c c c c c c c c c c }
               & $U/\lvert t_1 \rvert$  & $t_1$ &  $t_2$ & $t_3$ & $t_4$  & $t_5$ & $t_6$ & $t_7$ & $t_8$ & $t_9$  \\
                \colrule
                CaCuO$_2$ & 8.10 &-0.5214 & 0.1322 & -0.0467 & 0.0078 & -0.0002 & -0.0139 & 0.0041 & 0.0045 & -0.0009  \\
                Hg1201    & 7.35 &-0.5441 & 0.1112 & -0.0434 & 0.0096 & -0.0004 & -0.0044 & 0.0081 & -0.0028 & -0.0011  \\
                Bi2212    & 9.36 &-0.4516 & 0.1345 & -0.0528 & -0.0014 & 0.0071 & 0.0008 & -0.0015 & 0.0001 & 0.0002  \\
                Bi2201    & 8.34 &-0.5266 & 0.1402 & -0.0424 & 0.0087 & -0.0066 & -0.0017 & 0.0045 & -0.0024 & -0.0023  \\
                \colrule
                & $U$ & $V_1$  & $V_2$  & $V_3$  & $V_4$ & $V_5$ & $V_6$ & $V_7$ & $V_8$ & $V_9$   \\
            \colrule
            CaCuO$_2$  & 4.221 & 0.969 & 0.539 & 0.380 & 0.316 & 0.241 & 0.139 & 0.127 & 0.106 & 0.048\\
            Hg1201  & 3.999 & 1.002 & 0.596 & 0.448 & 0.389    & 0.320 & 0.174 & 0.165 & 0.147 & 0.069\\
            Bi2212  & 4.226 & 0.915 & 0.518 & 0.366 & 0.312   & 0.253 & 0.138 & 0.129 & 0.115 & 0.055\\
              Bi2201  & 4.393 & 1.030 & 0.602 & 0.450 & 0.395   & 0.334 & 0.178 & 0.170 & 0.156 & 0.075\\
        \end{tabular}
    \end{ruledtabular}
\end{table*}

\vskip 5mm
\noindent
{\bf S2.  Correlation functions of the $CmSn$ states in $CaCuO_2$} \\

Here we discuss the real space spin- and charge-configuration of the CmSn stripe states.
For this purpose, we calculate the following correlation functions
\begin{align}
    S_z(\bm r) &= ( n_{\bm r, \uparrow} - n_{\bm r, \downarrow})/2, \label{eq:Szr}\\
    n(\bm r)   &= (n_{\bm r, \uparrow} + n_{\bm r, \downarrow})/2, \label{eq:nr}
\end{align}
which are single-particle quantities.
Here the vector  $\bm r$ represents the two-dimensional lattice site coordinate $\bm r = (r_x, r_y)$, $S_z(\bm r)$ the spin configuration of each lattice site, and $n(\bm r)$ the corresponding charge configuration. 
Further we consider $S_s(\bm q)$, $S_c(\bm q)$, and $n(\bm k)$, as defined in equations~(2),~(3), and~(A1) in the main text.
Although the translational symmetry breaking does not occur in a finite-size systems in the true ground state, practically, ordered states are realized in a sufficiently large systems in the VMC results if the order is expected in the thermodynamic limit.

\vskip 5mm
\noindent
{\bf 1. C4S8} \\

\begin{figure}[tbh!]
    \includegraphics[width=\linewidth]{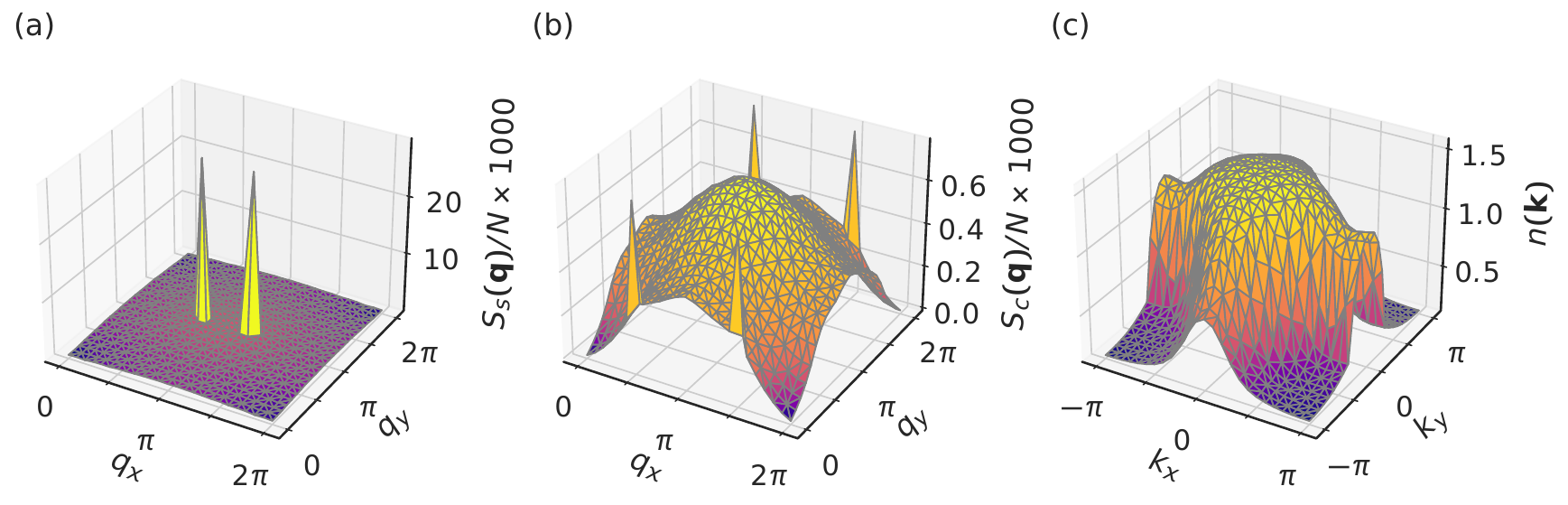} 
    \caption{Spin (panel (a)) and charge (panel (b)) structure factor ($S_s(\bm q)$, $S_c(\bm q)$) of the C4S8 state at $\delta = 0.125$ for CaCuO$_2$ on a $L = 24$ square lattice. (c) Momentum distribution $n(\bm k)$.} 
    \label{fig:CaCuO_C4S8_504_SSQSCQ}
\end{figure}

Starting with the C4S8 state, the spin- and charge structure factor is shown in Fig.~\ref{fig:CaCuO_C4S8_504_SSQSCQ}. 
To confirm the charge order of period four, we expect to see peaks in the charge structure factor. 
In panel~(b), indeed, there are peaks at $\bm q_{\mathrm{CDW}} = ( \pi/2, 0)$ and $( 3\pi/2, 0)$, confirming the charge order. 
Similarly we find peaks in the spin structure factor (see panel~(a)) at $\bm q_{\mathrm{SDW}} = ( 3\pi/4, \pi)$ and $( 5\pi/4, \pi)$ corresponding to a spin order of period eight.
In~(c) the corresponding momentum distribution is also shown. 

\begin{figure}[tbh!]
    \includegraphics[width=0.75\linewidth]{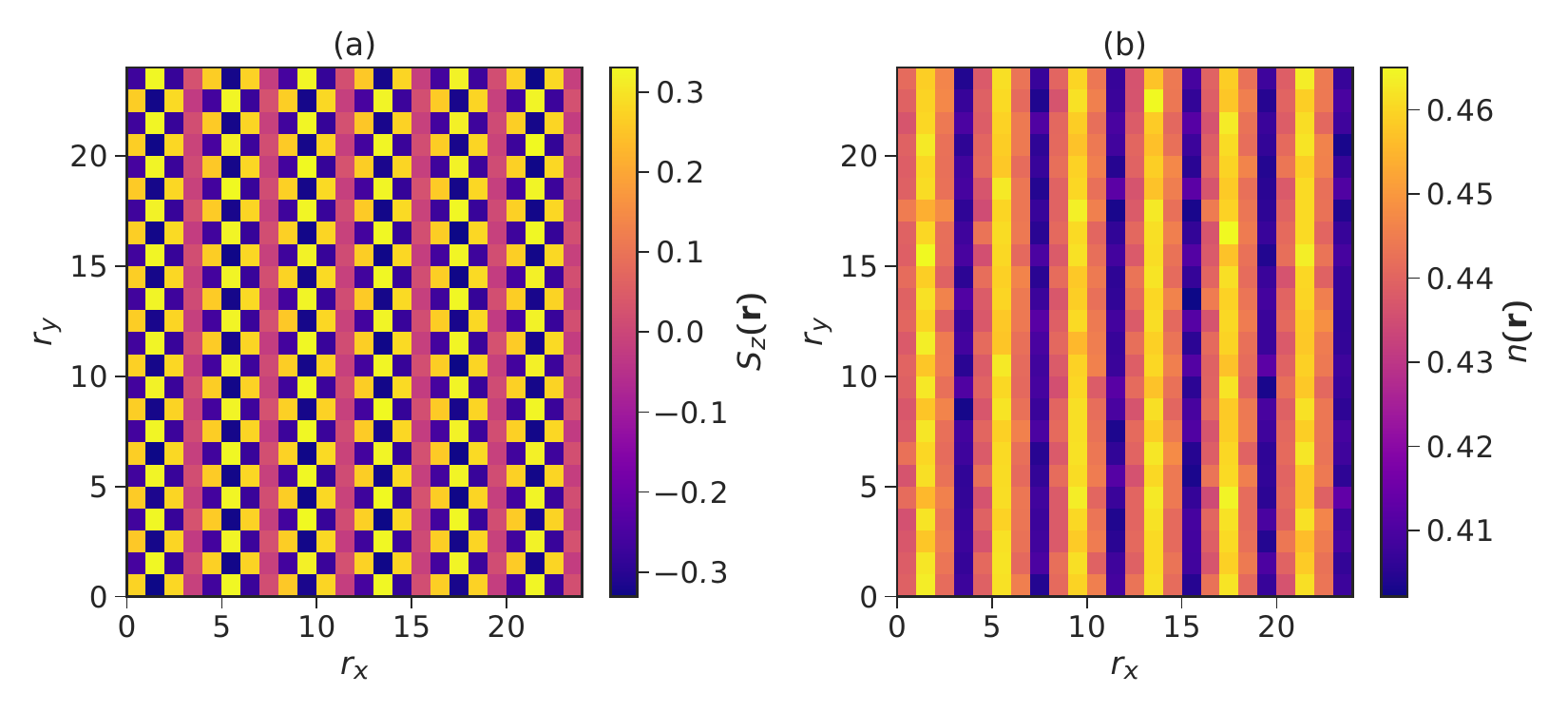} 
    \caption{Spin (panel (a)) and charge (panel (b)) configuration  ($S_z(\bm r)$, $n(\bm r)$) of the C4S8 state at $\delta = 0.125$ for CaCuO$_2$ on a $L = 24$ square lattice.} 
    \label{fig:CaCuO_C4S8_504_SzrNr}
\end{figure}

The C4S8 stripe sate can also be identified in the real space resolved quantities $S_z(\bm r)$ and $n(\bm r)$, when no additional momentum projections in mVMC are done. 
In Fig.~\ref{fig:CaCuO_C4S8_504_SzrNr} $S_z(\bm r)$~(a) and $n(\bm r)$~(b) are shown for the same case as a heatmap. 
The charge and spin stripes run along the $r_y$ direction, while periodically alternating in the $r_x$ direction with period four (charge) or eight (spin).

\vskip 5mm
\noindent
{\bf 2. C3S3} \\

\begin{figure}[tbh!]
    \includegraphics[width=\linewidth]{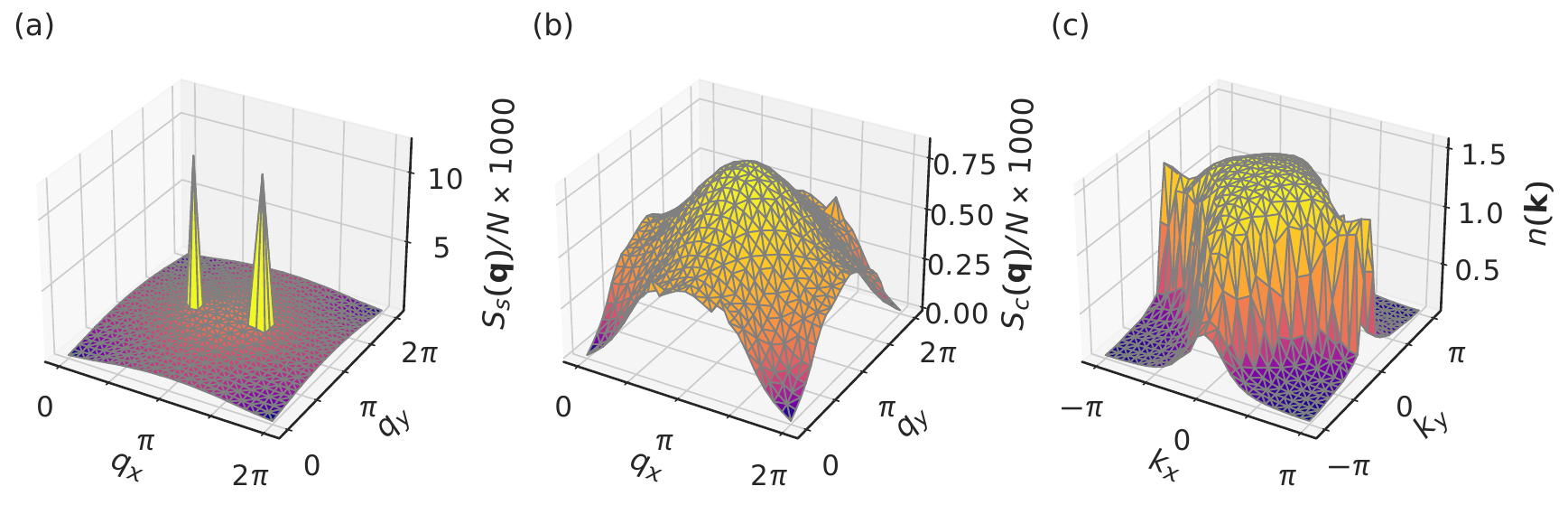} 
    \caption{Spin (panel (a)) and charge (panel (b)) structure factor ($S_s(\bm q)$, $S_c(\bm q)$) of the C3S3 state at $\delta = 0.207$ for CaCuO$_2$ on a $L = 24$ square lattice.} 
    \label{fig:CaCuO_C3S3_456_SSQ_SCQ}
\end{figure}

\begin{figure}[tbh!]
    \includegraphics[width=0.75\linewidth]{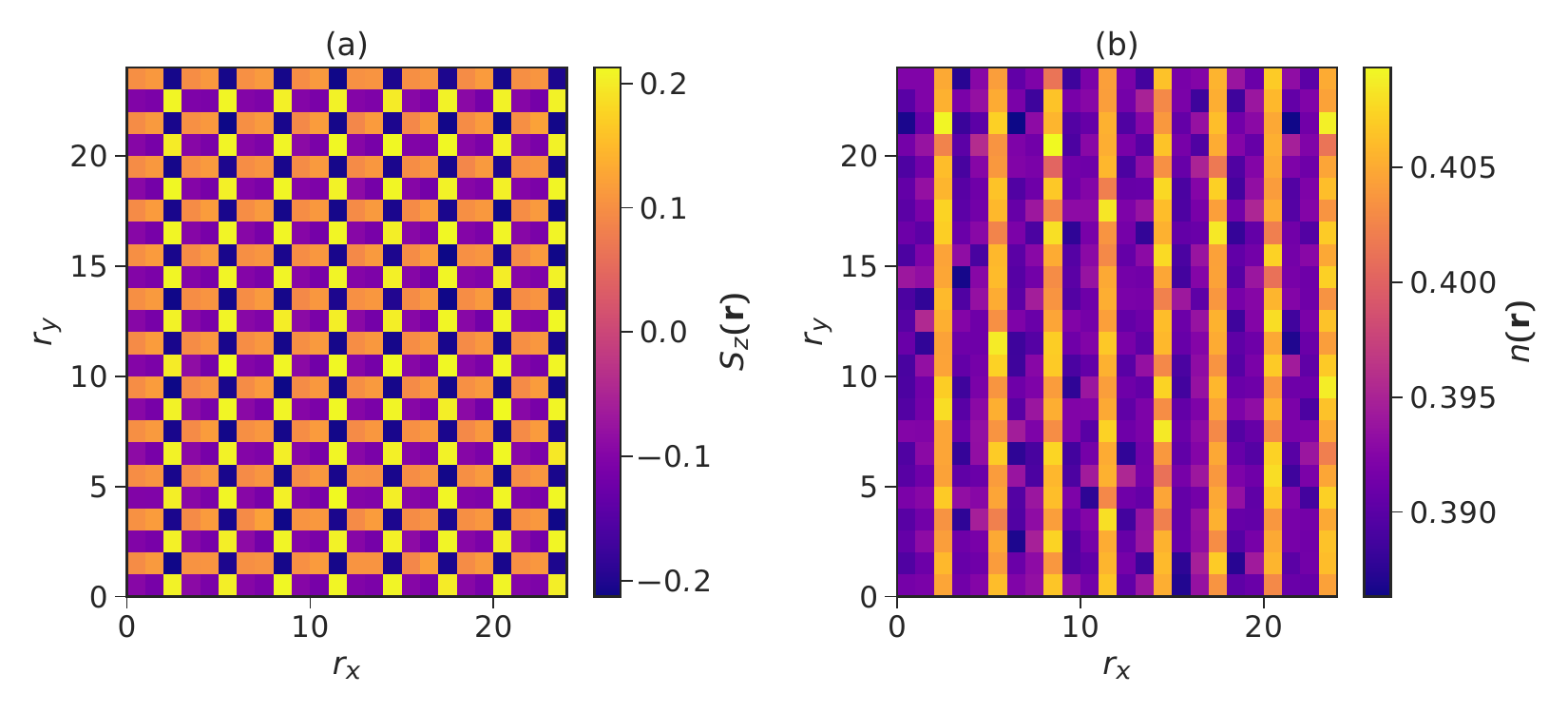} 
    \caption{Spin (panel (a)) and charge (panel (b)) configuration  ($S_z(\bm r)$, $n(\bm r)$) of the C3S3 state at $\delta = 0.207$ for CaCuO$_2$ on a $L = 24$ square lattice. (c) Momentum distribution $n(\bm k)$.} 
    \label{fig:CaCuO_C3S2_456_SzrNr}
\end{figure}

Using the same analysis as above the C3S3 state with charge period three and spin period three can be identified. The spin- and charge structure factors, depicted in Fig.~\ref{fig:CaCuO_C3S3_456_SSQ_SCQ} (a) or (b), show peaks at  $\bm q_{\mathrm{CDW}} = ( 2 \pi/3, 0)$ and $( 4 \pi/3, 0)$ for the charge structure factor, and $\bm q_{\mathrm{SDW}} = ( 2\pi/3, \pi)$ and $( 4\pi/3, \pi)$ for the spin structure factor.
In~(c) the corresponding momentum distribution is also shown. 
Fig.~\ref{fig:CaCuO_C3S2_456_SzrNr}~(a) or~(b) shows the real space configuration of spin and charge with the period of three in $r_x$ direction. 

Compared to the C4S8 state, the C3S3 state realized here has significantly smaller peaks at $\bm q_{\mathrm{CDW}}$ in $S_c(\bm q)$. 
As shown in Appendix~C of the main text, this may already indicate, that in the thermodynamic limit, this stripe state will fall into a paramagnetic state.

\vskip 5mm
\noindent
{\bf S3.  AF state for CaCuO$_2$ away from half filling} \\
\begin{figure}[tbh]
    \includegraphics[width=\linewidth]{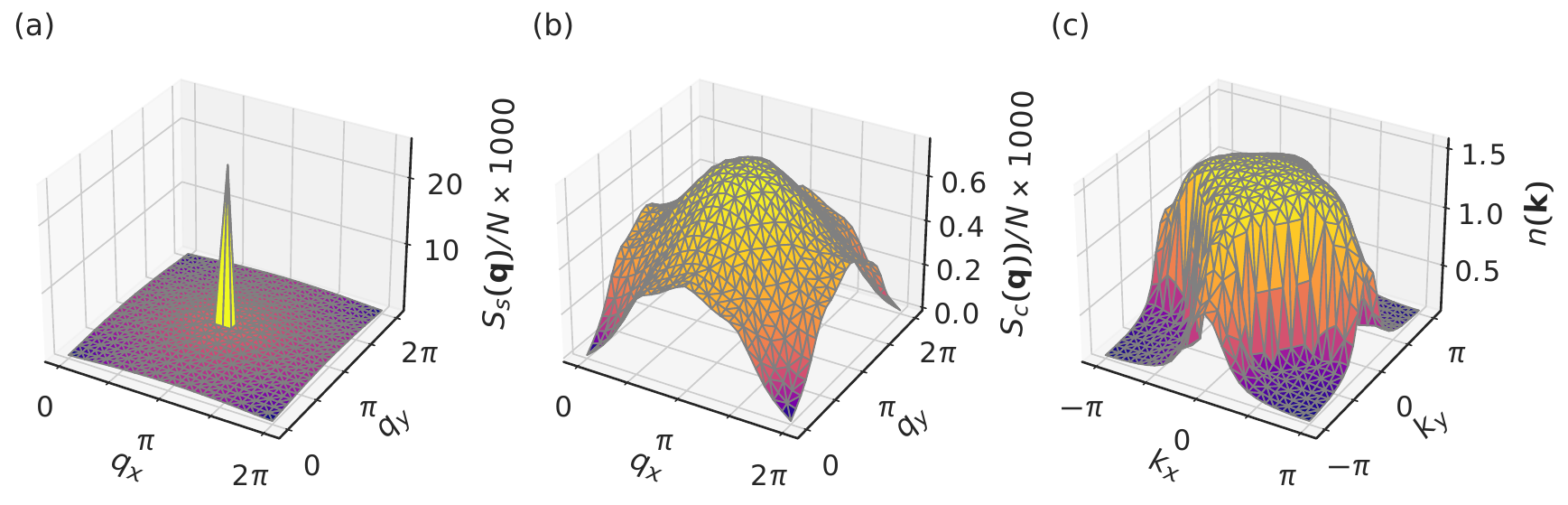}
    \caption{AF state at $\delta = 0.167$ hole doping on a $L = 24$ square lattice for CaCuO$_2$.  
    (a) Spin structure factor $10^3 \cdot S_s(\bm q)/N$. 
    (b) Charge structure factor $10^3 \cdot S_c(\bm q)/N$.
    (c) Momentum distribution $n(\bm k)$.}
    \label{fig:CaCuO_AF_480_SSQ_SCQ_Nk}
\end{figure}

For the AF state at $\delta = 0.167$ hole doping on a $L = 24$ square lattice for CaCuO$_2$, the spin- and charge structure factor, and momentum distribution is shown in Fig.~\ref{fig:CaCuO_AF_480_SSQ_SCQ_Nk}. 
In~(a) the strong peak at $\bm q = (\pi, \pi)$ in $S_s(\bm q)$ indicates a strong antiferromagnetic correlations, corresponding to the well known checkerboard spin configuration pattern. 
For the charge structure factor on the other hand (shown in~(b)), there are no sharp peaks.
Over the full range, the function is smooth, resulting in a charge homogeneous state.
The momentum distribution $n(\bm k)$ is  shown in~(c), suggesting non Fermi liquid behaviour.

\vskip 5mm
\noindent
{\bf S4. Uncertainty of the apical oxygen position in Bi2212 and Bi2201} \\

\begin{figure}[tbh]
    \includegraphics[width=\linewidth]{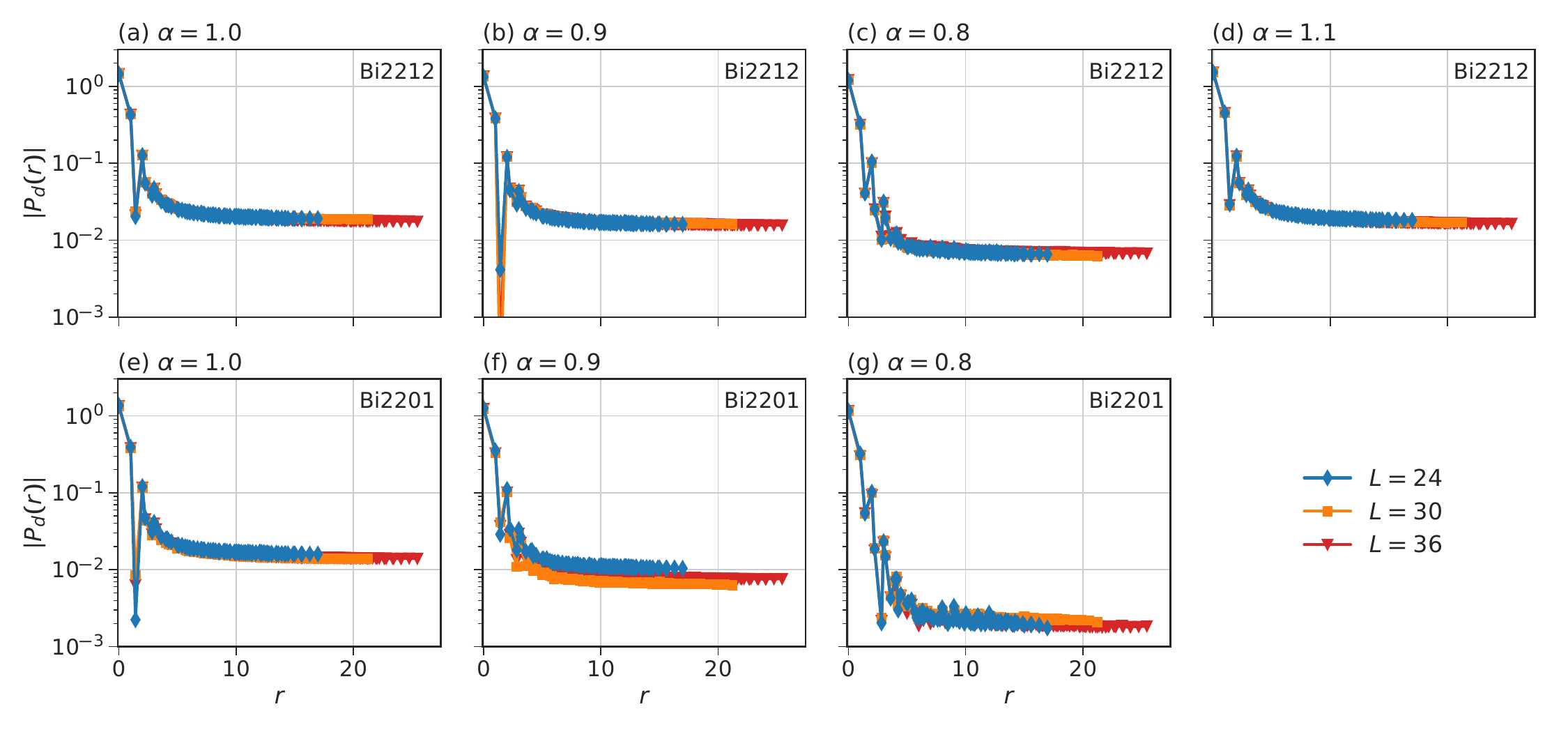}
    \caption{Size dependence of $P_d(r)$ under $\alpha$ scaling ($\mathcal{H}(\alpha,\alpha)$) in Bi2212 (upper row) and Bi2201 (bottom row) of the SC state at $\delta = 0.167$ hole doping.  
    Bi2212: (a)-(d) for $\alpha = 1.0, 0.9, 0.8, 1.1$. 
    Bi2201: (e)-(g) for $\alpha = 1.0, 0.9, 0.8$.}
    \label{fig:PddBiCompundAlphaScalingSC0167}
\end{figure}

SC correlations $P_d(r)$ as a function of distance $r$ for several choices of the scaling factor $\alpha=\xi$ with the Hamiltonian  $\mathcal{H}(\alpha,\alpha)$ are presented in Fig.~\ref{fig:PddBiCompundAlphaScalingSC0167} for Bi2212 and Bi2201.

\vskip 5mm
\noindent
{\bf S5.  Momentum distributions of CaCuO$_2$} \\

\begin{figure}[tb]
    \includegraphics[width=0.75\linewidth]{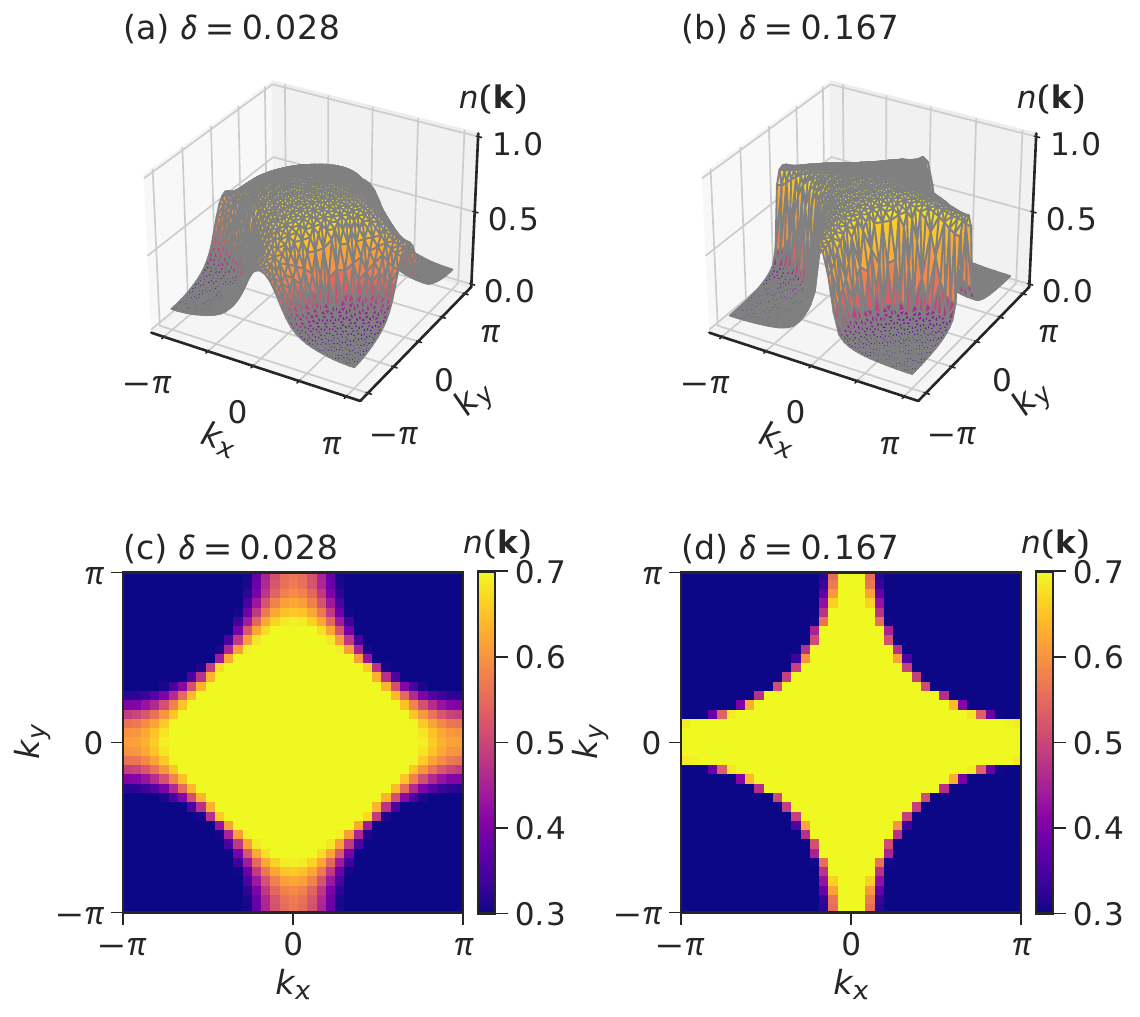} 
    \caption{Momentum distribution $n(\bm k)$ for CaCuO$_2$ in the Brillouin zone at: (a) $\delta = 0.028$ and (b) $\delta = 0.167$ on a $L = 36$ square lattice. 
    (c) and (d) represent  corresponding contour plots. } 
    \label{fig:CaCuONk0045And0167}
\end{figure}

We show $n(k)$ for the case of doped CaCuO$_2$ at $L=36$ lattice in Fig.~\ref{fig:CaCuONk0045And0167} for $\delta = 0.028$~(a),(c) and $\delta = 0.167$~(b),(d) for the SC state.


\begin{thebibliography}{10}%
\makeatletter
\providecommand \@ifxundefined [1]{%
 \ifx #1\undefined \expandafter \@firstoftwo
 \else \expandafter \@secondoftwo
\fi
}%
\providecommand \@ifnum [1]{%
 \ifnum #1\expandafter \@firstoftwo
 \else \expandafter \@secondoftwo
\fi
}%
\providecommand \enquote [1]{``#1''}%
\providecommand \bibnamefont  [1]{#1}%
\providecommand \bibfnamefont [1]{#1}%
\providecommand \citenamefont [1]{#1}%
\providecommand\href[0]{\@sanitize\@href}%
\providecommand\@href[1]{\endgroup\@@startlink{#1}\endgroup\@@href}%
\providecommand\@@href[1]{#1\@@endlink}%
\providecommand \@sanitize [0]{\begingroup\catcode`\&12\catcode`\#12\relax}%
\@ifxundefined \pdfoutput {\@firstoftwo}{%
 \@ifnum{\z@=\pdfoutput}{\@firstoftwo}{\@secondoftwo}%
}{%
 \providecommand\@@startlink[1]{\leavevmode}%
 \providecommand\@@endlink[0]{}%
}{%
 \providecommand\@@startlink[1]{%
  \leavevmode
  \pdfstartlink
   attr{/Border[0 0 1 ]/H/I/C[0 1 1]}%
   user{/Subtype/Link/A<</Type/Action/S/URI/URI(#1)>>}%
  \relax
 }%
 \providecommand\@@endlink[0]{\pdfendlink}%
}%
\providecommand \url  [0]{\begingroup\@sanitize \@url }%
\providecommand \@url [1]{\endgroup\@href {#1}{\urlprefix}}%
\providecommand \urlprefix [0]{URL }%
\providecommand \Eprint[0]{\href }%
\@ifxundefined \urlstyle {%
  \providecommand \doi [1]{doi:\discretionary{}{}{}#1}%
}{%
  \providecommand \doi [0]{doi:\discretionary{}{}{}\begingroup
  \urlstyle{rm}\Url }%
}%
\providecommand \doibase [0]{http://dx.doi.org/}%
\providecommand \Doi[1]{\href{\doibase#1}}%
\providecommand \bibAnnote [3]{%
  \BibitemShut{#1}%
  \begin{quotation}\noindent
    \textsc{Key:}\ #2\\\textsc{Annotation:}\ #3%
  \end{quotation}%
}%
\providecommand \bibAnnoteFile [2]{%
  \IfFileExists{#2}{\bibAnnote {#1} {#2} {\input{#2}}}{}%
}%
\providecommand \typeout [0]{\immediate \write \m@ne }%
\providecommand \selectlanguage [0]{\@gobble}%
\providecommand \bibinfo [0]{\@secondoftwo}%
\providecommand \bibfield [0]{\@secondoftwo}%
\providecommand \translation [1]{[#1]}%
\providecommand \BibitemOpen[0]{}%
\providecommand \bibitemStop [0]{}%
\providecommand \bibitemNoStop [0]{.\EOS\space}%
\providecommand \EOS [0]{\spacefactor3000\relax}%
\providecommand \BibitemShut [1]{\csname bibitem#1\endcsname}%
\bibitem{PhysRevLett.113.046402}%
  \BibitemOpen
  \bibfield{author}{%
  \bibinfo {author} {\bibfnamefont{Philippe}\ \bibnamefont{Corboz}}, \bibinfo
  {author} {\bibfnamefont{T.~M.}\ \bibnamefont{Rice}},\ and\ \bibinfo {author}
  {\bibfnamefont{Matthias}\ \bibnamefont{Troyer}},\ }%
  \bibfield{title}{%
  \enquote{\bibinfo {title} {{\it Competing States in the $t$-$J$ Model:
  Uniform $d$-Wave State versus Stripe State}},}\ }%
  \bibfield{journal}{%
  \Doi{10.1103/PhysRevLett.113.046402}{\bibinfo {journal} {Phys. Rev. Lett.}}\
  }%
  \textbf{\bibinfo {volume} {113}},\ \bibinfo {pages} {046402} (\bibinfo
  {month} {Jul}\ \bibinfo {year} {2014}),\
  \url{https://link.aps.org/doi/10.1103/PhysRevLett.113.046402}%
  \bibAnnoteFile{NoStop}{PhysRevLett.113.046402}%
\bibitem{PhysRevB.85.081110}%
  \BibitemOpen
  \bibfield{author}{%
  \bibinfo {author} {\bibfnamefont{Wen-Jun}\ \bibnamefont{Hu}}, \bibinfo
  {author} {\bibfnamefont{Federico}\ \bibnamefont{Becca}},\ and\ \bibinfo
  {author} {\bibfnamefont{Sandro}\ \bibnamefont{Sorella}},\ }%
  \bibfield{title}{%
  \enquote{\bibinfo {title} {{\it Absence of Static Stripes in the
  Two-Dimensional $t$-$J$ Model Determined Using an Accurate and Systematic
  Quantum Monte Carlo Approach}},}\ }%
  \bibfield{journal}{%
  \Doi{10.1103/PhysRevB.85.081110}{\bibinfo {journal} {Phys. Rev. B}}\ }%
  \textbf{\bibinfo {volume} {85}},\ \bibinfo {pages} {081110} (\bibinfo {month}
  {Feb}\ \bibinfo {year} {2012}),\
  \url{https://link.aps.org/doi/10.1103/PhysRevB.85.081110}%
  \bibAnnoteFile{NoStop}{PhysRevB.85.081110}%
\bibitem{zhao2017}%
  \BibitemOpen
  \bibfield{author}{%
  \bibinfo {author} {\bibfnamefont{Hui-Hai}\ \bibnamefont{Zhao}}, \bibinfo
  {author} {\bibfnamefont{Kota}\ \bibnamefont{Ido}}, \bibinfo {author}
  {\bibfnamefont{Satoshi}\ \bibnamefont{Morita}},\ and\ \bibinfo {author}
  {\bibfnamefont{Masatoshi}\ \bibnamefont{Imada}},\ }%
  \bibfield{title}{%
  \enquote{\bibinfo {title} {{\it Variational Monte Carlo Method for Fermionic
  Models Combined with Tensor Networks and Applications to the Hole-Doped
  Two-Dimensional Hubbard Model}},}\ }%
  \bibfield{journal}{%
  \Doi{10.1103/PhysRevB.96.085103}{\bibinfo {journal} {Phys. Rev. B}}\ }%
  \textbf{\bibinfo {volume} {96}},\ \bibinfo {pages} {085103} (\bibinfo {month}
  {Aug}\ \bibinfo {year} {2017}),\
  \url{https://link.aps.org/doi/10.1103/PhysRevB.96.085103}%
  \bibAnnoteFile{NoStop}{zhao2017}%
\bibitem{doi:10.1126/science.aam7127}%
  \BibitemOpen
  \bibfield{author}{%
  \bibinfo {author} {\bibfnamefont{Bo-Xiao}\ \bibnamefont{Zheng}}, \bibinfo
  {author} {\bibfnamefont{Chia-Min}\ \bibnamefont{Chung}}, \bibinfo {author}
  {\bibfnamefont{Philippe}\ \bibnamefont{Corboz}}, \bibinfo {author}
  {\bibfnamefont{Georg}\ \bibnamefont{Ehlers}}, \bibinfo {author}
  {\bibfnamefont{Ming-Pu}\ \bibnamefont{Qin}}, \bibinfo {author}
  {\bibfnamefont{Reinhard~M.}\ \bibnamefont{Noack}}, \bibinfo {author}
  {\bibfnamefont{Hao}\ \bibnamefont{Shi}}, \bibinfo {author}
  {\bibfnamefont{Steven~R.}\ \bibnamefont{White}}, \bibinfo {author}
  {\bibfnamefont{Shiwei}\ \bibnamefont{Zhang}},\ and\ \bibinfo {author}
  {\bibfnamefont{Garnet Kin-Lic}\ \bibnamefont{Chan}},\ }%
  \bibfield{title}{%
  \enquote{\bibinfo {title} {{\it Stripe Order in the Underdoped Region of the
  Two-Dimensional Hubbard Model}},}\ }%
  \bibfield{journal}{%
  \Doi{10.1126/science.aam7127}{\bibinfo {journal} {Science}}\ }%
  \textbf{\bibinfo {volume} {358}},\ \bibinfo {pages} {1155--1160} (\bibinfo
  {year} {2017}),\
  \url{https://www.science.org/doi/abs/10.1126/science.aam7127}%
  \bibAnnoteFile{NoStop}{doi:10.1126/science.aam7127}%
\bibitem{darmawan2018}%
  \BibitemOpen
  \bibfield{author}{%
  \bibinfo {author} {\bibfnamefont{Andrew~S.}\ \bibnamefont{Darmawan}},
  \bibinfo {author} {\bibfnamefont{Yusuke}\ \bibnamefont{Nomura}}, \bibinfo
  {author} {\bibfnamefont{Youhei}\ \bibnamefont{Yamaji}},\ and\ \bibinfo
  {author} {\bibfnamefont{Masatoshi}\ \bibnamefont{Imada}},\ }%
  \bibfield{title}{%
  \enquote{\bibinfo {title} {{\it Stripe and Superconducting Order Competing in
  the Hubbard Model on a Square Lattice Studied by a Combined Variational Monte
  Carlo and Tensor Network Method}},}\ }%
  \bibfield{journal}{%
  \Doi{10.1103/PhysRevB.98.205132}{\bibinfo {journal} {Phys. Rev. B}}\ }%
  \textbf{\bibinfo {volume} {98}},\ \bibinfo {pages} {205132} (\bibinfo {month}
  {Nov}\ \bibinfo {year} {2018}),\
  \url{https://link.aps.org/doi/10.1103/PhysRevB.98.205132}%
  \bibAnnoteFile{NoStop}{darmawan2018}%
\bibitem{ido2018}%
  \BibitemOpen
  \bibfield{author}{%
  \bibinfo {author} {\bibfnamefont{Kota}\ \bibnamefont{Ido}}, \bibinfo {author}
  {\bibfnamefont{Takahiro}\ \bibnamefont{Ohgoe}},\ and\ \bibinfo {author}
  {\bibfnamefont{Masatoshi}\ \bibnamefont{Imada}},\ }%
  \bibfield{title}{%
  \enquote{\bibinfo {title} {{\it Competition among Various
  Charge-Inhomogeneous States and $d$-Wave Superconducting State in Hubbard
  Models on Square Lattices}},}\ }%
  \bibfield{journal}{%
  \Doi{10.1103/PhysRevB.97.045138}{\bibinfo {journal} {Phys. Rev. B}}\ }%
  \textbf{\bibinfo {volume} {97}},\ \bibinfo {pages} {045138} (\bibinfo {month}
  {Jan}\ \bibinfo {year} {2018}),\
  \url{https://link.aps.org/doi/10.1103/PhysRevB.97.045138}%
  \bibAnnoteFile{NoStop}{ido2018}%
\bibitem{PhysRevX.13.011007}%
  \BibitemOpen
  \bibfield{author}{%
  \bibinfo {author} {\bibfnamefont{Bo}~\bibnamefont{Xiao}}, \bibinfo {author}
  {\bibfnamefont{Yuan-Yao}\ \bibnamefont{He}}, \bibinfo {author}
  {\bibfnamefont{Antoine}\ \bibnamefont{Georges}},\ and\ \bibinfo {author}
  {\bibfnamefont{Shiwei}\ \bibnamefont{Zhang}},\ }%
  \bibfield{title}{%
  \enquote{\bibinfo {title} {{\it Temperature Dependence of Spin and Charge
  Orders in the Doped Two-Dimensional Hubbard Model}},}\ }%
  \bibfield{journal}{%
  \Doi{10.1103/PhysRevX.13.011007}{\bibinfo {journal} {Phys. Rev. X}}\ }%
  \textbf{\bibinfo {volume} {13}},\ \bibinfo {pages} {011007} (\bibinfo {month}
  {Jan}\ \bibinfo {year} {2023}),\
  \url{https://link.aps.org/doi/10.1103/PhysRevX.13.011007}%
  \bibAnnoteFile{NoStop}{PhysRevX.13.011007}%
\bibitem{10.21468/SciPostPhys.12.6.180}%
  \BibitemOpen
  \bibfield{author}{%
  \bibinfo {author} {\bibfnamefont{Vito}\ \bibnamefont{Marino}}, \bibinfo
  {author} {\bibfnamefont{Federico}\ \bibnamefont{Becca}},\ and\ \bibinfo
  {author} {\bibfnamefont{Luca~F.}\ \bibnamefont{Tocchio}},\ }%
  \bibfield{title}{%
  \enquote{\bibinfo {title} {{\it Stripes in the Extended $t$-$t^\prime$
  Hubbard Model: A Variational Monte Carlo Analysis}},}\ }%
  \bibfield{journal}{%
  \Doi{10.21468/SciPostPhys.12.6.180}{\bibinfo {journal} {SciPost Phys.}}\ }%
  \textbf{\bibinfo {volume} {12}},\ \bibinfo {pages} {180} (\bibinfo {year}
  {2022}),\ \url{https://scipost.org/10.21468/SciPostPhys.12.6.180}%
  \bibAnnoteFile{NoStop}{10.21468/SciPostPhys.12.6.180}%
\bibitem{yokoyama2013}%
  \BibitemOpen
  \bibfield{author}{%
  \bibinfo {author} {\bibfnamefont{Hisatoshi}\ \bibnamefont{Yokoyama}},
  \bibinfo {author} {\bibfnamefont{Masao}\ \bibnamefont{Ogata}}, \bibinfo
  {author} {\bibfnamefont{Yukio}\ \bibnamefont{Tanaka}}, \bibinfo {author}
  {\bibfnamefont{Kenji}\ \bibnamefont{Kobayashi}},\ and\ \bibinfo {author}
  {\bibfnamefont{Hiroki}\ \bibnamefont{Tsuchiura}},\ }%
  \bibfield{title}{%
  \enquote{\bibinfo {title} {{\it Crossover between BCS Superconductor and
  Doped Mott Insulator of $d$-Wave Pairing State in Two-Dimensional Hubbard
  Model}},}\ }%
  \bibfield{journal}{%
  \Doi{10.7566/JPSJ.82.014707}{\bibinfo {journal} {J. Phys. Soc. Jpn.}}\ }%
  \textbf{\bibinfo {volume} {82}},\ \bibinfo {pages} {014707} (\bibinfo {year}
  {2013}),\ \url{https://doi.org/10.7566/JPSJ.82.014707}%
  \bibAnnoteFile{NoStop}{yokoyama2013}%
\bibitem{ohgoe2020}%
  \BibitemOpen
  \bibfield{author}{%
  \bibinfo {author} {\bibfnamefont{Takahiro}\ \bibnamefont{Ohgoe}}, \bibinfo
  {author} {\bibfnamefont{Motoaki}\ \bibnamefont{Hirayama}}, \bibinfo {author}
  {\bibfnamefont{Takahiro}\ \bibnamefont{Misawa}}, \bibinfo {author}
  {\bibfnamefont{Kota}\ \bibnamefont{Ido}}, \bibinfo {author}
  {\bibfnamefont{Youhei}\ \bibnamefont{Yamaji}},\ and\ \bibinfo {author}
  {\bibfnamefont{Masatoshi}\ \bibnamefont{Imada}},\ }%
  \bibfield{title}{%
  \enquote{\bibinfo {title} {{\it {\it Ab initio} Study of Superconductivity
  and Inhomogeneity in a Hg-Based Cuprate Superconductor}},}\ }%
  \bibfield{journal}{%
  \Doi{10.1103/PhysRevB.101.045124}{\bibinfo {journal} {Phys. Rev. B}}\ }%
  \textbf{\bibinfo {volume} {101}},\ \bibinfo {pages} {045124} (\bibinfo
  {month} {Jan}\ \bibinfo {year} {2020}),\
  \url{https://link.aps.org/doi/10.1103/PhysRevB.101.045124}%
  \bibAnnoteFile{NoStop}{ohgoe2020}%
\bibitem{PhysRevB.99.075135}%
  \BibitemOpen
  \bibfield{author}{%
  \bibinfo {author} {\bibfnamefont{F.}~\bibnamefont{Nilsson}}, \bibinfo
  {author} {\bibfnamefont{K.}~\bibnamefont{Karlsson}},\ and\ \bibinfo {author}
  {\bibfnamefont{F.}~\bibnamefont{Aryasetiawan}},\ }%
  \bibfield{title}{%
  \enquote{\bibinfo {title} {{\it Dynamically Screened Coulomb Interaction in
  the Parent Compounds of Hole-Doped Cuprates: Trends and Exceptions}},}\ }%
  \bibfield{journal}{%
  \Doi{10.1103/PhysRevB.99.075135}{\bibinfo {journal} {Phys. Rev. B}}\ }%
  \textbf{\bibinfo {volume} {99}},\ \bibinfo {pages} {075135} (\bibinfo {month}
  {Feb}\ \bibinfo {year} {2019}),\
  \url{https://link.aps.org/doi/10.1103/PhysRevB.99.075135}%
  \bibAnnoteFile{NoStop}{PhysRevB.99.075135}%
\bibitem{moree2022}%
  \BibitemOpen
  \bibfield{author}{%
  \bibinfo {author} {\bibfnamefont{Jean-Baptiste}\ \bibnamefont{Mor\'ee}},
  \bibinfo {author} {\bibfnamefont{Motoaki}\ \bibnamefont{Hirayama}}, \bibinfo
  {author} {\bibfnamefont{Michael~Thobias}\ \bibnamefont{Schmid}}, \bibinfo
  {author} {\bibfnamefont{Youhei}\ \bibnamefont{Yamaji}},\ and\ \bibinfo
  {author} {\bibfnamefont{Masatoshi}\ \bibnamefont{Imada}},\ }%
  \bibfield{title}{%
  \enquote{\bibinfo {title} {{\it {\it Ab initio} low-energy effective
  Hamiltonians for the high-temperature superconducting cuprates
  Bi$_{2}$Sr$_2$CuO$_{6}$, Bi$_{2}$Sr$_{2}$CaCu$_{2}$O$_{8}$,
  HgBa$_2$CuO$_{4}$, and CaCuO$_{2}$}},}\ }%
  \bibfield{journal}{%
  \Doi{10.1103/PhysRevB.106.235150}{\bibinfo {journal} {Phys. Rev. B}}\ }%
  \textbf{\bibinfo {volume} {106}},\ \bibinfo {pages} {235150} (\bibinfo
  {month} {Dec}\ \bibinfo {year} {2022}),\
  \url{https://link.aps.org/doi/10.1103/PhysRevB.106.235150}%
  \bibAnnoteFile{NoStop}{moree2022}%
\bibitem{Weber_2012}%
  \BibitemOpen
  \bibfield{author}{%
  \bibinfo {author} {\bibfnamefont{C.}~\bibnamefont{Weber}}, \bibinfo {author}
  {\bibfnamefont{C.}~\bibnamefont{Yee}}, \bibinfo {author}
  {\bibfnamefont{K.}~\bibnamefont{Haule}},\ and\ \bibinfo {author}
  {\bibfnamefont{G.}~\bibnamefont{Kotliar}},\ }%
  \bibfield{title}{%
  \enquote{\bibinfo {title} {{\it Scaling of the Transition Temperature of
  Hole-Doped Cuprate Superconductors with the Charge-Transfer Energy}},}\ }%
  \bibfield{journal}{%
  \Doi{10.1209/0295-5075/100/37001}{\bibinfo {journal} {Europhysics Letters}}\
  }%
  \textbf{\bibinfo {volume} {100}},\ \bibinfo {pages} {37001} (\bibinfo {month}
  {nov}\ \bibinfo {year} {2012}),\
  \url{https://dx.doi.org/10.1209/0295-5075/100/37001}%
  \bibAnnoteFile{NoStop}{Weber_2012}%
\bibitem{PhysRevResearch.3.033157}%
  \BibitemOpen
  \bibfield{author}{%
  \bibinfo {author} {\bibfnamefont{Hiroshi}\ \bibnamefont{Watanabe}}, \bibinfo
  {author} {\bibfnamefont{Tomonori}\ \bibnamefont{Shirakawa}}, \bibinfo
  {author} {\bibfnamefont{Kazuhiro}\ \bibnamefont{Seki}}, \bibinfo {author}
  {\bibfnamefont{Hirofumi}\ \bibnamefont{Sakakibara}}, \bibinfo {author}
  {\bibfnamefont{Takao}\ \bibnamefont{Kotani}}, \bibinfo {author}
  {\bibfnamefont{Hiroaki}\ \bibnamefont{Ikeda}},\ and\ \bibinfo {author}
  {\bibfnamefont{Seiji}\ \bibnamefont{Yunoki}},\ }%
  \bibfield{title}{%
  \enquote{\bibinfo {title} {{\it Unified Description of Cuprate
  Superconductors Using a Four-Band $d$-$p$ Model}},}\ }%
  \bibfield{journal}{%
  \Doi{10.1103/PhysRevResearch.3.033157}{\bibinfo {journal} {Phys. Rev. Res.}}\
  }%
  \textbf{\bibinfo {volume} {3}},\ \bibinfo {pages} {033157} (\bibinfo {month}
  {Aug}\ \bibinfo {year} {2021}),\
  \url{https://link.aps.org/doi/10.1103/PhysRevResearch.3.033157}%
  \bibAnnoteFile{NoStop}{PhysRevResearch.3.033157}%
\bibitem{Misawa2014-ym}%
  \BibitemOpen
  \bibfield{author}{%
  \bibinfo {author} {\bibfnamefont{Takahiro}\ \bibnamefont{Misawa}}\ and\
  \bibinfo {author} {\bibfnamefont{Masatoshi}\ \bibnamefont{Imada}},\ }%
  \bibfield{title}{%
  \enquote{\bibinfo {title} {{\it Superconductivity and Its Mechanism in an
  {\it ab initio} Model for Electron-Doped {LaFeAsO}}},}\ }%
  \bibfield{journal}{%
  \Doi{https://doi.org/10.1038/ncomms6738}{\bibinfo {journal} {Nature
  Communications}}\ }%
  \textbf{\bibinfo {volume} {5}},\ \bibinfo {pages} {5738} (\bibinfo {month}
  {Dec.}\ \bibinfo {year} {2014}),\
  \url{https://www.nature.com/articles/ncomms6738}%
  \bibAnnoteFile{NoStop}{Misawa2014-ym}%
\bibitem{Nomura15Ful}%
  \BibitemOpen
  \bibfield{author}{%
  \bibinfo {author} {\bibfnamefont{Yusuke}\ \bibnamefont{Nomura}}, \bibinfo
  {author} {\bibfnamefont{Shiro}\ \bibnamefont{Sakai}}, \bibinfo {author}
  {\bibfnamefont{Massimo}\ \bibnamefont{Capone}},\ and\ \bibinfo {author}
  {\bibfnamefont{Ryotaro}\ \bibnamefont{Arita}},\ }%
  \bibfield{title}{%
  \enquote{\bibinfo {title} {{\it Unified Understanding of Superconductivity
  and Mott Transition in Alkali-Doped Fullerides from First Principles}},}\ }%
  \bibfield{journal}{%
  \Doi{10.1126/sciadv.1500568}{\bibinfo {journal} {Science Advances}}\ }%
  \textbf{\bibinfo {volume} {1}},\ \bibinfo {pages} {e1500568} (\bibinfo {year}
  {2015}),\ \url{https://www.science.org/doi/abs/10.1126/sciadv.1500568}%
  \bibAnnoteFile{NoStop}{Nomura15Ful}%
\bibitem{Kitatani2020-ra}%
  \BibitemOpen
  \bibfield{author}{%
  \bibinfo {author} {\bibfnamefont{Motoharu}\ \bibnamefont{Kitatani}}, \bibinfo
  {author} {\bibfnamefont{Liang}\ \bibnamefont{Si}}, \bibinfo {author}
  {\bibfnamefont{Oleg}\ \bibnamefont{Janson}}, \bibinfo {author}
  {\bibfnamefont{Ryotaro}\ \bibnamefont{Arita}}, \bibinfo {author}
  {\bibfnamefont{Zhicheng}\ \bibnamefont{Zhong}},\ and\ \bibinfo {author}
  {\bibfnamefont{Karsten}\ \bibnamefont{Held}},\ }%
  \bibfield{title}{%
  \enquote{\bibinfo {title} {{\it Nickelate Superconductors---a Renaissance of
  the One-Band Hubbard Model}},}\ }%
  \bibfield{journal}{%
  \Doi{https://doi.org/10.1038/s41535-020-00260-y}{\bibinfo {journal} {npj
  Quantum Mat.}}\ }%
  \textbf{\bibinfo {volume} {5}},\ \bibinfo {pages} {59} (\bibinfo {month}
  {Aug.}\ \bibinfo {year} {2020}),\
  \url{https://www.nature.com/articles/s41535-020-00260-y}%
  \bibAnnoteFile{NoStop}{Kitatani2020-ra}%
\bibitem{misawa2019}%
  \BibitemOpen
  \bibfield{author}{%
  \bibinfo {author} {\bibfnamefont{Takahiro}\ \bibnamefont{Misawa}}, \bibinfo
  {author} {\bibfnamefont{Satoshi}\ \bibnamefont{Morita}}, \bibinfo {author}
  {\bibfnamefont{Kazuyoshi}\ \bibnamefont{Yoshimi}}, \bibinfo {author}
  {\bibfnamefont{Mitsuaki}\ \bibnamefont{Kawamura}}, \bibinfo {author}
  {\bibfnamefont{Yuichi}\ \bibnamefont{Motoyama}}, \bibinfo {author}
  {\bibfnamefont{Kota}\ \bibnamefont{Ido}}, \bibinfo {author}
  {\bibfnamefont{Takahiro}\ \bibnamefont{Ohgoe}}, \bibinfo {author}
  {\bibfnamefont{Masatoshi}\ \bibnamefont{Imada}},\ and\ \bibinfo {author}
  {\bibfnamefont{Takeo}\ \bibnamefont{Kato}},\ }%
  \bibfield{title}{%
  \enquote{\bibinfo {title} {{\it mVMC  -Open-Source Software for Many-Variable
  Variational Monte Carlo Method}},}\ }%
  \bibfield{journal}{%
  \Doi{https://doi.org/10.1016/j.cpc.2018.08.014}{\bibinfo {journal} {Computer
  Physics Communications}}\ }%
  \textbf{\bibinfo {volume} {235}},\ \bibinfo {pages} {447--462} (\bibinfo
  {year} {2019}),\ ISSN \bibinfo {issn} {0010-4655},\
  \url{https://www.sciencedirect.com/science/article/pii/S0010465518303102}%
  \bibAnnoteFile{NoStop}{misawa2019}%
\bibitem{tahara2008}%
  \BibitemOpen
  \bibfield{author}{%
  \bibinfo {author} {\bibfnamefont{Daisuke}\ \bibnamefont{Tahara}}\ and\
  \bibinfo {author} {\bibfnamefont{Masatoshi}\ \bibnamefont{Imada}},\ }%
  \bibfield{title}{%
  \enquote{\bibinfo {title} {{\it Variational Monte Carlo Method Combined with
  Quantum-Number Projection and Multi-Variable Optimization}},}\ }%
  \bibfield{journal}{%
  \Doi{10.1143/JPSJ.77.114701}{\bibinfo {journal} {J. Phys. Soc. Jpn.}}\ }%
  \textbf{\bibinfo {volume} {77}},\ \bibinfo {pages} {114701} (\bibinfo {year}
  {2008}),\ \url{https://doi.org/10.1143/JPSJ.77.114701}%
  \bibAnnoteFile{NoStop}{tahara2008}%
\bibitem{nomura2017}%
  \BibitemOpen
  \bibfield{author}{%
  \bibinfo {author} {\bibfnamefont{Yusuke}\ \bibnamefont{Nomura}}, \bibinfo
  {author} {\bibfnamefont{Andrew~S.}\ \bibnamefont{Darmawan}}, \bibinfo
  {author} {\bibfnamefont{Youhei}\ \bibnamefont{Yamaji}},\ and\ \bibinfo
  {author} {\bibfnamefont{Masatoshi}\ \bibnamefont{Imada}},\ }%
  \bibfield{title}{%
  \enquote{\bibinfo {title} {{\it Restricted Boltzmann Machine Learning for
  Solving Strongly Correlated Quantum Systems}},}\ }%
  \bibfield{journal}{%
  \Doi{10.1103/PhysRevB.96.205152}{\bibinfo {journal} {Phys. Rev. B}}\ }%
  \textbf{\bibinfo {volume} {96}},\ \bibinfo {pages} {205152} (\bibinfo {month}
  {Nov}\ \bibinfo {year} {2017}),\
  \url{https://link.aps.org/doi/10.1103/PhysRevB.96.205152}%
  \bibAnnoteFile{NoStop}{nomura2017}%
\bibitem{nomura2021}%
  \BibitemOpen
  \bibfield{author}{%
  \bibinfo {author} {\bibfnamefont{Yusuke}\ \bibnamefont{Nomura}}\ and\
  \bibinfo {author} {\bibfnamefont{Masatoshi}\ \bibnamefont{Imada}},\ }%
  \bibfield{title}{%
  \enquote{\bibinfo {title} {{\it Dirac-Type Nodal Spin Liquid Revealed by
  Refined Quantum Many-Body Solver Using Neural-Network Wave Function,
  Correlation Ratio, and Level Spectroscopy}},}\ }%
  \bibfield{journal}{%
  \Doi{10.1103/PhysRevX.11.031034}{\bibinfo {journal} {Phys. Rev. X}}\ }%
  \textbf{\bibinfo {volume} {11}},\ \bibinfo {pages} {031034} (\bibinfo {month}
  {Aug}\ \bibinfo {year} {2021})%
  \bibAnnoteFile{NoStop}{nomura2021}%
\bibitem{YokoyamaShiba1987}%
  \BibitemOpen
  \bibfield{author}{%
  \bibinfo {author} {\bibfnamefont{Hisatoshi}\ \bibnamefont{Yokoyama}}\ and\
  \bibinfo {author} {\bibfnamefont{Hiroyuki}\ \bibnamefont{Shiba}},\ }%
  \bibfield{title}{%
  \enquote{\bibinfo {title} {{\it Variational Monte-Carlo Studies of Hubbard
  Model. {I}}},}\ }%
  \bibfield{journal}{%
  \Doi{10.1143/JPSJ.56.1490}{\bibinfo {journal} {J. Phys. Soc. Jpn.}}\ }%
  \textbf{\bibinfo {volume} {56}},\ \bibinfo {pages} {1490--1506} (\bibinfo
  {year} {1987}),\ \url{https://doi.org/10.1143/JPSJ.56.1490}%
  \bibAnnoteFile{NoStop}{YokoyamaShiba1987}%
\bibitem{GrossJoyntRice1987}%
  \BibitemOpen
  \bibfield{author}{%
  \bibinfo {author} {\bibfnamefont{C.}~\bibnamefont{Gros}}, \bibinfo {author}
  {\bibfnamefont{R.}~\bibnamefont{Joynt}},\ and\ \bibinfo {author}
  {\bibfnamefont{T.~M.}\ \bibnamefont{Rice}},\ }%
  \bibfield{title}{%
  \enquote{\bibinfo {title} {{\it Antiferromagnetic Correlations in
  Almost-Localized Fermi Liquids}},}\ }%
  \bibfield{journal}{%
  \Doi{10.1103/PhysRevB.36.381}{\bibinfo {journal} {Phys. Rev. B}}\ }%
  \textbf{\bibinfo {volume} {36}},\ \bibinfo {pages} {381--393} (\bibinfo
  {month} {Jul}\ \bibinfo {year} {1987}),\
  \url{https://link.aps.org/doi/10.1103/PhysRevB.36.381}%
  \bibAnnoteFile{NoStop}{GrossJoyntRice1987}%
\bibitem{Gross1988}%
  \BibitemOpen
  \bibfield{author}{%
  \bibinfo {author} {\bibfnamefont{Claudius}\ \bibnamefont{Gros}},\ }%
  \bibfield{title}{%
  \enquote{\bibinfo {title} {{\it Superconductivity in Correlated Wave
  Functions}},}\ }%
  \bibfield{journal}{%
  \Doi{10.1103/PhysRevB.38.931}{\bibinfo {journal} {Phys. Rev. B}}\ }%
  \textbf{\bibinfo {volume} {38}},\ \bibinfo {pages} {931--934} (\bibinfo
  {month} {Jul}\ \bibinfo {year} {1988}),\
  \url{https://link.aps.org/doi/10.1103/PhysRevB.38.931}%
  \bibAnnoteFile{NoStop}{Gross1988}%
\bibitem{Capriotti2001}%
  \BibitemOpen
  \bibfield{author}{%
  \bibinfo {author} {\bibfnamefont{Luca}\ \bibnamefont{Capriotti}}, \bibinfo
  {author} {\bibfnamefont{Federico}\ \bibnamefont{Becca}}, \bibinfo {author}
  {\bibfnamefont{Alberto}\ \bibnamefont{Parola}},\ and\ \bibinfo {author}
  {\bibfnamefont{Sandro}\ \bibnamefont{Sorella}},\ }%
  \bibfield{title}{%
  \enquote{\bibinfo {title} {{\it Resonating Valence Bond Wave Functions for
  Strongly Frustrated Spin Systems}},}\ }%
  \bibfield{journal}{%
  \Doi{10.1103/PhysRevLett.87.097201}{\bibinfo {journal} {Phys. Rev. Lett.}}\
  }%
  \textbf{\bibinfo {volume} {87}},\ \bibinfo {pages} {097201} (\bibinfo {month}
  {Aug}\ \bibinfo {year} {2001}),\
  \url{https://link.aps.org/doi/10.1103/PhysRevLett.87.097201}%
  \bibAnnoteFile{NoStop}{Capriotti2001}%
\bibitem{Azuma1992}%
  \BibitemOpen
  \bibfield{author}{%
  \bibinfo {author} {\bibfnamefont{M.}~\bibnamefont{Azuma}}, \bibinfo {author}
  {\bibfnamefont{Z.}~\bibnamefont{Hiroi}}, \bibinfo {author}
  {\bibfnamefont{M.}~\bibnamefont{Takano}}, \bibinfo {author}
  {\bibfnamefont{Y.}~\bibnamefont{Bando}},\ and\ \bibinfo {author}
  {\bibfnamefont{Y.}~\bibnamefont{Takeda}},\ }%
  \bibfield{title}{%
  \enquote{\bibinfo {title} {{\it Superconductivity at 110 {K} in the
  Infinite-Layer Compound ({Sr}$_{1-x}${Ca}$_x$)$_{1-y}${CuO}$_2$}},}\ }%
  \bibfield{journal}{%
  \Doi{https://doi.org/10.1038/356775a0}{\bibinfo {journal} {Nature}}\ }%
  \textbf{\bibinfo {volume} {356}},\ \bibinfo {pages} {775--776} (\bibinfo
  {year} {1992}),\ \url{https://doi.org/10.1038/356775a0}%
  \bibAnnoteFile{NoStop}{Azuma1992}%
\bibitem{yamamoto2000}%
  \BibitemOpen
  \bibfield{author}{%
  \bibinfo {author} {\bibfnamefont{Ayako}\ \bibnamefont{Yamamoto}}, \bibinfo
  {author} {\bibfnamefont{Wei-Zhi}\ \bibnamefont{Hu}},\ and\ \bibinfo {author}
  {\bibfnamefont{Setsuko}\ \bibnamefont{Tajima}},\ }%
  \bibfield{title}{%
  \enquote{\bibinfo {title} {{\it Thermoelectric power and resistivity of
  HgBa$_{2}$CuO$_{4+\ensuremath{\delta}}$ over a wide doping range}},}\ }%
  \bibfield{journal}{%
  \Doi{10.1103/PhysRevB.63.024504}{\bibinfo {journal} {Phys. Rev. B}}\ }%
  \textbf{\bibinfo {volume} {63}},\ \bibinfo {pages} {024504} (\bibinfo {month}
  {Dec}\ \bibinfo {year} {2000}),\
  \url{https://link.aps.org/doi/10.1103/PhysRevB.63.024504}%
  \bibAnnoteFile{NoStop}{yamamoto2000}%
\bibitem{koike1989}%
  \BibitemOpen
  \bibfield{author}{%
  \bibinfo {author} {\bibfnamefont{Y.}~\bibnamefont{Koike}}, \bibinfo {author}
  {\bibfnamefont{Y.}~\bibnamefont{Iwabuchi}}, \bibinfo {author}
  {\bibfnamefont{S.}~\bibnamefont{Hosoya}}, \bibinfo {author}
  {\bibfnamefont{N.}~\bibnamefont{Kobayashi}},\ and\ \bibinfo {author}
  {\bibfnamefont{T.}~\bibnamefont{Fukase}},\ }%
  \bibfield{title}{%
  \enquote{\bibinfo {title} {{\it Correlation between T, and hole concentration
  in the cation-substituted {Bi}$_2${Sr}$_2${CaCu}$_2${O}$_{8+\delta}$
  system}},}\ }%
  \bibfield{journal}{%
  \Doi{10.1016/0921-4534(89)90110-X}{\bibinfo {journal} {Physica C}}\ }%
  \textbf{\bibinfo {volume} {159}},\ \bibinfo {pages} {105} (\bibinfo {year}
  {1989}),\ \url{https://doi.org/10.1016/0921-4534(89)90110-X}%
  \bibAnnoteFile{NoStop}{koike1989}%
\bibitem{fang1992}%
  \BibitemOpen
  \bibfield{author}{%
  \bibinfo {author} {\bibfnamefont{FANG}\ \bibnamefont{Minghu}}, \bibinfo
  {author} {\bibfnamefont{XU}~\bibnamefont{Zhuan}}, \bibinfo {author}
  {\bibfnamefont{WEI}\ \bibnamefont{Hongbin}}, \bibinfo {author}
  {\bibfnamefont{ZENG}\ \bibnamefont{Xingbin}}, \bibinfo {author}
  {\bibfnamefont{HU}~\bibnamefont{Gangjin}}, \bibinfo {author}
  {\bibfnamefont{ZHANG}\ \bibnamefont{Xuanjia}}, \bibinfo {author}
  {\bibfnamefont{ZHANG}\ \bibnamefont{Qirui}}, \bibinfo {author}
  {\bibfnamefont{WU}~\bibnamefont{Yuming}}, \bibinfo {author}
  {\bibfnamefont{WANG}\ \bibnamefont{Qidong}}, \bibinfo {author}
  {\bibfnamefont{SHA}\ \bibnamefont{Jian}},\ and\ \bibinfo {author}
  {\bibfnamefont{CAO}\ \bibnamefont{Liezhao}},\ }%
  \bibfield{title}{%
  \enquote{\bibinfo {title} {{\it Hole concentration dependence of $T_c$ in
  Bi$_2$Sr$_2$BaCu$_2$O$_{8+y}$ system}},}\ }%
  \bibfield{journal}{%
  \bibinfo {journal} {Chinese Physics Letters}\ }%
  \textbf{\bibinfo {volume} {9}},\ \bibinfo {eid} {159-161} (\bibinfo {year}
  {1992}),\
  \url{http://cpl.iphy.ac.cn/Y1992/V9/I3/159?utm_source=TrendMD&utm_medium=cpc&utm_campaign=Chinese_Physics_Letters_TrendMD_0}%
  \bibAnnoteFile{NoStop}{fang1992}%
\bibitem{Hobou2009}%
  \BibitemOpen
  \bibfield{author}{%
  \bibinfo {author} {\bibfnamefont{H.}~\bibnamefont{Hobou}}, \bibinfo {author}
  {\bibfnamefont{S.}~\bibnamefont{Ishida}}, \bibinfo {author}
  {\bibfnamefont{K.}~\bibnamefont{Fujita}}, \bibinfo {author}
  {\bibfnamefont{M.}~\bibnamefont{Ishikado}}, \bibinfo {author}
  {\bibfnamefont{K.~M.}\ \bibnamefont{Kojima}}, \bibinfo {author}
  {\bibfnamefont{H.}~\bibnamefont{Eisaki}},\ and\ \bibinfo {author}
  {\bibfnamefont{S.}~\bibnamefont{Uchida}},\ }%
  \bibfield{title}{%
  \enquote{\bibinfo {title} {{\it Enhancement of the superconducting critical
  temperature in Bi$_2$Sr$_2$BaCu$_2$O$_{8+y}$ by controlling disorder outside
  CuO$_{2}$ planes}},}\ }%
  \bibfield{journal}{%
  \Doi{10.1103/PhysRevB.79.064507}{\bibinfo {journal} {Phys. Rev. B}}\ }%
  \textbf{\bibinfo {volume} {79}},\ \bibinfo {pages} {064507} (\bibinfo {month}
  {Feb}\ \bibinfo {year} {2009}),\
  \url{https://link.aps.org/doi/10.1103/PhysRevB.79.064507}%
  \bibAnnoteFile{NoStop}{Hobou2009}%
\bibitem{marzari1997maximally}%
  \BibitemOpen
  \bibfield{author}{%
  \bibinfo {author} {\bibfnamefont{Nicola}\ \bibnamefont{Marzari}}\ and\
  \bibinfo {author} {\bibfnamefont{David}\ \bibnamefont{Vanderbilt}},\ }%
  \bibfield{title}{%
  \enquote{\bibinfo {title} {{\it Maximally Localized Generalized Wannier
  Functions for Composite Energy Bands}},}\ }%
  \bibfield{journal}{%
  \Doi{10.1103/PhysRevB.56.12847}{\bibinfo {journal} {Phys. Rev. B}}\ }%
  \textbf{\bibinfo {volume} {56}},\ \bibinfo {pages} {12847--12865} (\bibinfo
  {month} {Nov}\ \bibinfo {year} {1997}),\
  \url{https://link.aps.org/doi/10.1103/PhysRevB.56.12847}%
  \bibAnnoteFile{NoStop}{marzari1997maximally}%
\bibitem{souza2001maximally}%
  \BibitemOpen
  \bibfield{author}{%
  \bibinfo {author} {\bibfnamefont{Ivo}\ \bibnamefont{Souza}}, \bibinfo
  {author} {\bibfnamefont{Nicola}\ \bibnamefont{Marzari}},\ and\ \bibinfo
  {author} {\bibfnamefont{David}\ \bibnamefont{Vanderbilt}},\ }%
  \bibfield{title}{%
  \enquote{\bibinfo {title} {{\it Maximally Localized Wannier Functions for
  Entangled Energy Bands}},}\ }%
  \bibfield{journal}{%
  \Doi{10.1103/PhysRevB.65.035109}{\bibinfo {journal} {Phys. Rev. B}}\ }%
  \textbf{\bibinfo {volume} {65}},\ \bibinfo {pages} {035109} (\bibinfo {month}
  {Dec}\ \bibinfo {year} {2001}),\
  \url{https://link.aps.org/doi/10.1103/PhysRevB.65.035109}%
  \bibAnnoteFile{NoStop}{souza2001maximally}%
\bibitem{SM_Michael}%
  \BibitemOpen
  \enquote{\bibinfo {title} {See supplemental material},}\ %
  \bibAnnoteFile{NoStop}{SM_Michael}%
\bibitem{hirayama2019}%
  \BibitemOpen
  \bibfield{author}{%
  \bibinfo {author} {\bibfnamefont{Motoaki}\ \bibnamefont{Hirayama}}, \bibinfo
  {author} {\bibfnamefont{Takahiro}\ \bibnamefont{Misawa}}, \bibinfo {author}
  {\bibfnamefont{Takahiro}\ \bibnamefont{Ohgoe}}, \bibinfo {author}
  {\bibfnamefont{Youhei}\ \bibnamefont{Yamaji}},\ and\ \bibinfo {author}
  {\bibfnamefont{Masatoshi}\ \bibnamefont{Imada}},\ }%
  \bibfield{title}{%
  \enquote{\bibinfo {title} {{\it Effective Hamiltonian for Cuprate
  Superconductors Derived from Multiscale {\it ab initio} Scheme with Level
  Renormalization}},}\ }%
  \bibfield{journal}{%
  \Doi{10.1103/PhysRevB.99.245155}{\bibinfo {journal} {Phys. Rev. B}}\ }%
  \textbf{\bibinfo {volume} {99}},\ \bibinfo {pages} {245155} (\bibinfo {month}
  {Jun}\ \bibinfo {year} {2019}),\
  \url{https://link.aps.org/doi/10.1103/PhysRevB.99.245155}%
  \bibAnnoteFile{NoStop}{hirayama2019}%
\bibitem{gutzwiller1963}%
  \BibitemOpen
  \bibfield{author}{%
  \bibinfo {author} {\bibfnamefont{Martin~C.}\ \bibnamefont{Gutzwiller}},\ }%
  \bibfield{title}{%
  \enquote{\bibinfo {title} {{\it Effect of Correlation on the Ferromagnetism
  of Transition Metals}},}\ }%
  \bibfield{journal}{%
  \Doi{10.1103/PhysRevLett.10.159}{\bibinfo {journal} {Phys. Rev. Lett.}}\ }%
  \textbf{\bibinfo {volume} {10}},\ \bibinfo {pages} {159--162} (\bibinfo
  {month} {Mar}\ \bibinfo {year} {1963}),\
  \url{https://link.aps.org/doi/10.1103/PhysRevLett.10.159}%
  \bibAnnoteFile{NoStop}{gutzwiller1963}%
\bibitem{jastrow1955}%
  \BibitemOpen
  \bibfield{author}{%
  \bibinfo {author} {\bibfnamefont{Robert}\ \bibnamefont{Jastrow}},\ }%
  \bibfield{title}{%
  \enquote{\bibinfo {title} {{\it Many-Body Problem with Strong Forces}},}\ }%
  \bibfield{journal}{%
  \Doi{10.1103/PhysRev.98.1479}{\bibinfo {journal} {Phys. Rev.}}\ }%
  \textbf{\bibinfo {volume} {98}},\ \bibinfo {pages} {1479--1484} (\bibinfo
  {month} {Jun}\ \bibinfo {year} {1955}),\
  \url{https://link.aps.org/doi/10.1103/PhysRev.98.1479}%
  \bibAnnoteFile{NoStop}{jastrow1955}%
\bibitem{capello2005}%
  \BibitemOpen
  \bibfield{author}{%
  \bibinfo {author} {\bibfnamefont{Manuela}\ \bibnamefont{Capello}}, \bibinfo
  {author} {\bibfnamefont{Federico}\ \bibnamefont{Becca}}, \bibinfo {author}
  {\bibfnamefont{Michele}\ \bibnamefont{Fabrizio}}, \bibinfo {author}
  {\bibfnamefont{Sandro}\ \bibnamefont{Sorella}},\ and\ \bibinfo {author}
  {\bibfnamefont{Erio}\ \bibnamefont{Tosatti}},\ }%
  \bibfield{title}{%
  \enquote{\bibinfo {title} {{\it Variational Description of Mott
  Insulators}},}\ }%
  \bibfield{journal}{%
  \Doi{10.1103/PhysRevLett.94.026406}{\bibinfo {journal} {Phys. Rev. Lett.}}\
  }%
  \textbf{\bibinfo {volume} {94}},\ \bibinfo {pages} {026406} (\bibinfo {month}
  {Jan}\ \bibinfo {year} {2005}),\
  \url{https://link.aps.org/doi/10.1103/PhysRevLett.94.026406}%
  \bibAnnoteFile{NoStop}{capello2005}%
\bibitem{yokoyama1990}%
  \BibitemOpen
  \bibfield{author}{%
  \bibinfo {author} {\bibfnamefont{Hisatoshi}\ \bibnamefont{Yokoyama}}\ and\
  \bibinfo {author} {\bibfnamefont{Hiroyuki}\ \bibnamefont{Shiba}},\ }%
  \bibfield{title}{%
  \enquote{\bibinfo {title} {{\it Variational Monte-Carlo Studies of Hubbard
  Model. III. Intersite Correlation Effects}},}\ }%
  \bibfield{journal}{%
  \Doi{10.1143/JPSJ.59.3669}{\bibinfo {journal} {J. Phys. Soc. Jpn.}}\ }%
  \textbf{\bibinfo {volume} {59}},\ \bibinfo {pages} {3669--3686} (\bibinfo
  {year} {1990})%
  \bibAnnoteFile{NoStop}{yokoyama1990}%
\bibitem{imada2000}%
  \BibitemOpen
  \bibfield{author}{%
  \bibinfo {author} {\bibfnamefont{Masatoshi}\ \bibnamefont{Imada}}\ and\
  \bibinfo {author} {\bibfnamefont{Tsuyoshi}\ \bibnamefont{Kashima}},\ }%
  \bibfield{title}{%
  \enquote{\bibinfo {title} {{\it Path-Integral Renormalization Group Method
  for Numerical Study of Strongly Correlated Electron Systems}},}\ }%
  \bibfield{journal}{%
  \Doi{10.1143/JPSJ.69.2723}{\bibinfo {journal} {J. Phys. Soc. Jpn.}}\ }%
  \textbf{\bibinfo {volume} {69}},\ \bibinfo {pages} {2723} (\bibinfo {year}
  {2000}),\ \url{https://doi.org/10.1143/JPSJ.69.2723}%
  \bibAnnoteFile{NoStop}{imada2000}%
\bibitem{kashima2001}%
  \BibitemOpen
  \bibfield{author}{%
  \bibinfo {author} {\bibfnamefont{Tsuyoshi}\ \bibnamefont{Kashima}}\ and\
  \bibinfo {author} {\bibfnamefont{Masatoshi}\ \bibnamefont{Imada}},\ }%
  \bibfield{title}{%
  \enquote{\bibinfo {title} {{\it Path-Integral Renormalization Group Method
  for Numerical Study on Ground States of Strongly Correlated Electron
  Systems}},}\ }%
  \bibfield{journal}{%
  \Doi{10.1143/JPSJ.70.2287}{\bibinfo {journal} {J. Phys. Soc. Jpn.}}\ }%
  \textbf{\bibinfo {volume} {70}},\ \bibinfo {pages} {2287} (\bibinfo {year}
  {2001}),\ \url{https://doi.org/10.1143/JPSJ.70.2287}%
  \bibAnnoteFile{NoStop}{kashima2001}%
\bibitem{sorella2001}%
  \BibitemOpen
  \bibfield{author}{%
  \bibinfo {author} {\bibfnamefont{Sandro}\ \bibnamefont{Sorella}},\ }%
  \bibfield{title}{%
  \enquote{\bibinfo {title} {{\it Generalized Lanczos algorithm for variational
  quantum Monte Carlo}},}\ }%
  \bibfield{journal}{%
  \Doi{10.1103/PhysRevB.64.024512}{\bibinfo {journal} {Phys. Rev. B}}\ }%
  \textbf{\bibinfo {volume} {64}},\ \bibinfo {pages} {024512} (\bibinfo {month}
  {Jun}\ \bibinfo {year} {2001}),\
  \url{https://link.aps.org/doi/10.1103/PhysRevB.64.024512}%
  \bibAnnoteFile{NoStop}{sorella2001}%
\bibitem{ido2022}%
  \BibitemOpen
  \bibfield{author}{%
  \bibinfo {author} {\bibfnamefont{Kota}\ \bibnamefont{Ido}}, \bibinfo {author}
  {\bibfnamefont{Kazuyoshi}\ \bibnamefont{Yoshimi}}, \bibinfo {author}
  {\bibfnamefont{Takahiro}\ \bibnamefont{Misawa}},\ and\ \bibinfo {author}
  {\bibfnamefont{Masatoshi}\ \bibnamefont{Imada}},\ }%
  \bibfield{title}{%
  \enquote{\bibinfo {title} {{\it Unconventional Dual 1D--2D Quantum Spin
  Liquid Revealed by {\it ab initio} Studies on Organic Solids Family}},}\ }%
  \bibfield{journal}{%
  \Doi{10.1038/s41535-022-00452-8}{\bibinfo {journal} {npj Quantum Materials}}\
  }%
  \textbf{\bibinfo {volume} {7}},\ \bibinfo {pages} {48} (\bibinfo {year}
  {2022}),\ \url{https://doi.org/10.1038/s41535-022-00452-8}%
  \bibAnnoteFile{NoStop}{ido2022}%
\bibitem{hiroi1993}%
  \BibitemOpen
  \bibfield{author}{%
  \bibinfo {author} {\bibfnamefont{Z.}~\bibnamefont{Hiroi}}, \bibinfo {author}
  {\bibfnamefont{M.}~\bibnamefont{Azuma}}, \bibinfo {author}
  {\bibfnamefont{M.}~\bibnamefont{Takano}},\ and\ \bibinfo {author}
  {\bibfnamefont{Y.}~\bibnamefont{Takeda}},\ }%
  \bibfield{title}{%
  \enquote{\bibinfo {title} {{\it Structure and Superconductivity of the
  Infinite-Layer Compound (Ca$_{1-y}$Sr$_y$)$_{1-x}$CuO$_{2-z}$}},}\ }%
  \bibfield{journal}{%
  \Doi{https://doi.org/10.1016/0921-4534(93)90200-A}{\bibinfo {journal}
  {Physica C: Superconductivity}}\ }%
  \textbf{\bibinfo {volume} {208}},\ \bibinfo {pages} {286--296} (\bibinfo
  {year} {1993}),\ ISSN \bibinfo {issn} {0921-4534},\
  \url{https://www.sciencedirect.com/science/article/pii/092145349390200A}%
  \bibAnnoteFile{NoStop}{hiroi1993}%
\bibitem{Tanaka2006}%
  \BibitemOpen
  \bibfield{author}{%
  \bibinfo {author} {\bibfnamefont{K.}~\bibnamefont{Tanaka}}, \bibinfo {author}
  {\bibfnamefont{W.~S.}\ \bibnamefont{Lee}}, \bibinfo {author}
  {\bibfnamefont{D.~H.}\ \bibnamefont{Lu}}, \bibinfo {author}
  {\bibfnamefont{A.}~\bibnamefont{Fujimor}}, \bibinfo {author}
  {\bibfnamefont{T.}~\bibnamefont{Fujii}}, \bibinfo {author}
  {\bibnamefont{Risdiana}}, \bibinfo {author}
  {\bibfnamefont{I.}~\bibnamefont{Terasaki}}, \bibinfo {author}
  {\bibfnamefont{D.~J.}\ \bibnamefont{Scalapino}}, \bibinfo {author}
  {\bibfnamefont{T.~P.}\ \bibnamefont{Devereaux}}, \bibinfo {author}
  {\bibfnamefont{Z.}~\bibnamefont{Hussain}},\ and\ \bibinfo {author}
  {\bibfnamefont{Z.-X.}\ \bibnamefont{Shen}},\ }%
  \bibfield{title}{%
  \enquote{\bibinfo {title} {{\it Distinct Fermi-Momentum Dependent Energy
  Gaps in Deeply Underdoped Bi2212}},}\ }%
  \bibfield{journal}{%
  \Doi{10.1126/science.1133411}{\bibinfo {journal} {Science}}\ }%
  \textbf{\bibinfo {volume} {314}},\ \bibinfo {pages} {1910--1913} (\bibinfo
  {year} {2006})%
  \bibAnnoteFile{NoStop}{Tanaka2006}%
\bibitem{Alldredge2008}%
  \BibitemOpen
  \bibfield{author}{%
  \bibinfo {author} {\bibfnamefont{J.W.}\ \bibnamefont{Alldredge}}, \bibinfo
  {author} {\bibfnamefont{J.}~\bibnamefont{Lee}}, \bibinfo {author}
  {\bibfnamefont{K.}~\bibnamefont{McElroy}}, \bibinfo {author}
  {\bibfnamefont{M.}~\bibnamefont{Wang}}, \bibinfo {author}
  {\bibfnamefont{K.}~\bibnamefont{Fujita}}, \bibinfo {author}
  {\bibfnamefont{Y}~\bibnamefont{Kohsaka}}, \bibinfo {author}
  {\bibfnamefont{C.}~\bibnamefont{Taylor}}, \bibinfo {author}
  {\bibfnamefont{H.}~\bibnamefont{Eisaki}}, \bibinfo {author}
  {\bibfnamefont{S.}~\bibnamefont{Uchida}}, \bibinfo {author}
  {\bibfnamefont{P.J.}\ \bibnamefont{Hirschfeld}},\ and\ \bibinfo {author}
  {\bibfnamefont{J.C.}\ \bibnamefont{Davis}},\ }%
  \bibfield{title}{%
  \enquote{\bibinfo {title} {{\it Evolution of the Electronic Excitation
  Spectrum with Strongly Diminishing Hole Density in Superconducting
  Bi$_2$Sr$_2$CaCu$_2$O$_{8+\delta}$}},}\ }%
  \bibfield{journal}{%
  \Doi{https://doi.org/10.1038/nphys917}{\bibinfo {journal} {Nat. Phys.}}\ }%
  \textbf{\bibinfo {volume} {4}},\ \bibinfo {pages} {319--326} (\bibinfo {year}
  {2008})%
  \bibAnnoteFile{NoStop}{Alldredge2008}%
\bibitem{Sakai2018}%
  \BibitemOpen
  \bibfield{author}{%
  \bibinfo {author} {\bibfnamefont{Shiro}\ \bibnamefont{Sakai}}, \bibinfo
  {author} {\bibfnamefont{Marcello}\ \bibnamefont{Civelli}},\ and\ \bibinfo
  {author} {\bibfnamefont{Masatoshi}\ \bibnamefont{Imada}},\ }%
  \bibfield{title}{%
  \enquote{\bibinfo {title} {{\it Direct Connection Between Mott Insulators and
  $d$-Wave High-Temperature Superconductors Revealed by Continuous Evolution of
  Self-Energy Poles}},}\ }%
  \bibfield{journal}{%
  \Doi{10.1103/PhysRevB.98.195109}{\bibinfo {journal} {Phys. Rev. B}}\ }%
  \textbf{\bibinfo {volume} {98}},\ \bibinfo {pages} {195109} (\bibinfo {month}
  {Nov}\ \bibinfo {year} {2018}),\
  \url{https://link.aps.org/doi/10.1103/PhysRevB.98.195109}%
  \bibAnnoteFile{NoStop}{Sakai2018}%
\bibitem{Uemura1989}%
  \BibitemOpen
  \bibfield{author}{%
  \bibinfo {author} {\bibfnamefont{Y.~J.}\ \bibnamefont{Uemura}}, \bibinfo
  {author} {\bibfnamefont{G.~M.}\ \bibnamefont{Luke}}, \bibinfo {author}
  {\bibfnamefont{B.~J.}\ \bibnamefont{Sternlieb}}, \bibinfo {author}
  {\bibfnamefont{J.~H.}\ \bibnamefont{Brewer}}, \bibinfo {author}
  {\bibfnamefont{J.~F.}\ \bibnamefont{Carolan}}, \bibinfo {author}
  {\bibfnamefont{W.~N.}\ \bibnamefont{Hardy}}, \bibinfo {author}
  {\bibfnamefont{R.}~\bibnamefont{Kadono}}, \bibinfo {author}
  {\bibfnamefont{J.~R.}\ \bibnamefont{Kempton}}, \bibinfo {author}
  {\bibfnamefont{R.~F.}\ \bibnamefont{Kiefl}}, \bibinfo {author}
  {\bibfnamefont{S.~R.}\ \bibnamefont{Kreitzman}}, \bibinfo {author}
  {\bibfnamefont{P.}~\bibnamefont{Mulhern}}, \bibinfo {author}
  {\bibfnamefont{T.~M.}\ \bibnamefont{Riseman}}, \bibinfo {author}
  {\bibfnamefont{D.~Ll.}\ \bibnamefont{Williams}}, \bibinfo {author}
  {\bibfnamefont{B.~X.}\ \bibnamefont{Yang}}, \bibinfo {author}
  {\bibfnamefont{S.}~\bibnamefont{Uchida}}, \bibinfo {author}
  {\bibfnamefont{H.}~\bibnamefont{Takagi}}, \bibinfo {author}
  {\bibfnamefont{J.}~\bibnamefont{Gopalakrishnan}}, \bibinfo {author}
  {\bibfnamefont{A.~W.}\ \bibnamefont{Sleight}}, \bibinfo {author}
  {\bibfnamefont{M.~A.}\ \bibnamefont{Subramanian}}, \bibinfo {author}
  {\bibfnamefont{C.~L.}\ \bibnamefont{Chien}}, \bibinfo {author}
  {\bibfnamefont{M.~Z.}\ \bibnamefont{Cieplak}}, \bibinfo {author}
  {\bibfnamefont{Gang}\ \bibnamefont{Xiao}}, \bibinfo {author}
  {\bibfnamefont{V.~Y.}\ \bibnamefont{Lee}}, \bibinfo {author}
  {\bibfnamefont{B.~W.}\ \bibnamefont{Statt}}, \bibinfo {author}
  {\bibfnamefont{C.~E.}\ \bibnamefont{Stronach}}, \bibinfo {author}
  {\bibfnamefont{W.~J.}\ \bibnamefont{Kossler}},\ and\ \bibinfo {author}
  {\bibfnamefont{X.~H.}\ \bibnamefont{Yu}},\ }%
  \bibfield{title}{%
  \enquote{\bibinfo {title} {{\it Universal Correlations between ${T}_{c}$ and
  ${n}_{s}/{m}^{*}$ (Carrier Density over Effective Mass) in High-${T}_{c}$
  Cuprate Superconductors}},}\ }%
  \bibfield{journal}{%
  \Doi{10.1103/PhysRevLett.62.2317}{\bibinfo {journal} {Phys. Rev. Lett.}}\ }%
  \textbf{\bibinfo {volume} {62}},\ \bibinfo {pages} {2317--2320} (\bibinfo
  {month} {May}\ \bibinfo {year} {1989}),\
  \url{https://link.aps.org/doi/10.1103/PhysRevLett.62.2317}%
  \bibAnnoteFile{NoStop}{Uemura1989}%
\bibitem{Torrance1988}%
  \BibitemOpen
  \bibfield{author}{%
  \bibinfo {author} {\bibfnamefont{J.B.}\ \bibnamefont{Torrance}}, \bibinfo
  {author} {\bibfnamefont{Y.}~\bibnamefont{Tokura}}, \bibinfo {author}
  {\bibfnamefont{S.J.}\ \bibnamefont{LaPlaca}}, \bibinfo {author}
  {\bibfnamefont{T.C.}\ \bibnamefont{Huang}}, \bibinfo {author}
  {\bibfnamefont{R.J.}\ \bibnamefont{Savoy}},\ and\ \bibinfo {author}
  {\bibfnamefont{A.I.}\ \bibnamefont{Nazzal}},\ }%
  \bibfield{title}{%
  \enquote{\bibinfo {title} {{\it New Class of High $T_c$ Structures:
  Intergrowth of Multiple Copper Oxide Perovskite-like Layers with Double
  Sheets of {BiO}}},}\ }%
  \bibfield{journal}{%
  \Doi{https://doi.org/10.1016/0038-1098(88)90987-8}{\bibinfo {journal} {Solid
  State Communications}}\ }%
  \textbf{\bibinfo {volume} {66}},\ \bibinfo {pages} {703--706} (\bibinfo
  {year} {1988}),\ ISSN \bibinfo {issn} {0038-1098},\
  \url{https://www.sciencedirect.com/science/article/pii/0038109888909878}%
  \bibAnnoteFile{NoStop}{Torrance1988}%
\bibitem{Torardi1988}%
  \BibitemOpen
  \bibfield{author}{%
  \bibinfo {author} {\bibfnamefont{C.~C.}\ \bibnamefont{Torardi}}, \bibinfo
  {author} {\bibfnamefont{M.~A.}\ \bibnamefont{Subramanian}}, \bibinfo {author}
  {\bibfnamefont{J.~C.}\ \bibnamefont{Calabrese}}, \bibinfo {author}
  {\bibfnamefont{J.}~\bibnamefont{Gopalakrishnan}}, \bibinfo {author}
  {\bibfnamefont{E.~M.}\ \bibnamefont{McCarron}}, \bibinfo {author}
  {\bibfnamefont{K.~J.}\ \bibnamefont{Morrissey}}, \bibinfo {author}
  {\bibfnamefont{T.~R.}\ \bibnamefont{Askew}}, \bibinfo {author}
  {\bibfnamefont{R.~B.}\ \bibnamefont{Flippen}}, \bibinfo {author}
  {\bibfnamefont{U.}~\bibnamefont{Chowdhry}},\ and\ \bibinfo {author}
  {\bibfnamefont{A.~W.}\ \bibnamefont{Sleight}},\ }%
  \bibfield{title}{%
  \enquote{\bibinfo {title} {{\it Structures of the Superconducting Oxides
  Tl$_{2}$Ba$_{2}$CuO$_{6}$ and Bi$_{2}$Sr$_{2}$CuO$_{6}$}},}\ }%
  \bibfield{journal}{%
  \Doi{10.1103/PhysRevB.38.225}{\bibinfo {journal} {Phys. Rev. B}}\ }%
  \textbf{\bibinfo {volume} {38}},\ \bibinfo {pages} {225--231} (\bibinfo
  {month} {Jul}\ \bibinfo {year} {1988}),\
  \url{https://link.aps.org/doi/10.1103/PhysRevB.38.225}%
  \bibAnnoteFile{NoStop}{Torardi1988}%
\bibitem{Ito1998}%
  \BibitemOpen
  \bibfield{author}{%
  \bibinfo {author} {\bibfnamefont{Yoshiaki}\ \bibnamefont{Ito}}, \bibinfo
  {author} {\bibfnamefont{Aurel-Mihai}\ \bibnamefont{Vlaicu}}, \bibinfo
  {author} {\bibfnamefont{Takeshi}\ \bibnamefont{Mukoyama}}, \bibinfo {author}
  {\bibfnamefont{Shoichi}\ \bibnamefont{Sato}}, \bibinfo {author}
  {\bibfnamefont{Sinzo}\ \bibnamefont{Yoshikado}}, \bibinfo {author}
  {\bibfnamefont{Cristian}\ \bibnamefont{Julien}}, \bibinfo {author}
  {\bibfnamefont{Iksu}\ \bibnamefont{Chong}}, \bibinfo {author}
  {\bibfnamefont{Yasunori}\ \bibnamefont{Ikeda}}, \bibinfo {author}
  {\bibfnamefont{Mikio}\ \bibnamefont{Takano}},\ and\ \bibinfo {author}
  {\bibfnamefont{Evgeny~Ya.}\ \bibnamefont{Sherman}},\ }%
  \bibfield{title}{%
  \enquote{\bibinfo {title} {{\it Detailed Structure of a Pb-doped
  Bi$_2$Sr$_2$CuO$_{6}$ Superconductor}},}\ }%
  \bibfield{journal}{%
  \Doi{10.1103/PhysRevB.58.2851}{\bibinfo {journal} {Phys. Rev. B}}\ }%
  \textbf{\bibinfo {volume} {58}},\ \bibinfo {pages} {2851--2858} (\bibinfo
  {month} {Aug}\ \bibinfo {year} {1998}),\
  \url{https://link.aps.org/doi/10.1103/PhysRevB.58.2851}%
  \bibAnnoteFile{NoStop}{Ito1998}%
\bibitem{Schloegl1993}%
  \BibitemOpen
  \bibfield{author}{%
  \bibinfo {author} {\bibfnamefont{A.E.}\ \bibnamefont{Schlgl}}, \bibinfo
  {author} {\bibfnamefont{J.J.}\ \bibnamefont{Neumeier}}, \bibinfo {author}
  {\bibfnamefont{J.}~\bibnamefont{Diederichs}}, \bibinfo {author}
  {\bibfnamefont{C.}~\bibnamefont{Allgeier}},\ and\ \bibinfo {author}
  {\bibfnamefont{J.S.}\ \bibnamefont{Schilling}},\ }%
  \bibfield{title}{%
  \enquote{\bibinfo {title} {{\it Transport, Structuralm and Magnetic
  Properties of the Single-Copper-Oxygen Layer Bi$_2$Sr$_{2}$La$_x$CuO$_y$
  System}},}\ }%
  \bibfield{journal}{%
  \Doi{https://doi.org/10.1016/0921-4534(93)90084-4}{\bibinfo {journal}
  {Physica C: Superconductivity}}\ }%
  \textbf{\bibinfo {volume} {216}},\ \bibinfo {pages} {417--431} (\bibinfo
  {year} {1993}),\ ISSN \bibinfo {issn} {0921-4534},\
  \url{https://www.sciencedirect.com/science/article/pii/0921453493900844}%
  \bibAnnoteFile{NoStop}{Schloegl1993}%
\bibitem{Beskrovnyi1990}%
  \BibitemOpen
  \bibfield{author}{%
  \bibinfo {author} {\bibfnamefont{AI}~\bibnamefont{Beskrovnyi}}, \bibinfo
  {author} {\bibfnamefont{M}~\bibnamefont{Dlouh{\'a}}}, \bibinfo {author}
  {\bibfnamefont{Z}~\bibnamefont{Jir{\'a}k}},\ and\ \bibinfo {author}
  {\bibfnamefont{S}~\bibnamefont{Vratislav}},\ }%
  \bibfield{title}{%
  \enquote{\bibinfo {title} {{\it Study of the Modulated Structure of
  Bi$_2$(Sr, Ca)$_3$Cu$_2$O$_{8+\gamma}$ in the Range 8--920 K}},}\ }%
  \bibfield{journal}{%
  \Doi{https://doi.org/10.1016/0921-4534(90)90450-S}{\bibinfo {journal}
  {Physica C: Superconductivity}}\ }%
  \textbf{\bibinfo {volume} {171}},\ \bibinfo {pages} {19--24} (\bibinfo {year}
  {1990}),\ ISSN \bibinfo {issn} {0921-4534},\
  \url{https://www.sciencedirect.com/science/article/pii/092145349090450S}%
  \bibAnnoteFile{NoStop}{Beskrovnyi1990}%
\bibitem{Cicco1993}%
  \BibitemOpen
  \bibfield{author}{%
  \bibinfo {author} {\bibfnamefont{Andrea}\ \bibnamefont{{di Cicco}}}\ and\
  \bibinfo {author} {\bibfnamefont{Mario}\ \bibnamefont{Berrettoni}},\ }%
  \bibfield{title}{%
  \enquote{\bibinfo {title} {{\it X-ray Absorption Multiple-Scattering Study of
  Angle Distribution in High-Tc Superconductors}},}\ }%
  \bibfield{journal}{%
  \Doi{https://doi.org/10.1016/0375-9601(93)90936-T}{\bibinfo {journal}
  {Physics Letters A}}\ }%
  \textbf{\bibinfo {volume} {176}},\ \bibinfo {pages} {375--381} (\bibinfo
  {year} {1993}),\ ISSN \bibinfo {issn} {0375-9601},\
  \url{https://www.sciencedirect.com/science/article/pii/037596019390936T}%
  \bibAnnoteFile{NoStop}{Cicco1993}%
\bibitem{yamamoto1990}%
  \BibitemOpen
  \bibfield{author}{%
  \bibinfo {author} {\bibfnamefont{Akiji}\ \bibnamefont{Yamamoto}}, \bibinfo
  {author} {\bibfnamefont{Mitsuko}\ \bibnamefont{Onoda}}, \bibinfo {author}
  {\bibfnamefont{Eiji}\ \bibnamefont{Takayama-Muromachi}}, \bibinfo {author}
  {\bibfnamefont{Fujio}\ \bibnamefont{Izumi}}, \bibinfo {author}
  {\bibfnamefont{Toru}\ \bibnamefont{Ishigaki}},\ and\ \bibinfo {author}
  {\bibfnamefont{Hajime}\ \bibnamefont{Asano}},\ }%
  \bibfield{title}{%
  \enquote{\bibinfo {title} {{\it Rietveld Analysis of the Modulated Structure
  in the Superconducting Oxide
  ${\mathrm{Bi}}_{2}$(Sr,Ca${)}_{3}$${\mathrm{Cu}}_{2}$${\mathrm{O}}_{8+\mathit{x}}$}},}\
  }%
  \bibfield{journal}{%
  \Doi{10.1103/PhysRevB.42.4228}{\bibinfo {journal} {Phys. Rev. B}}\ }%
  \textbf{\bibinfo {volume} {42}},\ \bibinfo {pages} {4228--4239} (\bibinfo
  {month} {Sep}\ \bibinfo {year} {1990}),\
  \url{https://link.aps.org/doi/10.1103/PhysRevB.42.4228}%
  \bibAnnoteFile{NoStop}{yamamoto1990}%
\bibitem{slezak2008}%
  \BibitemOpen
  \bibfield{author}{%
  \bibinfo {author} {\bibfnamefont{J.~A.}\ \bibnamefont{Slezak}}, \bibinfo
  {author} {\bibfnamefont{Jinho}\ \bibnamefont{Lee}}, \bibinfo {author}
  {\bibfnamefont{M.}~\bibnamefont{Wang}}, \bibinfo {author}
  {\bibfnamefont{K.}~\bibnamefont{McElroy}}, \bibinfo {author}
  {\bibfnamefont{K.}~\bibnamefont{Fujita}}, \bibinfo {author}
  {\bibfnamefont{B.~M.}\ \bibnamefont{Andersen}}, \bibinfo {author}
  {\bibfnamefont{P.~J.}\ \bibnamefont{Hirschfeld}}, \bibinfo {author}
  {\bibfnamefont{H.}~\bibnamefont{Eisaki}}, \bibinfo {author}
  {\bibfnamefont{S.}~\bibnamefont{Uchida}},\ and\ \bibinfo {author}
  {\bibfnamefont{J.~C.}\ \bibnamefont{Davis}},\ }%
  \bibfield{title}{%
  \enquote{\bibinfo {title} {{\it Imaging the Impact on Cuprate
  Superconductivity of Varying the Interatomic Distances within Individual
  Crystal Unit Cells}},}\ }%
  \bibfield{journal}{%
  \Doi{10.1073/pnas.0706795105}{\bibinfo {journal} {Proceedings of the National
  Academy of Sciences}}\ }%
  \textbf{\bibinfo {volume} {105}},\ \bibinfo {pages} {3203--3208} (\bibinfo
  {year} {2008}),\ \url{https://www.pnas.org/doi/abs/10.1073/pnas.0706795105}%
  \bibAnnoteFile{NoStop}{slezak2008}%
\bibitem{iwano2022}%
  \BibitemOpen
  \bibfield{author}{%
  \bibinfo {author} {\bibfnamefont{Akito}\ \bibnamefont{Iwano}}\ and\ \bibinfo
  {author} {\bibfnamefont{Youhei}\ \bibnamefont{Yamaji}},\ }%
  \bibfield{title}{%
  \enquote{\bibinfo {title} {{\it Superconductivity in Bilayer $t$-$t'$
  Hubbard Models}},}\ }%
  \bibfield{journal}{%
  \Doi{10.7566/JPSJ.91.094702}{\bibinfo {journal} {J. Phys. Soc. Jpn.}}\ }%
  \textbf{\bibinfo {volume} {91}},\ \bibinfo {pages} {094702} (\bibinfo {year}
  {2022}),\ \url{https://doi.org/10.7566/JPSJ.91.094702}%
  \bibAnnoteFile{NoStop}{iwano2022}%
\bibitem{imada_suzuki2019}%
  \BibitemOpen
  \bibfield{author}{%
  \bibinfo {author} {\bibfnamefont{Masatoshi}\ \bibnamefont{Imada}}\ and\
  \bibinfo {author} {\bibfnamefont{Takafumi~J.}\ \bibnamefont{Suzuki}},\ }%
  \bibfield{title}{%
  \enquote{\bibinfo {title} {{\it Excitons and Dark Fermions as Origins of Mott
  Gap, Pseudogap and Superconductivity in Cuprate Superconductors -General
  Concept and Basic Formalism Based on Gap Physics}},}\ }%
  \bibfield{journal}{%
  \Doi{10.7566/JPSJ.88.024701}{\bibinfo {journal} {J. Phys. Soc. Jpn.}}\ }%
  \textbf{\bibinfo {volume} {88}},\ \bibinfo {pages} {024701} (\bibinfo {year}
  {2019}),\ \url{https://doi.org/10.7566/JPSJ.88.024701}%
  \bibAnnoteFile{NoStop}{imada_suzuki2019}%
\bibitem{imada_review2021}%
  \BibitemOpen
  \bibfield{author}{%
  \bibinfo {author} {\bibfnamefont{Masatoshi}\ \bibnamefont{Imada}},\ }%
  \bibfield{title}{%
  \enquote{\bibinfo {title} {{\it Charge Order and Superconductivity as
  Competing Brothers in Cuprate High-Tc Superconductors}},}\ }%
  \bibfield{journal}{%
  \Doi{10.7566/JPSJ.90.111009}{\bibinfo {journal} {J. Phys. Soc. Jpn.}}\ }%
  \textbf{\bibinfo {volume} {90}},\ \bibinfo {pages} {111009} (\bibinfo {year}
  {2021}),\ \url{https://doi.org/10.7566/JPSJ.90.111009}%
  \bibAnnoteFile{NoStop}{imada_review2021}%
\bibitem{PhysRevResearch.3.043099}%
  \BibitemOpen
  \bibfield{author}{%
  \bibinfo {author} {\bibfnamefont{Youhei}\ \bibnamefont{Yamaji}}, \bibinfo
  {author} {\bibfnamefont{Teppei}\ \bibnamefont{Yoshida}}, \bibinfo {author}
  {\bibfnamefont{Atsushi}\ \bibnamefont{Fujimori}},\ and\ \bibinfo {author}
  {\bibfnamefont{Masatoshi}\ \bibnamefont{Imada}},\ }%
  \bibfield{title}{%
  \enquote{\bibinfo {title} {{\it Hidden Self-Energies as Origin of Cuprate
  Superconductivity Revealed by Machine Learning}},}\ }%
  \bibfield{journal}{%
  \Doi{10.1103/PhysRevResearch.3.043099}{\bibinfo {journal} {Phys. Rev. Res.}}\
  }%
  \textbf{\bibinfo {volume} {3}},\ \bibinfo {pages} {043099} (\bibinfo {month}
  {Nov}\ \bibinfo {year} {2021}),\
  \url{https://link.aps.org/doi/10.1103/PhysRevResearch.3.043099}%
  \bibAnnoteFile{NoStop}{PhysRevResearch.3.043099}%
\bibitem{pavarini2001}%
  \BibitemOpen
  \bibfield{author}{%
  \bibinfo {author} {\bibfnamefont{E.}~\bibnamefont{Pavarini}}, \bibinfo
  {author} {\bibfnamefont{I.}~\bibnamefont{Dasgupta}}, \bibinfo {author}
  {\bibfnamefont{T.}~\bibnamefont{Saha-Dasgupta}}, \bibinfo {author}
  {\bibfnamefont{O.}~\bibnamefont{Jepsen}},\ and\ \bibinfo {author}
  {\bibfnamefont{O.~K.}\ \bibnamefont{Andersen}},\ }%
  \bibfield{title}{%
  \enquote{\bibinfo {title} {{\it Band-Structure Trend in Hole-Doped Cuprates
  and Correlation with $T_{c{\rm max}}$}},}\ }%
  \bibfield{journal}{%
  \Doi{10.1103/PhysRevLett.87.047003}{\bibinfo {journal} {Phys. Rev. Lett.}}\
  }%
  \textbf{\bibinfo {volume} {87}},\ \bibinfo {pages} {047003} (\bibinfo {month}
  {Jul}\ \bibinfo {year} {2001}),\
  \url{https://link.aps.org/doi/10.1103/PhysRevLett.87.047003}%
  \bibAnnoteFile{NoStop}{pavarini2001}%
\bibitem{mori2008}%
  \BibitemOpen
  \bibfield{author}{%
  \bibinfo {author} {\bibfnamefont{Michiyasu}\ \bibnamefont{Mori}}, \bibinfo
  {author} {\bibfnamefont{Giniyat}\ \bibnamefont{Khaliullin}}, \bibinfo
  {author} {\bibfnamefont{Takami}\ \bibnamefont{Tohyama}},\ and\ \bibinfo
  {author} {\bibfnamefont{Sadamichi}\ \bibnamefont{Maekawa}},\ }%
  \bibfield{title}{%
  \enquote{\bibinfo {title} {{\it Origin of the Spatial Variation of the
  Pairing Gap in Bi-Based High Temperature Cuprate Superconductors}},}\ }%
  \bibfield{journal}{%
  \Doi{10.1103/PhysRevLett.101.247003}{\bibinfo {journal} {Phys. Rev. Lett.}}\
  }%
  \textbf{\bibinfo {volume} {101}},\ \bibinfo {pages} {247003} (\bibinfo
  {month} {Dec}\ \bibinfo {year} {2008}),\
  \url{https://link.aps.org/doi/10.1103/PhysRevLett.101.247003}%
  \bibAnnoteFile{NoStop}{mori2008}%
\bibitem{weber2010}%
  \BibitemOpen
  \bibfield{author}{%
  \bibinfo {author} {\bibfnamefont{C\'edric}\ \bibnamefont{Weber}}, \bibinfo
  {author} {\bibfnamefont{Kristjan}\ \bibnamefont{Haule}},\ and\ \bibinfo
  {author} {\bibfnamefont{Gabriel}\ \bibnamefont{Kotliar}},\ }%
  \bibfield{title}{%
  \enquote{\bibinfo {title} {{\it Apical Oxygens and Correlation Strength in
  Electron- and Hole-Doped Copper Oxides}},}\ }%
  \bibfield{journal}{%
  \Doi{10.1103/PhysRevB.82.125107}{\bibinfo {journal} {Phys. Rev. B}}\ }%
  \textbf{\bibinfo {volume} {82}},\ \bibinfo {pages} {125107} (\bibinfo {month}
  {Sep}\ \bibinfo {year} {2010}),\
  \url{https://link.aps.org/doi/10.1103/PhysRevB.82.125107}%
  \bibAnnoteFile{NoStop}{weber2010}%
\bibitem{sakakibara2010}%
  \BibitemOpen
  \bibfield{author}{%
  \bibinfo {author} {\bibfnamefont{Hirofumi}\ \bibnamefont{Sakakibara}},
  \bibinfo {author} {\bibfnamefont{Hidetomo}\ \bibnamefont{Usui}}, \bibinfo
  {author} {\bibfnamefont{Kazuhiko}\ \bibnamefont{Kuroki}}, \bibinfo {author}
  {\bibfnamefont{Ryotaro}\ \bibnamefont{Arita}},\ and\ \bibinfo {author}
  {\bibfnamefont{Hideo}\ \bibnamefont{Aoki}},\ }%
  \bibfield{title}{%
  \enquote{\bibinfo {title} {{\it Two-Orbital Model Explains the Higher
  Transition Temperature of the Single-Layer Hg-Cuprate Superconductor Compared
  to That of the La-Cuprate Superconductor}},}\ }%
  \bibfield{journal}{%
  \Doi{10.1103/PhysRevLett.105.057003}{\bibinfo {journal} {Phys. Rev. Lett.}}\
  }%
  \textbf{\bibinfo {volume} {105}},\ \bibinfo {pages} {057003} (\bibinfo
  {month} {Jul}\ \bibinfo {year} {2010}),\
  \url{https://link.aps.org/doi/10.1103/PhysRevLett.105.057003}%
  \bibAnnoteFile{NoStop}{sakakibara2010}%
\bibitem{hu2014}%
  \BibitemOpen
  \bibfield{author}{%
  \bibinfo {author} {\bibfnamefont{Wanzheng}\ \bibnamefont{Hu}}, \bibinfo
  {author} {\bibfnamefont{Stefan}\ \bibnamefont{Kaiser}}, \bibinfo {author}
  {\bibfnamefont{Daniele}\ \bibnamefont{Nicoletti}}, \bibinfo {author}
  {\bibfnamefont{Cassandra~R}\ \bibnamefont{Hunt}}, \bibinfo {author}
  {\bibfnamefont{Isabella}\ \bibnamefont{Gierz}}, \bibinfo {author}
  {\bibfnamefont{Matthias~C}\ \bibnamefont{Hoffmann}}, \bibinfo {author}
  {\bibfnamefont{M}~\bibnamefont{Le~Tacon}}, \bibinfo {author}
  {\bibfnamefont{T}~\bibnamefont{Loew}}, \bibinfo {author}
  {\bibfnamefont{B}~\bibnamefont{Keimer}},\ and\ \bibinfo {author}
  {\bibfnamefont{Andrea}\ \bibnamefont{Cavalleri}},\ }%
  \bibfield{title}{%
  \enquote{\bibinfo {title} {{\it Optically Enhanced Coherent Transport in
  YBa$_2$Cu$_3$O$_{6.5}$ by Ultrafast Redistribution of Interlayer
  Coupling}},}\ }%
  \bibfield{journal}{%
  \Doi{10.1038/nmat3963}{\bibinfo {journal} {Nature Mater.}}\ }%
  \textbf{\bibinfo {volume} {13}},\ \bibinfo {pages} {705--711} (\bibinfo
  {month} {July}\ \bibinfo {year} {2014}),\
  \url{https://doi.org/10.1038/nmat3963}%
  \bibAnnoteFile{NoStop}{hu2014}%
\bibitem{Himeda2002}%
  \BibitemOpen
  \bibfield{author}{%
  \bibinfo {author} {\bibfnamefont{A.}~\bibnamefont{Himeda}}, \bibinfo {author}
  {\bibfnamefont{T.}~\bibnamefont{Kato}},\ and\ \bibinfo {author}
  {\bibfnamefont{M.}~\bibnamefont{Ogata}},\ }%
  \bibfield{title}{%
  \enquote{\bibinfo {title} {{\it Stripe States with Spatially Oscillating
  $\mathit{d}$-Wave Superconductivity in the Two-Dimensional
  $\mathit{t}$-${\mathit{t}}^{\ensuremath{'}}$-$\mathit{J}$ Model}},}\ }%
  \bibfield{journal}{%
  \Doi{10.1103/PhysRevLett.88.117001}{\bibinfo {journal} {Phys. Rev. Lett.}}\
  }%
  \textbf{\bibinfo {volume} {88}},\ \bibinfo {pages} {117001} (\bibinfo {month}
  {Feb}\ \bibinfo {year} {2002}),\
  \url{https://link.aps.org/doi/10.1103/PhysRevLett.88.117001}%
  \bibAnnoteFile{NoStop}{Himeda2002}%
\bibitem{carleo2017}%
  \BibitemOpen
  \bibfield{author}{%
  \bibinfo {author} {\bibfnamefont{Giuseppe}\ \bibnamefont{Carleo}}\ and\
  \bibinfo {author} {\bibfnamefont{Matthias}\ \bibnamefont{Troyer}},\ }%
  \bibfield{title}{%
  \enquote{\bibinfo {title} {{\it Solving the Quantum Many-Body Problem with
  Artificial Neural Networks}},}\ }%
  \bibfield{journal}{%
  \Doi{10.1126/science.aag2302}{\bibinfo {journal} {Science}}\ }%
  \textbf{\bibinfo {volume} {355}},\ \bibinfo {pages} {602--606} (\bibinfo
  {year} {2017}),\
  \url{https://www.science.org/doi/abs/10.1126/science.aag2302}%
  \bibAnnoteFile{NoStop}{carleo2017}%
\bibitem{Ponsioen2019}%
  \BibitemOpen
  \bibfield{author}{%
  \bibinfo {author} {\bibfnamefont{Boris}\ \bibnamefont{Ponsioen}}, \bibinfo
  {author} {\bibfnamefont{Sangwoo~S.}\ \bibnamefont{Chung}},\ and\ \bibinfo
  {author} {\bibfnamefont{Philippe}\ \bibnamefont{Corboz}},\ }%
  \bibfield{title}{%
  \enquote{\bibinfo {title} {{\it Period 4 Stripe in the Extended
  Two-Dimensional Hubbard Model}},}\ }%
  \bibfield{journal}{%
  \Doi{10.1103/PhysRevB.100.195141}{\bibinfo {journal} {Phys. Rev. B}}\ }%
  \textbf{\bibinfo {volume} {100}},\ \bibinfo {pages} {195141} (\bibinfo
  {month} {Nov}\ \bibinfo {year} {2019}),\
  \url{https://link.aps.org/doi/10.1103/PhysRevB.100.195141}%
  \bibAnnoteFile{NoStop}{Ponsioen2019}%
\bibitem{Tocchio2019}%
  \BibitemOpen
  \bibfield{author}{%
  \bibinfo {author} {\bibfnamefont{Luca~F.}\ \bibnamefont{Tocchio}}, \bibinfo
  {author} {\bibfnamefont{Arianna}\ \bibnamefont{Montorsi}},\ and\ \bibinfo
  {author} {\bibfnamefont{Federico}\ \bibnamefont{Becca}},\ }%
  \bibfield{title}{%
  \enquote{\bibinfo {title} {{\it Metallic and Insulating Stripes and Their
  Relation with Superconductivity in the Doped Hubbard Model}},}\ }%
  \bibfield{journal}{%
  \Doi{10.21468/SciPostPhys.7.2.021}{\bibinfo {journal} {SciPost Phys.}}\ }%
  \textbf{\bibinfo {volume} {7}},\ \bibinfo {pages} {021} (\bibinfo {year}
  {2019}),\ \url{https://scipost.org/10.21468/SciPostPhys.7.2.021}%
  \bibAnnoteFile{NoStop}{Tocchio2019}%
\bibitem{Qin2020}%
  \BibitemOpen
  \bibfield{author}{%
  \bibinfo {author} {\bibfnamefont{Mingpu}\ \bibnamefont{Qin}}, \bibinfo
  {author} {\bibfnamefont{Chia-Min}\ \bibnamefont{Chung}}, \bibinfo {author}
  {\bibfnamefont{Hao}\ \bibnamefont{Shi}}, \bibinfo {author}
  {\bibfnamefont{Ettore}\ \bibnamefont{Vitali}}, \bibinfo {author}
  {\bibfnamefont{Claudius}\ \bibnamefont{Hubig}}, \bibinfo {author}
  {\bibfnamefont{Ulrich}\ \bibnamefont{Schollw\"ock}}, \bibinfo {author}
  {\bibfnamefont{Steven~R.}\ \bibnamefont{White}},\ and\ \bibinfo {author}
  {\bibfnamefont{Shiwei}\ \bibnamefont{Zhang}} (\bibinfo {collaboration}
  {Simons Collaboration on the Many-Electron Problem}),\ }%
  \bibfield{title}{%
  \enquote{\bibinfo {title} {{\it Absence of Superconductivity in the Pure
  Two-Dimensional Hubbard Model}},}\ }%
  \bibfield{journal}{%
  \Doi{10.1103/PhysRevX.10.031016}{\bibinfo {journal} {Phys. Rev. X}}\ }%
  \textbf{\bibinfo {volume} {10}},\ \bibinfo {pages} {031016} (\bibinfo {month}
  {Jul}\ \bibinfo {year} {2020}),\
  \url{https://link.aps.org/doi/10.1103/PhysRevX.10.031016}%
  \bibAnnoteFile{NoStop}{Qin2020}%
\bibitem{Sorella2021}%
  \BibitemOpen
  \bibfield{author}{%
  \bibinfo {author} {\bibfnamefont{Sandro}\ \bibnamefont{Sorella}},\ }%
  \enquote{\bibinfo {title} {{\it The Phase Diagram of the Hubbard Model by
  Variational Auxiliary Field Quantum Monte Carlo}},}\  (\bibinfo {year}
  {2021}),\ \Eprint{http://arxiv.org/abs/2101.07045}{arXiv:2101.07045
  [cond-mat.str-el]}%
  \bibAnnoteFile{NoStop}{Sorella2021}%
\bibitem{Marino2022}%
  \BibitemOpen
  \bibfield{author}{%
  \bibinfo {author} {\bibfnamefont{Vito}\ \bibnamefont{Marino}}, \bibinfo
  {author} {\bibfnamefont{Federico}\ \bibnamefont{Becca}},\ and\ \bibinfo
  {author} {\bibfnamefont{Luca~F.}\ \bibnamefont{Tocchio}},\ }%
  \bibfield{title}{%
  \enquote{\bibinfo {title} {{\it Stripes in the Extended $t$-$t^\prime$
  Hubbard Model: A Variational Monte Carlo Analysis}},}\ }%
  \bibfield{journal}{%
  \Doi{10.21468/SciPostPhys.12.6.180}{\bibinfo {journal} {SciPost Phys.}}\ }%
  \textbf{\bibinfo {volume} {12}},\ \bibinfo {pages} {180} (\bibinfo {year}
  {2022}),\ \url{https://scipost.org/10.21468/SciPostPhys.12.6.180}%
  \bibAnnoteFile{NoStop}{Marino2022}%
\bibitem{Xu2023}%
  \BibitemOpen
  \bibfield{author}{%
  \bibinfo {author} {\bibfnamefont{H.}~\bibnamefont{Xu}}, \bibinfo {author}
  {\bibfnamefont{C.-M.}\ \bibnamefont{Chung}}, \bibinfo {author}
  {\bibfnamefont{M.}~\bibnamefont{Qin}}, \bibinfo {author}
  {\bibnamefont{Schollwöck}}, \bibinfo {author} {\bibfnamefont{S.~R.}\
  \bibnamefont{White}},\ and\ \bibinfo {author} {\bibfnamefont{Shiwei}\
  \bibnamefont{Zhang}},\ }%
  \enquote{\bibinfo {title} {{\it Coexistence of Superconductivity with
  Partially Filled Stripes in the Hubbard Model}},}\  (\bibinfo {year}
  {2023}),\ \Eprint{http://arxiv.org/abs/2303.08376}{arXiv:2303.08376
  [cond-mat.str-el]}%
  \bibAnnoteFile{NoStop}{Xu2023}%
\bibitem{Wu2023}%
  \BibitemOpen
  \bibfield{author}{%
  \bibinfo {author} {\bibfnamefont{Dian}\ \bibnamefont{Wu}}\ and\ \bibinfo
  {author} {\bibnamefont{{\it et al.}}},\ }%
  \enquote{\bibinfo {title} {{\it Variational Benchmarks for Quantum Many-Body
  Problems}},}\  (\bibinfo {year} {2023}),\
  \Eprint{http://arxiv.org/abs/2302.04919}{arXiv:2302.04919 [cond-mat.str-el]}%
  \bibAnnoteFile{NoStop}{Wu2023}%
\bibitem{Panagopoulos1999}%
  \BibitemOpen
  \bibfield{author}{%
  \bibinfo {author} {\bibfnamefont{C.}~\bibnamefont{Panagopoulos}}, \bibinfo
  {author} {\bibfnamefont{B.~D.}\ \bibnamefont{Rainford}}, \bibinfo {author}
  {\bibfnamefont{J.~R.}\ \bibnamefont{Cooper}}, \bibinfo {author}
  {\bibfnamefont{W.}~\bibnamefont{Lo}}, \bibinfo {author}
  {\bibfnamefont{J.~L.}\ \bibnamefont{Tallon}}, \bibinfo {author}
  {\bibfnamefont{J.~W.}\ \bibnamefont{Loram}}, \bibinfo {author}
  {\bibfnamefont{J.}~\bibnamefont{Betouras}}, \bibinfo {author}
  {\bibfnamefont{Y.~S.}\ \bibnamefont{Wang}},\ and\ \bibinfo {author}
  {\bibfnamefont{C.~W.}\ \bibnamefont{Chu}},\ }%
  \bibfield{title}{%
  \enquote{\bibinfo {title} {{\it Effects of Carrier Concentration on the
  Superfluid Density of High-${T}_{c}$ Cuprates}},}\ }%
  \bibfield{journal}{%
  \Doi{10.1103/PhysRevB.60.14617}{\bibinfo {journal} {Phys. Rev. B}}\ }%
  \textbf{\bibinfo {volume} {60}},\ \bibinfo {pages} {14617--14620} (\bibinfo
  {month} {Dec}\ \bibinfo {year} {1999}),\
  \url{https://link.aps.org/doi/10.1103/PhysRevB.60.14617}%
  \bibAnnoteFile{NoStop}{Panagopoulos1999}%
\bibitem{Kondo2011}%
  \BibitemOpen
  \bibfield{author}{%
  \bibinfo {author} {\bibfnamefont{Takeshi}\ \bibnamefont{Kondo}}, \bibinfo
  {author} {\bibfnamefont{Yoichiro}\ \bibnamefont{Hamaya}}, \bibinfo {author}
  {\bibfnamefont{Ari~D}\ \bibnamefont{Palczewski}}, \bibinfo {author}
  {\bibfnamefont{Tsunehiro}\ \bibnamefont{Takeuchi}}, \bibinfo {author}
  {\bibfnamefont{JS}~\bibnamefont{Wen}}, \bibinfo {author}
  {\bibfnamefont{ZJ}~\bibnamefont{Xu}}, \bibinfo {author}
  {\bibfnamefont{Genda}\ \bibnamefont{Gu}}, \bibinfo {author}
  {\bibfnamefont{J{\"o}rg}\ \bibnamefont{Schmalian}},\ and\ \bibinfo {author}
  {\bibfnamefont{Adam}\ \bibnamefont{Kaminski}},\ }%
  \bibfield{title}{%
  \enquote{\bibinfo {title} {{\it Disentangling Cooper-Pair Formation above the
  Transition Temperature from the Pseudogap State in the Cuprates}},}\ }%
  \bibfield{journal}{%
  \Doi{10.1038/nphys1851}{\bibinfo {journal} {Nature Physics}}\ }%
  \textbf{\bibinfo {volume} {7}},\ \bibinfo {pages} {21} (\bibinfo {year}
  {2011}),\ \url{https://www.nature.com/articles/nphys1851}%
  \bibAnnoteFile{NoStop}{Kondo2011}%
\bibitem{LeTacon2006}%
  \BibitemOpen
  \bibfield{author}{%
  \bibinfo {author} {\bibfnamefont{M}~\bibnamefont{Le~Tacon}}, \bibinfo
  {author} {\bibfnamefont{A}~\bibnamefont{Sacuto}}, \bibinfo {author}
  {\bibfnamefont{A}~\bibnamefont{Georges}}, \bibinfo {author}
  {\bibfnamefont{G}~\bibnamefont{Kotliar}}, \bibinfo {author}
  {\bibfnamefont{Y}~\bibnamefont{Gallais}}, \bibinfo {author}
  {\bibfnamefont{D}~\bibnamefont{Colson}},\ and\ \bibinfo {author}
  {\bibfnamefont{A}~\bibnamefont{Forget}},\ }%
  \bibfield{title}{%
  \enquote{\bibinfo {title} {{\it Two Energy Scales and Two Distinct
  Quasiparticle Dynamics in the Superconducting State of Underdoped
  Cuprates}},}\ }%
  \bibfield{journal}{%
  \Doi{10.1038/nphys362}{\bibinfo {journal} {Nature Physics}}\ }%
  \textbf{\bibinfo {volume} {2}},\ \bibinfo {pages} {537--543} (\bibinfo {year}
  {2006}),\ \url{https://doi.org/10.1038/nphys362}%
  \bibAnnoteFile{NoStop}{LeTacon2006}%
\bibitem{Teranishi2018}%
  \BibitemOpen
  \bibfield{author}{%
  \bibinfo {author} {\bibfnamefont{Shingo}\ \bibnamefont{Teranishi}}, \bibinfo
  {author} {\bibfnamefont{Kazutaka}\ \bibnamefont{Nishiguchi}},\ and\ \bibinfo
  {author} {\bibfnamefont{Koichi}\ \bibnamefont{Kusakabe}},\ }%
  \bibfield{title}{%
  \enquote{\bibinfo {title} {{\it Material-Dependent Screening of Coulomb
  Interaction in Single-Layer Cuprates}},}\ }%
  \bibfield{journal}{%
  \Doi{10.7566/JPSJ.87.114701}{\bibinfo {journal} {J. Phys. Soc. Jpn.}}\ }%
  \textbf{\bibinfo {volume} {87}},\ \bibinfo {pages} {114701} (\bibinfo {year}
  {2018}),\ \url{https://doi.org/10.7566/JPSJ.87.114701}%
  \bibAnnoteFile{NoStop}{Teranishi2018}%
\bibitem{Jang2019}%
  \BibitemOpen
  \bibfield{author}{%
  \bibinfo {author} {\bibfnamefont{Seung~Woo}\ \bibnamefont{Jang}}, \bibinfo
  {author} {\bibfnamefont{Hirofumi}\ \bibnamefont{Sakakibara}}, \bibinfo
  {author} {\bibfnamefont{Hiori}\ \bibnamefont{Kino}}, \bibinfo {author}
  {\bibfnamefont{Takao}\ \bibnamefont{Kotani}}, \bibinfo {author}
  {\bibfnamefont{Kazuhiko}\ \bibnamefont{Kuroki}},\ and\ \bibinfo {author}
  {\bibfnamefont{Myung~Joon}\ \bibnamefont{Han}},\ }%
  \bibfield{title}{%
  \enquote{\bibinfo {title} {{\it Direct Theoretical Evidence for Weaker
  Correlations in Electron-Doped and Hg-Based Hole-Doped Cuprates}},}\ }%
  \bibfield{journal}{%
  \Doi{10.1038/srep33397}{\bibinfo {journal} {Sci. Rep.}}\ }%
  \textbf{\bibinfo {volume} {6}},\ \bibinfo {pages} {33397} (\bibinfo {year}
  {2019}),\ \url{https://doi.org/10.1038/srep33397}%
  \bibAnnoteFile{NoStop}{Jang2019}%
\bibitem{Kowalski2021}%
  \BibitemOpen
  \bibfield{author}{%
  \bibinfo {author} {\bibfnamefont{Nicolas}\ \bibnamefont{Kowalski}}, \bibinfo
  {author} {\bibfnamefont{Sidhartha}\ \bibnamefont{Dash}}, \bibinfo {author}
  {\bibfnamefont{Patrick~Sémon}\ \bibnamefont{Sémon}},\ and\ \bibinfo
  {author} {\bibfnamefont{André-Marie}\ \bibnamefont{Tremblay}},\ }%
  \bibfield{title}{%
  \enquote{\bibinfo {title} {{\it Oxygen Hole Content, Charge-Transfer Gap,
  Covalency, and Cuprate Superconductivity}},}\ }%
  \bibfield{journal}{%
  \Doi{10.1073/pnas.2106476118}{\bibinfo {journal} {Proc. Natl. Acad. Sci.}}\
  }%
  \textbf{\bibinfo {volume} {118}},\ \bibinfo {pages} {e2106476118} (\bibinfo
  {year} {2021}),\ \url{https://doi.org/10.1073/pnas.2106476118}%
  \bibAnnoteFile{NoStop}{Kowalski2021}%
\bibitem{Cui2022}%
  \BibitemOpen
  \bibfield{author}{%
  \bibinfo {author} {\bibfnamefont{Z.~H.}\ \bibnamefont{Cui}}, \bibinfo
  {author} {\bibfnamefont{H.}~\bibnamefont{Zhai}}, \bibinfo {author}
  {\bibfnamefont{X.}~\bibnamefont{Zhang}},\ and\ \bibinfo {author}
  {\bibfnamefont{Garnet Kin-Lic}\ \bibnamefont{Chan}},\ }%
  \bibfield{title}{%
  \enquote{\bibinfo {title} {{\it Systematic Electronic Structure in the
  Cuprate Parent State from Quantum Many-Body Simulations}},}\ }%
  \bibfield{journal}{%
  \Doi{10.1126/science.abm2295}{\bibinfo {journal} {Science}}\ }%
  \textbf{\bibinfo {volume} {377}},\ \bibinfo {pages} {1192--1198} (\bibinfo
  {year} {2022}),\ \url{https://www.science.org/doi/10.1126/science.abm2295}%
  \bibAnnoteFile{NoStop}{Cui2022}%
\bibitem{Cui2023}%
  \BibitemOpen
  \bibfield{author}{%
  \bibinfo {author} {\bibfnamefont{Z.~H.}\ \bibnamefont{Cui}}, \bibinfo
  {author} {\bibfnamefont{J.}~\bibnamefont{Yang}}, \bibinfo {author}
  {\bibfnamefont{J.}~\bibnamefont{Tolle}}, \bibinfo {author}
  {\bibfnamefont{H.~Z.}\ \bibnamefont{Ye}}, \bibinfo {author}
  {\bibfnamefont{H.}~\bibnamefont{Zhai}}, \bibinfo {author}
  {\bibfnamefont{R.}~\bibnamefont{Kim}}, \bibinfo {author}
  {\bibfnamefont{X.}~\bibnamefont{Zhang}}, \bibinfo {author}
  {\bibfnamefont{L.}~\bibnamefont{Lin}}, \bibinfo {author}
  {\bibfnamefont{T.~C.}\ \bibnamefont{Berkelbach}},\ and\ \bibinfo {author}
  {\bibfnamefont{G.~K.}\ \bibnamefont{Chan}},\ }%
  \enquote{\bibinfo {title} {{\it Variational Benchmarks for Quantum Many-Body
  Problems}},}\  (\bibinfo {year} {2023}),\
  \Eprint{http://arxiv.org/abs/2306.16561}{arXiv:2306.16561 [cond-mat.str-el]}%
  \bibAnnoteFile{NoStop}{Cui2023}%
\end{thebibliography}
\end{document}